\catcode`\@=11
\newif\if@fewtab\@fewtabtrue

{\count255=\time\divide\count255 by 60
\xdef\hourmin{\number\count255}
\multiply\count255 by-60\advance\count255 by\time
\xdef\hourmin{\hourmin:\ifnum\count255<10 0\fi\the\count255}}
\def\ps@draft{\let\@mkboth\@gobbletwo
    \def\@oddhead{}
    \def\@oddfoot
       {\hbox to 7 cm{$\scriptstyle Draft\ version:\ \draftdate$
       \hfil}\hskip -7cm\hfil\rm\thepage \hfil}
    \def\@evenhead{}\let\@evenfoot\@oddfoot}

\def\ceqno{\global\@fewtabfalse
    \ifcase\@eqcnt \def\@tempa{& & &}\or \def\@tempa{& &}
      \or \def\@tempa{&}
      \or\def\@tempa{}\fi\@tempa
{\rm(\theequation)}}

\def\aeqno#1{\global\@fewtabfalse
    \ifcase\@eqcnt \def\@tempa{& & &}\or \def\@tempa{& &}
      \or \def\@tempa{&}
      \or\def\@tempa{}\fi\@tempa
{\rm(\theequation,#1)}}

\def\label#1{\ifnum\draftcontrol=1
 \global\def\draftnote{$\scriptstyle #1$}\fi
 \@bsphack\if@filesw {\let\thepage\relax
   \def\protect{\noexpand\noexpand\noexpand}%
\xdef\@gtempa{\write\@auxout{\string
      \newlabel{#1}{{\@currentlabel}{\thepage}}}}}\@gtempa
   \if@nobreak \ifvmode\nobreak\fi\fi\fi
  \@esphack}

\def\alabel#1#2{\label{#1}\global\@fewtabfalse
    \ifcase\@eqcnt \def\@tempa{& & &}\or \def\@tempa{& &}
      \or \def\@tempa{&}
      \or\def\@tempa{}\fi\@tempa
{\hbox to 3cm{\phantom{\rm(\theequation,#2)}
\draftnote \hfil}\hskip -3cm {\rm(\theequation,#2)}}}

\def\clabel#1{\label{#1}\global\@fewtabfalse
    \ifcase\@eqcnt \def\@tempa{& & &}\or \def\@tempa{& &}
      \or \def\@tempa{&}
      \or\def\@tempa{}\fi\@tempa
{\hbox to 3cm{\phantom{\rm(\theequation)}
\draftnote \hfil}\hskip -3cm{\rm(\theequation)}}}

\def\eqnarray{\def\draftnote{{}}\global\@fewtabtrue
\stepcounter{equation}\let\@currentlabel=\theequation
\global\@eqnswtrue
\global\@eqcnt\z@\tabskip\@centering\let\\=\@eqncr
$$\halign to \displaywidth\bgroup\@eqnsel\hskip\@centering\@eqcnt\z@
  $\displaystyle\tabskip\z@{##}$&\global\@eqcnt\@ne
  \hskip 1\arraycolsep \hfil${##}$\hfil
  &\global\@eqcnt\tw@ \hskip 1\arraycolsep
$\displaystyle\tabskip\z@{##}$
\hfil  \tabskip\@centering&\global\@eqcnt\thr@@\llap{##}\tabskip\z@
\cr}

\def\endeqnarray{\@@eqncr\egroup
      \global\advance\c@equation\m@ne$$\global\@ignoretrue}

\def\@eqnnum{\hbox to 3cm{\phantom{\rm(\theequation)} \draftnote
                         \hfil}\hskip -3cm {\rm(\theequation)}}

\def\@@eqncr{\let\@tempa\relax
    \ifcase\@eqcnt \def\@tempa{& & &}\or \def\@tempa{& &}
      \or \def\@tempa{&}
      \or\def\@tempa{}
\fi\@tempa
\if@eqnsw
\if@fewtab\@eqnnum\fi
\stepcounter{equation}\fi\global
\@eqnswtrue\global\@eqcnt\z@\global\@fewtabtrue\cr}

\def\draftcite#1{\ifnum\draftcontrol=1#1\else{}\fi}

\def\@lbibitem[#1]#2{\item{}\hskip -3cm \hbox to 2cm
{\hfil$\scriptstyle\draftcite{#2}$}\hskip
1cm[\@biblabel{#1}]\if@filesw
     {\def\protect##1{\string ##1\space}\immediate
      \write\@auxout{\string\bibcite{#2}{#1}}}\fi\ignorespaces}

\def\@bibitem#1{\item\hskip -3cm \hbox to 2cm
{\hfil $\scriptstyle\draftcite{#1}$}\hskip 1cm
\if@filesw \immediate\write\@auxout
       {\string\bibcite{#1}{\the\value{\@listctr}}}\fi\ignorespaces}

     \def\nsection#1{\section{#1}\setcounter{equation}{0}}

\newfam\dlfam  
\newfam\glfam  

\def\draftdate{\number\month/\number\day/\number\year\ \ \ \hourmin }
 \global\def\draftcontrol{0}
\catcode`\@=12 \def\tilde{\widetilde} \def\hat{\widehat}
\documentstyle[12pt,epsf]{article}
\def\theequation{{\thesection.\arabic{equation}}}

\newcommand{\be}{\begin{eqnarray}} \newcommand{\en}{\end{eqnarray}\vs
  0.5 cm} \newcommand{\non}{\nonumber} 
\newcommand{\vs}{\vskip} \newcommand{\hs}{\hspace}
 
 \newcommand{\un}{\underline}
\newcommand{\ov}{\overline}
\newcommand{\NR}{{{\bf R}}}
\newcommand{\NC}{{{\bf C}}}
\newcommand{\NT}{{{\bf T}}}
\newcommand{\NZ}{{{\bf Z}}}
\newcommand{\NW}{{{\bf W}}}
 
\newcommand{\Ng}{{{\bf g}}} \newcommand{\Ntt}{{{\bf t}}}
\newcommand{\Nh}{{{\bf h}}}

\newcommand{\Nt}{{{\bf t}}}

\newcommand{\qq}{\begin{eqnarray}} \newcommand{\de}{\bar\partial}
  \newcommand{\da}{\partial} \newcommand{\ee}{{\rm e}}
   \newcommand{\qqq}{\end{eqnarray}}

 \newcommand{\tr}{\hbox{tr}}
 
 \newcommand{\CA}{{\cal A}}
 \newcommand{\CC}{{\cal C}}
\newcommand{\CD}{{\cal D}} 
 \newcommand{\CG}{{\cal G}}
\newcommand{\CH}{{\cal H}} \newcommand{\CI}{{\cal I}}
 
\newcommand{\CL}{{\cal L}} 
\newcommand{\CN}{{\cal N}} \newcommand{\CO}{{\cal O}}
\newcommand{\CP}{{\cal P}} 
\newcommand{\CR}{{\cal R}} 
\newcommand{\CT}{{\cal T}} \newcommand{\CU}{{\cal U}}
 \newcommand{\CW}{{\cal W}}
\newcommand{\CX}{{\cal X}} 
\newcommand{\CZ}{{\cal Z}} \newcommand{\s}{\hspace{0.05cm}}
\newcommand{\m}{\hspace{0.025cm}} \newcommand{\ch}{{\rm ch}}
 
\newcommand{\hf}{{_1\over^2}}

\begin{document}
\title{\bf{Conformal field theory:\\ 
a case study}}

\author{ \\Krzysztof Gaw\c{e}dzki\\C.N.R.S., I.H.E.S.,
  91440 Bures-sur-Yvette, France\\ }

\date{ } 
\maketitle
\vskip 0.2cm
\begin{abstract}
\vskip 0.2cm
\noindent This is a set of introductory lecture notes 
devoted to the Wess-Zumino-Witten model of two-dimensional 
conformal field theory. We review the construction 
of the exact solution of the model from the functional 
integral point of view. The boundary version of the theory
is also briefly discussed. 
\vskip 0.3cm

\hfill
\end{abstract}



\vskip 1cm

\nsection{Introduction}

Quantum field theory is a structure at the root of our
understanding of physical world from the subnuclear scales
to the astrophysical and the cosmological ones. The concept 
of a quantum field is very rich and still poorly
understood although much progress have been achieved
over some 70 years of its history. The main problem
is that, among various formulations of quantum field theory 
it is still the original Lagrangian approach which is by far 
the most insightful, but it is also the least precise way to 
talk about quantum fields. The strong point of the Lagrangian 
approach is that it is rooted in the classical theory.
As such, it permits a perturbative analysis of the field 
theory in powers of the Planck constant and also captures 
some semi-classical non-perturbative effects (solitons, 
instantons). On the other hand, however, it masks genuinely 
non-perturbative effects. In the quest for a deeper 
understanding of quantum field theory 
an important role has been played by two dimensional models. 
Much of what we have learned about nonperturbative phenomena 
in quantum field theory has its origin in such models.
One could cite the Thirring model with its anomalous
dimensions and the fermion-boson equivalence to the
sine-Gordon model, the Schwinger model with the confinement 
of electric charge, the non-linear sigma model with the
non-perturbative mass generation, and so on. 
\vskip 0.3cm

The two-dimensional models exhibiting conformal invariance 
have played a specially important role. On one side, 
they are not without direct physical importance, describing, 
in their Euclidean versions, the long-distance behavior 
of the two-dimensional statistical-mechanical systems, 
like the Ising or the Potts models, at the second order phase 
transitions. On the other hand, the (quantum) conformal field 
theory (CFT) models constitute the essential building blocks 
of the classical vacua of string theory, a candidate ``theory
of everything'', including quantum gravity. The two-dimensional
space-time plays simply the role of a string world sheet
parametrizing the string evolution, similarly as the 
one-dimensional time axis plays the role of a world line 
of point particles. The recent developments seem to indicate 
that string theory, or what will eventually 
emerge from it, provides the appropriate language to talk
about general quantum fields, whence the central place
of two-dimensional CFT in the quantum field theory edifice.
\vskip 0.3cm

Due to the infinite-dimensional nature of the conformal symmetry 
in two space-time dimensions, the two-dimensional models of CFT 
lend themselves to a genuinely non-perturbative approach 
based on the infinite symmetries and the concept of the operator 
product expansion \cite{3}. In the present lectures, we shall 
discuss a specific model of two-dimensional CFT, the so called 
Wess-Zumino-(Novikov)-Witten model (WZW) \cite{4}\cite{5}\cite{6}. 
It is an example of a non-linear sigma model 
with the classical fields on the space-time taking values 
in a manifold which for the WZW model is taken as a group 
manifold of a compact Lie group $G$. We shall root our
treatment in the Lagrangian approach and will work slowly
our way towards a non-perturbative formulation. 
This will, hopefully, provide a better understanding of
the emerging structure which, to a large extent, is common 
to all CFT models. In fact, the WZW theory is a prototype
of general (rational) CFT models which may be obtained 
from the WZW one by different variants of the so called 
coset construction. In view of the stringy applications, 
where the perturbation expansion is built by considering 
two-dimensional conformal theories on surfaces of arbitrary 
topology, we shall define and study the WZW model on 
a general Riemann surface.
\vskip 0.3cm

A word of warning is due to a more advanced audience.
The purpose of these notes is not to present a complete 
up-to-date account of the WZW theory, even less of CFT. 
That would largely overpass the scope of a summer-school
lecture notes. As a result, we limit ourselves to 
the simplest version of the model leaving completely 
aside the ramifications involving models with non-simply 
connected groups, orbifolds, etc, as well as applications
to string theory. We profit, however, from this simple example 
to introduce on the way some of the main concepts 
of two-dimensional CFT.
Much of the material presented is not new, 
even old, by the time-scale standard of the subject, with 
the possible exception of the last section devoted to 
the boundary WZW models. The author still hopes that 
the following exposition, which he failed to present 
at the 1998 Istanbul summer school, may be useful to 
a young reader starting in the field. 
\vskip 0.3cm

The notes are organized as follows. In Sect.\,\,2, 
we discuss a simple quantum-mechanical version of the WZW
model: the quantum particle on a group manifold.
This simple model, exactly solvable by harmonic analysis
on the group, permits to describe many structures similar 
to the ones present in the two-dimensional theory
and to understand better the origin of those.
Sect.\,\,3 is devoted to the definition of the action
functional of the WZW model. The action contains
a topological term, which requires a special treatment.
We discuss separately the case of the surfaces without
and with boundary, in the latter case postponing
the discussion of local boundary conditions
to the last section. In Sect.\,\,4, we introduce
the basic objects of the (Euclidean) quantum WZW theory:
the quantum amplitudes taking values in the 
spaces of states of the theory and the correlation
functions. We state the infinite-dimensional symmetry 
properties of the theory related to the chiral gauge
transformations and to the conformal transformations. 
The symmetries give rise to the action of the two copies 
of the current and Virasoro algebras in the Hilbert 
space of states of the theory constructed with the
use of the representation theory of those algebras.
We discuss briefly the operator product expansions
which encode the symmetry properties of the correlation 
functions. Sect.\,\,5 is devoted to the relation
between the WZW theory and the Schr\"{o}dinger picture 
quantum states of the topological three-dimensional 
Chern-Simons theory. The relation is established
via the Ward identities expressing the behavior 
of the WZW correlation functions, coupled to external 
gauge field, under the chiral gauge transformations.
We discuss the structure of the spaces of the Chern-Simons                                             
states, the fusion ring giving rise to the Verlinde 
formula for their dimensions and their Hilbert-space 
scalar product, as well as the Knizhnik-Zamolodchikov
connection which permits to compare the states
for different complex structures.
In particular, we explain how the knowledge of
the scalar product of the Chern-Simons states permits 
to obtain exact expressions for the correlation 
functions of the WZW theory. 
In Sect.\,\,6 we give a brief account of the coset 
construction of a large family of CFT models which may be 
solved exactly, given the exact solution of the WZW
model. Finally, Sect.\,\,7 is devoted to the WZW theory 
with local boundary conditions. Again, for the sake
of simplicity, we restrict ourselves to a simple family
of the conditions that do not break the infinite-dimensional
symmetries of the theory. We discuss how to define the
action functional of the model in the presence of such  
boundary conditions and what are the elementary properties
of the corresponding spaces of states, quantum amplitudes
and correlation functions.

\nsection{Quantum mechanics of a particle
on a group}
\subsection{The geodesic flow on a group}

Non-linear sigma models describe field theories with fields 
taking values in manifolds. These lectures will be devoted
to a special type of sigma models, known under the name 
of Wess-Zumino(-Novikov)-Witten (or WZW) models. They are 
prototypes of conformal field theories in two-dimensional
space-time. As such they play a role in string theory 
whose classical solutions are built out of two-dimensional 
quantum conformal field theory models by a cohomological
construction. Before we plunge, however, into the details 
of the WZW theory, we shall discuss a simpler but largely 
parallel model in one dimension, i.e. in the domain of mechanics 
rather than of field theory.
\vskip 0.3cm

One-dimensional sigma models describe the geodesic flows 
on manifolds $M$ endowed with a Riemannian or a pseudo-Riemannian 
metric $\gamma$. The classical action for the trajectory $[0,T]\ni
\,\mapsto\,x(t)\in M$ of such a system is
\qq
S(x)\ =\ \hf\int_{_0}^{^T}\hspace{-0.1cm}
\gamma_{\mu\nu}(x)\,
{_{dx^\mu}\over^{dt}}\,{_{dx^\nu}\over^{dt}}\,\m dt
\label{act}
\qqq
and the classical solutions $\delta S=0$ correspond to
the geodesic curves in $M$ parametrized by a rescaled length.
If $M=\NR^n$, for example, and $\gamma_{\mu\nu}=m\,\delta_{\mu\nu}$,
we obtain the action of the free, non-relativistic particle
of mass $m$ undergoing linear classical motions 
$x(t)=x_0+{p\over m}\m t$. 
The action (\ref{act}) is not parametrization-invariant but it
may be viewed as a gauged-fixed version of the action
\qq
S_{_P}(x)\ =\ \hf\int_{_0}^{^T}\hspace{-0.1cm}
\gamma_{\mu\nu}(x)\,
{_{dx^\mu}\over^{dt}}\,{_{dx^\nu}\over^{dt}}\,\eta^{-\hf}\,dt\ +\ 
\hf\int_{_0}^{^T}\hspace{-0.1cm}\eta^\hf\, dt\,,
\non
\qqq
where the reparametrization invariance is restored by
coupling the system {\it \`{a} la} Polyakov to the metric
$\eta(t)\m (dt)^2$ on the word-line of the particle. The
$\eta=1$ gauge reproduces then the action (\ref{act}),
whereas extremizing over $\eta$, one obtains the relativistic
action 
\qq
S_r(x)\ =\ \int_{_0}^{^T}\hspace{-0.07cm}
\left(\gamma_{\mu\nu}(x)\,
{_{dx^\mu}\over^{dt}}\,{_{dx^\nu}\over^{dt}}\right)^{\hf}dt
\non
\qqq
given by the geodesic length of the trajectory.
\vskip 0.3cm

The exact solvability of the geodesic equations can be
achieved in sufficiently symmetric situations. In particular,
we shall be interested in the case when $M$ is a manifold
of a compact Lie group $G$ and when $\gamma$ is a left-right 
invariant metric on $G$ given by ${k\over 2}$ times a positive 
bilinear $ad$-invariant form $\tr\m(XY)$ 
on the Lie algebra\footnote{we use 
the physicists' convention in which the exponential map between
the Lie algebra and the group is $X\mapsto\ee^{iX}$} $\Ng$.
For matrix algebras, as the algebra $su(N)$ of the hermitian
$n\times n$ traceless matrices, the form is given by the 
matrix trace in the defining representation, hence
the notation. The positive constant $k$ will play the role 
of a coupling constant.
The action (\ref{act}) may be then rewritten as
\qq
S(g)\ =\ -\m{_k\over^4}\int_{_0}^{^T}\hspace{-0.1cm}
\tr\,\, (g^{-1}{_d\over^{dt}}\m g)^2\m dt\m.
\non
\qqq
The variation of the action under the infinitesimal
change of $g$ vanishing on the boundary is
\qq
\delta S(g)\ =\ {_k\over^2}\int_{_0}^{^T}\hspace{-0.1cm}
\tr\,\,(g^{-1}\delta g)\,\,{_d\over^{dt}}(g^{-1}
{_d\over^{dt}}\m g))\,\m dt\,.
\non
\qqq
Consequently, the classical trajectories are solutions of the
equations
\qq
{_d\over^{dt}}(g^{-1}{_d\over^{dt}}\m g)\ =\ 0\m.
\label{clequ}
\qqq
The case $G=\NT^n$, where $\NT^n=\NR^n/\NZ^n$ is the n-dimensional
torus, is the prototype of an integrable system whose
trajectories are periodic or quasiperiodic motions with the angles 
evolving linearly in time. The case $G=SO(3)$ corresponds to the 
symmetric top whose positions are parametrized by rigid rotations. 
The classical trajectories solving Eq.\,\,(\ref{clequ})
have a simple form:
\qq
g(t)=g_\ell\,\m\ee^{it \lambda/k}\m g_r^{-1}\m,
\label{alts}
\qqq
where $g_{\ell,r}$ are fixed elements in $G$ and $\lambda$
may be taken in the Cartan subalgebra $\Ntt\subset\Ng$.
For the later convenience, we have introduced the factor
${1\over k}$ in the exponential.
\vskip 0.3cm

The space $\CP$ of classical solutions forms the phase space 
of the system. It may be parametrized by the initial data 
$(g(0),\m p(0))$ where the momentum\footnote{we identify $\Ng$
with its dual using the bilinear form $\m\tr\m(XY)$} 
$\m p(t)={k\over{2i}}\m{d\over{dt}}\m g$. \m As usually, 
the phase space $\CP$ comes equipped with the symplectic form
\qq
\Omega\ =\ {_1\over^i}\, d\,\tr\,(p\,g^{-1}dg)\m,
\qqq
where the right hand side may be calculated at any instance 
of time giving a result independent of time. The symplectic 
structure on $\CP$ allows to associate the vector fields $\CX_f$ 
to functions $f$ on $\CP$ by the relation $\,-\m df=\iota_{\CX_f}
\Omega\m$, \,where $\iota_{X}\alpha$ denotes the contraction 
of a vector field $\CX$ with a differential form $\alpha$. 
These are the Hamiltonian vector fields that preserve 
the symplectic form: 
$\,\CL_{X_f}\Omega=0\m,$ \,where $\CL_{\CX}$ is the Lie 
derivative that acts on differential forms by $\,\CL_{\CX}\alpha
=\iota_{\CX}d\alpha+d\iota_{\CX}\alpha\m$. The Poisson bracket 
of functions is defined by: $\{f,g\}=\CX_f(g)$. In particular, 
the time evolution is induced by the vector field associated 
with the classical Hamiltonian
\qq
h\ =\ {_1\over^k}\,\tr\,\, p^2\ =\ -\m{_k\over^4}\,\tr\,(g^{-1}
\hspace{-0.05cm}{_d\over^{dt}}\m g)^2
\non
\qqq
which stays constant during the evolution.
In the alternative way (\ref{alts}) to parametrize the solutions,
$\,h={1\over{4k}}\m\tr\m\lambda^2\,$ and the symplectic
structure splits:
\qq
\Omega\ =\ \Omega_\ell\m-\,\Omega_r\,,\quad\ \ {\rm where}
\quad\ \Omega_\ell\ =\ {_{i}\over^2}\,\tr\,
[\m\lambda\m(g_\ell^{-1}dg_\ell)^2
\m-\m d\lambda\, g_\ell^{-1}dg_\ell\m]
\label{clsp}
\qqq
and $\Omega_r$ is given by the same formula with the subscript 
$\ell$ replaced by $r$. 
\vskip 0.3cm

There are two commuting actions of the group $G$
on $\CP$: from the left $g(t)\mapsto g_0g(t)$ and from
the right $g(t)\mapsto g(t)g_0^{-1}$. Both preserve
the symplectic structure and the Hamiltonian $h$.
The vector fields corresponding to the left and right actions 
of the infinitesimal generators $t^a\in\Ng$ are induced
by the functions
\qq
j^a&=&{_{ki}\over^2}\,\tr\,(t^a\m g\m{_d\over^{dt}}g^{-1})\,=\,
\hf\,\tr\,(t^a\m g_\ell\lambda\m g_\ell^{-1})\cr
\tilde j^a&=&{_{ki}\over^2}\,\tr\,(t^a\m 
g^{-1}{_d\over^{dt}}\m g)\,=\,-\m\hf\,\tr\,(t^a\m g_r
\lambda\m g_r^{-1}),
\non
\qqq
respectively. Note that, if we normalize $t^a$'s so that
$\tr\,(t^a t^b)=\hf\m\delta^{ab}$, then
\qq
h\ =\ {_2\over^k}\, j^aj^a\ =\ {_2\over^k}\,\tilde j^a\tilde j^a
\non
\qqq
(summation convention!). The symplectic form $\Omega_\ell$
gives, for $\lambda$ fixed, the canonical symplectic form 
on the (co)adjoint orbit $\,\{\m g_\ell\lambda g_\ell^{-1}\,|
\,g_\ell\in G\m\}\,$ 
passing through $\lambda$. The left action of the group is 
$\,g_\ell\mapsto g_0g_\ell\,$ so that it coincides with the 
(co)adjoint action on the orbit. As is well known, upon geometric 
quantization of the coadjoint orbits for appropriate $\lambda$, 
this action gives rise to irreducible representations of $G$.
\vskip 0.4cm

\subsection{The quantization}

The geodesic motion on a group is easy to quantize.
As the Hilbert space $\CH$ one takes the space 
$L^2(G,dg)$ of functions on $G$ square integrable 
with respect to the normalized Haar measure $dg$.
The two commuting actions of $G$ in $\CH$:
\qq
f\,\mapsto\,{}^{^{h}}\hspace{-0.14cm}f\m=\m f(h^{-1}\m\cdot\,)\m,
\qquad f\,\mapsto\,f^{^{h}}\m=\m f(\,\cdot\,h)\m,
\non
\qqq
give rise to the actions 
\qq
J^af\ =\ {_1\over^i}\,{_d\over^{d\epsilon}}\bigg\vert_{_{\epsilon=0}}
{}^{^{\ee^{i\epsilon t^a}}}\hspace{-0.14cm}f\,,\qquad
\tilde J^af\ =\ {_1\over^i}\,{_d\over^{d\epsilon}}
\bigg\vert_{_{\epsilon=0}}f^{^{\hspace{0.07cm}\ee^{i\epsilon t^a}}}
\non
\qqq
of the infinitesimal generators $t^a$ of $\Ng$.
The commutation relations
\qq
[J^a,\m J^b]\ =\ i\m f^{abc}\, J^c\m\qquad
[\tilde J^a,\m\tilde J^b]\ =\ i\m f^{abc}\,\tilde J^c\m,
\non
\qqq
reflect the relation $[t^a,\m t^b]=i\m f^{abc}\m t^c$
in the Lie algebra $\Ng$. The quantum Hamiltonian
\qq
H\ =\ {_2\over^k}\, J^aJ^a\ =\ {_2\over^k}\,\tilde J^a\tilde J^a
\non
\qqq
coincides with $-\m{2\over k}\m$ times the Laplace-Beltrami
operator on $G$ and is a positive self-adjoint operator. 
\vskip 0.3cm

The irreducible representations $R$ of the compact Lie group $G$
are finite dimensional and are necessarily unitarizable
so that we may assume that they act in finite-dimensional
vector spaces $V_{_R}$ preserving their scalar product.
We shall denote by $g_{_R}$ and $X_{_R}$ the endomorphisms 
of $V_{_R}$ representing $g\in G$ and $X\in\Ng$. 
Up to isomorphism, the irreducible representation 
of $G$ may be characterized by their {\bf highest weights}. 
Let us recall what this means. 
The complexified Lie algebra may be decomposed into 
the eigenspaces of the adjoint action of its Cartan subalgebra 
$\Ntt$ as
\qq
\Ng^\NC\ =\ \Ntt^\NC\oplus(\mathop{\oplus}\limits_{\alpha}\NC\m
e_{\alpha})
\qqq
where \,$[X,e_\alpha]\m=\m\tr\,(\alpha\m X)\, e_\alpha\,$
for all $X\in\Ntt$. The set of the roots $\alpha\in\Ntt$
may be divided into the positive roots and their negatives.
We shall normalize the invariant form $\tr$ on $\Ng$
so that the long roots have the length squared 2 (this 
agrees with the normalization of the matrix trace for
$\Ng=su(N)$). The ``step generators'' $e_{\pm\alpha}$ may 
be chosen so that $[e_{\alpha},e_{-\alpha}]$ is equal to the
coroot \,$\alpha^\vee\equiv{2\m\alpha\over\hbox{tr}\,\alpha^2}\,$
corresponding to $\alpha$. The elements $\lambda\in\Ntt$ such
that $\tr\,(\alpha^\vee\lambda)$ is integer for all roots
are called weights. A non-zero vector $v\in V_{_R}$ 
(unique up to normalization) is called a highest weight (HW) 
vector if it is an eigen-vector of the action of the Cartan
algebra: $\,X_{_R}v=\tr\,(\lambda_{_R}X)\, v$ and if 
$(e_\alpha)_{_R}v=0$ for all positive roots $\alpha$.
The element $\lambda_{_R}$ of the Cartan algebra, a weight,
is called the highest weight (HW) of the representation $R$ 
and it determines completely $R$. All weights $\lambda\in\Ntt$ 
such that $\,\tr\,(\alpha^\vee\lambda)$ is a non-negative integer 
for each positive $\alpha$ appear as HW's of irreducible 
representations\footnote{such $\lambda$ are usually called 
dominant weights} of $G$. The representations R may be obtained
by the geometric quantization of the (co)adjoint orbit
passing through $\lambda_{_R}$. \,For the $su(2)$ Lie algebra 
spanned by the Pauli matrices 
$\sigma_i$, one usually takes $\sigma_3$ as the positive root
and the matrices $\sigma_\pm=\hf\m(\sigma_1\pm i\m\sigma_2)$
as the corresponding step generators. 
The HW's are of the form $j\sigma_3$ with $j=0,\hf,1,\dots$
called the {\bf spin} of the representation.
\vskip 0.3cm

With respect to the left-right action of $G\times G$,
the Hilbert space $L^2(G,dg)$ decomposes as
\qq
\CH\ \cong\ \mathop{\oplus}\limits_{R}\,V_{_R}\otimes V_{_{\ov{R}}}\,,
\label{decomp}
\qqq
where the (infinite) sum is over the (equivalence classes of)
irreducible representations of $G$ and $\overline{R}$ 
denotes the representation complex-conjugate to $R$, 
i.e. $V_{_{\overline{R}}}=\overline{V}_{_R}$ and
$g_{_{\ov{R}}}=\ov{g}_{_R}$. 
\,Recall that the complex conjugate vector space $\ov{V}$
is composed of the vectors $v\in V$, denoted for distinction
by $\ov{v}$, with the multiplication by scalars defined by
$\mu\m\ov{v}=\ov{\bar{\mu} v}$. A linear transformation
$A$ of $V$, when viewed as a transformation $\ov{A}$ of $\ov{V}$, 
is still linear. \,The above factorization of the
Hilbert space reflects the classical splitting (\ref{clsp}).
Let $(g_{_R}^{ij})$ be the (unitary) 
matrix of the endomorphism $g_{_R}$ with respect 
to a fixed orthonormal bases $(e_{_R}^i)$ in $V_{_R}$. 
The decomposition (\ref{decomp}) is given by the assignment
\qq
V_{_R}\otimes V_{_{\overline{R}}}\,\ni\,
e_{_R}^i\otimes\ov{e_{_R}^j}\ \ \ \mapsto\ \ \ d_{_R}^{\,\hf}\, 
\,\ov{g_{_R}^{ij}}\,\in\,L^2(G,dg)\,.
\label{hwq}
\qqq
The Schur orthogonality 
relations
\qq
\int_{_G}\overline{g_{_{R'}}^{ij}}\,\,g_{_R}^{rs}
\,\,dg\ =\ {_1\over^{d_{_R}}}\,\delta_{_{R'R}}\,\,\delta^{ir}\, 
\delta^{js} 
\non
\qqq
assure that this assignment preserves the scalar product. 
The matrix elements $g^{ij}_{_R}$ span a dense subspace
in $L^2(G,dg)$. 
\vskip 0.3cm

The function on $G$ invariant under the adjoint 
action $g\m\mapsto\m Ad_{g_0}(g)\equiv g_0\, g\, g_0^{-1}$
are called class functions. They are constant on the conjugacy 
classes 
\qq
\CC_{\lambda}=\{\m g_0\,\ee^{\m2\pi i\m \lambda/k}
g_0^{-1}\,\vert\,g_0\in G\m\}
\qqq
with $\lambda$ in the Cartan algebra $\Ntt$. 
The characters $\m\chi_{_R}(g) =\tr_{_{V_{_{_R}}}}g_{_R}\m$ of the 
irreducible representations $R$ are class functions. The Schur 
relations imply that
\qq
\int_{_G}\overline{\chi_{_{R'}}(g)}\,\,\chi_{_R}(g)\,\, dg\ 
=\ \delta_{_{R' R}}\m.
\non
\qqq
The class functions in $L^2(G,dg)$ form a closed  
subspace and the characters $\chi_{_R}$ form
an orthonormal bases of it. 
Note that under the isomorphism (\ref{decomp}), 
\qq
\ov{\chi_{_R}}\ \ \cong\ \ d_{_R}^{\m-\hf}\,
e_{_R}^i\otimes\ov{e_{_R}^i}
\label{chdec}
\qqq
(sum over $i$\,!).
\vskip 0.3cm

The Hamiltonian $H$ becomes diagonal in the decomposition 
(\ref{decomp}) of the Hilbert space. It acts on 
$V_{_R}\otimes V_{_{\ov{R}}}$ 
as the multiplication by ${2\over k}\m c_{_R}$ where $c_{_R}$ 
is the value of the quadratic Casimir $c=t^at^a$ in the 
representation $R$. In terms of the HW's, $c_{_R}=\hf\,\tr\,
(\lambda_{_R}(\lambda_{_R}+2\rho))\m$, \,where $\rho$, 
the {\bf Weyl vector}, 
is equal to half the sum of the positive roots. 
The Hamiltonian generates a 1-parameter family of
unitary transformation $\ee^{it\m H}$ describing the time
evolution of the quantum system. In the Euclidean spirit,
we shall be more interested, however, in the semigroup
of the thermal density matrices $\ee^{-\m\beta\m H}$
obtained by the Wick rotation of time $\beta=it\geq 0$. 
Their (heat) kernels are given by:
\qq
\ee^{-\m \beta\m H}(g_0,g_1)\ =\ \sum\limits_R\m d_{_R}\,\m
\ee^{-{_2\over^k}\m \beta\m c_{_R}}\,\,\chi_{_R}(g_0g_1^{-1})\m.
\non
\qqq
In particular, at $\beta=0$, we obtain a representation for
the delta-function concentrated at an alement $g_0\in G$:
\qq
\delta_{g_0}(g_1)\ =\ \sum\limits_R\m d_{_R}\,
\chi_{_R}(g_0g_1^{-1})\m.
\non
\qqq
We shall also need below the delta-functions concentrated on 
the conjugacy classes $\CC_{\lambda}$. They may be obtained
by smearing the delta-function $\delta_{g}$ over $\CC_\lambda\m$:
\qq
&&\delta_{_{\CC_{\lambda}}}\hspace{-0.07cm}(g_1)
\ =\ \int_{_G}\hspace{-0.05cm}
\delta_{g_0\,\ee^{\m2\pi i\m\lambda/k}g_0^{-1}}(g_1)\,\, dg_0
\ =\ \sum\limits_Rd_{_R}\int_{_G}\hspace{-0.05cm}
\chi_{_R}(g_0\,\ee^{\m2\pi i\m\lambda/k}g_0^{-1}g_1^{-1})\,\, 
dg_0\cr
&&=\ \sum\limits_Rd_{_R}\int_{_G}\hspace{-0.05cm}
(g_0)^{^{ij}}_{_R}\,(\ee^{\m2\pi i\m\lambda/k})^{^{j\ell}}_{_R}
\,\ov{(g_0)^{^{n\ell}}_{_R}}\,\m\ov{(g_1)^{^{in}}_{_R}}\,\,dg
\ =\ \sum\limits_R\chi_{_R}(\ee^{\m2\pi i\m\lambda/k})\,\m
\ov{\chi_{_R}(g_1)}\,,
\non
\qqq
where we have used the Schur relations.
It follows from the correspondence (\ref{chdec})
that in the language of the isomorphism (\ref{decomp}),
\qq
\delta_{_{\CC_{\lambda}}}\ \cong\ \sum\limits_R 
{\chi_{_R}(1)^{-\hf}}\,\,{\chi_{_R}(\ee^{\m 2\pi 
i\m\lambda/k})}\,\,e^{^i}_{_R}\otimes\ov{e^{^i}_{_R}}\,.
\label{Ishcl}
\qqq
More exactly, $\delta_{_{\CC_{\lambda}}}$ is not a normalizable
state in $\CH$ but it defines an antilinear functional 
on a dense subspace in $\CH$, e.g.\,\,the one of vectors 
with a finite number of components in the decomposition 
(\ref{decomp}).
\vskip 0.3cm

The delta-functions $\delta_{_{\CC_{\lambda}}}$ may be
used to disintegrate the Haar measure $dg$ into the
measures along the conjugacy classes and over the
set of different conjugacy classes:
\qq
dg\,=\,{_1\over^{\vert T\vert}}\,
\vert\Pi(\ee^{\m 2\pi i\m\lambda/k})\vert^2\,\m 
\delta_{_{\CC_\lambda}}(g)\m\,d\lambda\, dg\,.
\qqq
We shall choose the measure $d\lambda$ such
that it corresponds to the normalized Haar measure on
the Cartan group $T\subset G$ under the exponential map
$\m\lambda\mapsto\ee^{\m 2\pi i\m\lambda/k}$. Then
\qq
\Pi(\ee^{\m 2\pi i\m\lambda/k})=\prod\limits_{\alpha>0}(\ee^{\m\pi 
i\,\tr\,(\alpha\m\lambda)/k}-\m\ee^{-\pi i\,\tr\,(\alpha\m\lambda)
/k})
\label{WD}
\qqq
is the so called {\bf Weyl denominator}. In particular, for class 
functions constant on the conjugacy classes one obtains
the Weyl formula:
\qq
\int_{_G}\hspace{-0.07cm}f\,\m dg\ =\ 
\int f(\ee^{\m 2\pi i\m\lambda/k})\,\m\vert\Pi(\ee^{\m 2\pi i\m
\lambda/k})\vert^2\,\m d\lambda\,,
\label{WDcl}
\qqq
where each conjugacy class should be represented
ones. We shall employ this representation of the 
integral of class functions later.
\vskip 0.3cm

The Feynman-Kac formula allows to express the heat kernel
on the group as a path integral:
\qq
\ee^{-\m \beta\m H}(g_0,g_1)\ \ =\int\limits_{g:\m[0,\beta]\m
\rightarrow\m G\atop g(0)=g_0,\ g(\beta)=g_1}\hspace{-0.2cm}
\ee^{\m -S(g)}\,\,Dg\,,
\non
\qqq
where $Dg$ stands for the product of the Haar measures
$dg(t)$. The integral on the right hand side may be
given a rigorous meaning as the one with respect to
the Brownian bridge measure $dW_{g_0,g_1}(g)$ supported
by continuous paths in $G$.  
The path integral may be also used to define the
thermal {\bf correlation function}
\qq
<\prod\limits_ng^{^{i_nj_n}}_{_{R_n}}
(t_n)>_{_{\hspace{-0.08cm}\beta}}\ \ \equiv\ \ 
{{\int\prod\limits_{n=1}^N g^{^{i_nj_n}}_{_{R_n}}(t_n)
\,\,\ee^{-S(g)}\,\, Dg}\over 
{\int\ee^{-S(g)}\,\, Dg}}
\,,
\label{fkp}
\qqq
where on the right hand side one integrates over the 
periodic paths $g:[0,\beta]\rightarrow G$. Upon ordering
the (Euclidean) times $\m t_1\leq\dots\leq t_N\m$, 
the above path integral may be expressed in the operator 
language:
\qq
&&<\prod\limits_ng^{^{i_nj_n}}_{_{R_n}}
(t_n)>_{_{\hspace{-0.08cm}\beta}}\cr 
&&\hspace{0.7cm}=\ {{\tr_{_{\CH}}\,\left( 
\ee^{-t_1\CH}\m g^{^{i_1j_1}}_{_{R_1}}
\m\ee^{-(t_2-t_1)\CH}\m g^{^{i_2j_2}}_{_{R_2}}
\,\cdots\,\ee^{-(t_{_N}-t_{_{N-1}})\CH}
\m g^{^{i_{_N}j_{_N}}}_{_{R_N}}\m\ee^{-(\beta-t_{_N})\CH}
\right)}\over{\tr_{_\CH}\ \ee^{-\beta\m H}}}\,,
\hspace{0.7cm}
\label{corf}
\qqq
where the functions $g^{ij}_{_R}$ on $G$ are viewed 
as the multiplication operators in $L^2(G,dg)$.
\vskip 0.3cm

The right hand side of Eq.\,\,(\ref{corf}) may be calculated 
using harmonic analysis on $G$. Indeed, what is really needed
for such a computation are the matrix elements
\qq
(\m e_{_{R_1}}^{^{i_1}}\otimes
\ov{e_{_{R_1}}^{^{j_1}}}\m,
\,\m g_{_R}^{^{ij}}\,\, e_{_{R_2}}^{^{i_2}}
\otimes\ov{e_{_{R_2}}^{^{j_2}}}\m)
\ =\ d_{_{R_1}}^\hf d_{_{R_2}}^{\hf}
\int_{_G}\hspace{-0.07cm}
\ov{g_{_{R_2}}^{^{i_2j_2}}}\,\m g_{_{R_1}}^{^{i_1j_1}}
\,\m g_{_R}^{^{ij}}\,\, dg
\qqq
encoding the decomposition of the tensor product of 
the irreducible representations
\qq
V_{_R}\otimes V_{_{R_1}}\ \cong\ \mathop{\oplus}\limits_{R_2}
M_{_{R_1R}}^{^{R_2}}\otimes V_{_{R_2}}\,.
\label{tpd}
\qqq
In particular, the dimensions $N_{_{R_1R}}^{^{R_2}}$ 
of the multiplicity spaces $M_{_{R_1R}}^{^{R_2}}$
may be obtained from the traces of the above matrix elements:
\qq
N_{_{R_1R}}^{^{R_2}}&=&\int_{_G}\hspace{-0.07cm}
\ov{\chi_{_{R_2}}(g)}\,\m\chi_{_{R_1}}(g)\,
\m\chi_{_{R}}(g)\,\,dg\,.
\non
\qqq
The finite combinations with integer coefficients
$\sum n_i\chi_{_{R_i}}$ of characters of irreducible 
representations form a subring $\CR_{_G}$ in the commutative 
ring of class functions. The identity 
\qq
\chi_{_R}\,\chi_{_{R_1}}\ =\ \sum\limits_{R_2}
N_{_{R\, R_1}}^{^{R_2}}\,\chi_{_{R_2}}
\label{strf}
\qqq
shows that the integers $N_{_{R_1R}}^{^{R_2}}
=N_{_{R\, R_1}}^{^{R_2}}$  
play the role of structure constants of this ring.  
\vskip 0.3cm

One can define a version of the correlation functions
by replacing the integral
over the periodic paths $g:[0,\beta]\rightarrow G$
in Eq.\,\,(\ref{fkp})   
by the one over the paths constraint
to fixed conjugacy classes on the boundary
of the interval:
\qq
{{\int\limits_{g:[0,\beta]\rightarrow G}
\prod\limits_{n=1}^N g^{^{i_nj_n}}_{_{R_n}}(t_n) 
\,\,\delta_{_{\CC_{\lambda_1}}} 
\hspace{-0.07cm}(g(0))\,\,\delta_{_{\CC_{\lambda_2}}}
\hspace{-0.07cm}(g(\beta))\,\,\ee^{-S(g)}\,\,Dg}
\over
{\int\limits_{g:[0,\beta]\rightarrow G}
\delta_{_{\CC_{\lambda_1}}} 
\hspace{-0.07cm}(g(0))\,\,\delta_{_{\CC_{\lambda_2}}}
\hspace{-0.07cm}(g(\beta))\,\,\ee^{-S(g)}\,\,Dg}}
\ \equiv\ <\prod\limits_ng^{^{i_nj_n}}(t_n)
>_{_{\hspace{-0.08cm}\beta,
\lambda_1\lambda_2}}\,.\hspace{0.7cm}
\qqq
For $G=SU(2)\m=\m\{\m x_0+i x_i\sigma_i\,\vert\, 
x_0^2+x_i^2=1\}\cong S^3$, the conjugacy classes
are the 2-spheres with fixed $x_0$ so that
one integrates over the paths as in Fig.\,\,1.

\leavevmode\epsffile[-110 -20 310 193]{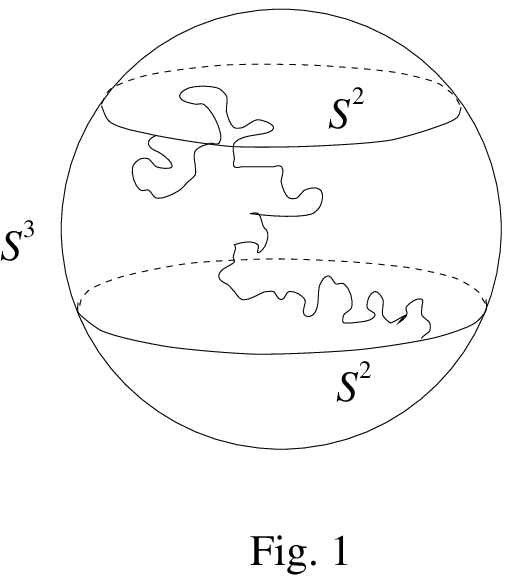}

The above functional integral may be rewritten 
in the operator language as
\qq
<\prod\limits_ng^{^{i_nj_n}}(t_n)>_{_{\hspace{-0.08cm}\beta,
\lambda_1\lambda_2}}
\,=\m{{\Big(\delta_{_{\CC_{\lambda_1}}}\m,\,\m
\ee^{-t_1\CH}\m g^{^{i_1j_1}}_{_{R_1}}
\,\cdots\,\ee^{-(t_{_N}-t_{_{N-1}})\CH}
\m g^{^{i_{_N}j_{_N}}}_{_{R_N}}\m\ee^{-(\beta-t_{N})\CH}\,\m
\delta_{_{\CC_{\lambda_2}}}\Big)}\over
{\Big(\delta_{_{\CC_{\lambda_1}}}\m,\,\m
\ee^{-\beta\m\CH}\,\m\delta_{_{\CC_{\lambda_2}}}\Big)}}
\,.\hspace{0.8cm}
\label{bdcf}
\qqq
Although $\m\delta_{_{\CC_{\lambda}}}$ are generalized 
functions rather than normalizable states in $L^2(G,dg)$, 
the matrix elements on the right hand side 
are finite and may again be computed by harmonic analysis 
on $G$. 
\vskip 0.3cm

We shall encounter field-theoretical generalization
of the above quantum-mechanical constructions below.
\vskip 0.6cm

\nsection{The WZW action}
\subsection{Two-dimensional sigma models}

The two-dimensional sigma models describe field theories
with fields mapping a surface $\Sigma$
to a target manifold $M$, both equipped with metric
structures. Such field configurations represent  
evolution of a string in the target $M$ with 
$\Sigma$ being the string world sheet. 
The (Euclidean) action functional of the field 
configuration $\m X:\Sigma\rightarrow M\,$ is
\qq
S^\gamma(X)\ =\ {_1\over^{4\pi}}\int_{_\Sigma}\hspace{-0.1cm}
\gamma_{\mu\nu}(X)\,\da_\alpha X^\mu\m\da_\beta X^\nu
\eta^{\alpha\beta}\,\sqrt{\eta}\,,
\qqq
where $\gamma_{\mu\nu}$ is the Riemannian 
metric on $M$,  $\m\eta_{\alpha\beta}$ the one 
on $\Sigma$ and $\sqrt{\eta}\equiv\sqrt{\det{
\eta_{\alpha\beta}}}$ is 
the Riemannian volume density on $\Sigma$. 
In particular, if $M=\NR^n$ with the standard metric,
we obtain the quadratic action of the free field
on a two-dimensional surface leading to linear 
classical equations. The general case, however, results
in a non-linear classical theory.
\vskip 0.3cm

The term $S^\gamma$ does not change under the local
rescalings $\eta_{\alpha\beta}\mapsto\ee^{2\sigma}
\eta_{\alpha\beta}$ of the metric on $\Sigma$ 
i.e. it possesses two-dimensional conformal invariance.
For oriented $\Sigma$, conformal classes of the
metric are in one to one correspondence with 
complex structures on $\Sigma$ such that $\eta_{zz}
=\eta_{\bar z\bar z}=0$ in the holomorphic coordinates 
and that the latter preserve the orientation. 
The action $S^\gamma$ may be written using
explicitly only the complex structure of $\Sigma\m$:
\qq
S^\gamma(X)\ =\ {_i\over^{2\pi}}\int_{_\Sigma}\hspace{-0.1cm}                                
\gamma_{\mu\nu}(X)\,\m\da X^\mu\,\de X^\nu\m,
\qqq
where $\da=dz\m\da_z$ and $\de=d\bar z\m\da_{\bar z}$.
One-dimensional complex manifolds are called Riemann surfaces.
It follows that the action $S^\gamma(X)$ may be defined
on such surfaces.
\vskip 0.3cm

To the $S^\gamma$ term, one may add the expression
\qq
S^\beta(X)\ =\ {_1\over^{4\pi i}}
\int_{_\Sigma}\hspace{-0.1cm}\beta_{\mu\nu}(X)\,
\da_\alpha X^\mu\m\da_\beta X^\nu \epsilon^{\alpha\beta}
\qqq
where $\beta_{\mu\nu}=-\beta_{\nu\mu}$ are the coefficients
of a 2-form $\beta$ on $M$. Geometrically, $S^\beta$ is
proportional to the integral of the pullback of $\beta$
by $X$:
\qq
S^\beta(X)\ =\ {_1\over^{4\pi i}}
\int_{_\Sigma}\hspace{-0.1cm}X^*\beta\,.
\qqq
The imaginary coefficient is required by the unitarity
of the theory after the Wick rotation to the Minkowski 
signature. The term $S^\beta$ does not use the metric 
on $\Sigma$ but only the orientation and is often called
a topological term. Hence the classical two-dimensional 
conformal invariance of the model with the action 
$S=S^\gamma+S^\beta$.
\vskip 0.3cm

On the quantum level, the sigma model requires a renormalization
which often imposes the addition to the action of further terms
\qq
S^{tach}(X)\ =\ {_1\over^{2\pi}}
\int_{_\Sigma}\hspace{-0.09cm}\CT(X)\,\sqrt{\eta}\quad\ \ 
{\rm and}\ \ \quad
S^{dil}(X)\ =\ {_1\over^{2\pi}}\int_{_\Sigma}\hspace{-0.08cm}
\CD(X)\,\,r\,\sqrt{\eta}\,,
\qqq
where 
where $\CT$ and $\CD$ are functions on $M$ called tachyonic
and dilatonic potentials, respectively, and $r$ is the scalar
curvature of the metric $\eta_{\alpha\beta}$. 
The renormalization breaks the conformal 
invariance (note that $S^{tach}$ and $S^{dil}$ are  
not conformal invariant). We shall be interested, however,  
in the case of the WZW model \cite{4}, an example of a CFT,  
where the classical conformal invariance is (almost)
not broken on the quantum level. 
\vskip 0.3cm

The WZW model is the two-dimensional counterpart of the 
particle on a group and may be thought of as describing
the movement of a string on a group manifold $M=G$ equipped 
with the invariant metric $\gamma$ described before.
We then have for $g:\Sigma\rightarrow G$,
\qq
S^\gamma(g)\ =\ {_k\over^{4\pi i}}\int_{_\Sigma}\hspace{-0.09cm}                
\tr\,\m(g^{-1}\da g)(g^{-1}\de g)\m,
\label{sgam}
\qqq
where $k$ is a positive constant. The quantization
of a model with such an action leads, however, to
a theory without conformal invariance. To restore
the latter, one adds to the $S^\gamma$ term, following Witten 
\cite{4}, a topological term, the so called 
Wess-Zumino (WZ) term $S^{WZ}$. In the first approximation,
$S^{WZ}=k\m S^\beta$ where $\beta$ is a 2-form on $G$ such that
$d\beta$ is equal to the canonical 3-form $\chi\equiv{1\over 3}
\m\tr\,(g^{-1}dg)^3$ on $G$. If the group $G$ is abelian,
such a description is indeed possible and the overall
action is a simple version of the free field one. 
In the non-abelian case, however, the difficulty comes from
the fact that the 3-form $\chi$ is closed but not globally
exact so that the forms $\beta$ exist only locally 
and are defined only up to closed 2-forms. Hence 
the definition of the WZ term of the action requires
a more refined discussion.
\vskip 0.4cm

\subsection{Particle in the field of a magnetic monopole}

It may be useful to recall a simpler situation where
one is confronted with a similar problem. Suppose that we
want to define the contribution $S^{Dir}(x)$ to the action 
of a mechanical particle of the term 
$$e\int_{_0}^{^T}\hspace{-0.07cm}A_\nu(x)\m {dx^\nu\over dt
\,\m}\m dt\ =\ e\int x^*A$$ describing
the coupling to the abelian gauge field $A=A_\nu dx^\nu$ 
with the field strength $F_{\nu\lambda}=\hf(\da_\nu A_\lambda
-\da_\lambda A_\nu)$, or in the language of differential forms, 
with $F=dA\m$, \m where $F=F_{\nu\lambda}\m dx^\nu dx^\lambda$. 
The constant $\,e\m$ stands for the electric charge 
of the particle.
For concreteness, suppose that $F_{\mu\nu}$ corresponds to 
the magnetic field of a monopole of magnetic charge $\mu$ 
placed at the origin of $\NR^3$:
\qq
F_{\nu\lambda}\,=\,\hf\m \mu\,\epsilon_{\nu\lambda\kappa}\,
{{x^\kappa}\over{\vert x\vert^3}}\,.
\non
\qqq
There is no global 1-form $A$ on $\NR^3$ without the origin
such that $dA=F$. For a closed trajectory $t\mapsto x(t)$,
however, we may pose
\qq
S^{Dir}(x)\ =\ e\int_{_D}\hspace{-0.08cm}\tilde x^*F\,,
\qqq
where $\tilde x$ is a map of a disc $D$ into 
$\NR^3\setminus\{0\}$ coinciding on the boundary of the disc
with $x$. For two different extensions $\tilde x$, however,
the above prescription may give different results.
Their difference may be written as the integral
\qq
e\int_{_{S^2}}\hspace{-0.08cm}\tilde x^*F
\label{inF}
\qqq
over the 2-sphere $S^2$ obtained by gluing the two disc $D$, one 
with the inverted orientation, along the boundary and for the map
$\,\tilde x:S^2\rightarrow\NR^3\setminus\{0\}\,$ glued from 
the two extensions of $x$ to the respective discs, see Fig.\,\,2.

\leavevmode\epsffile[-150 -15 270 156]{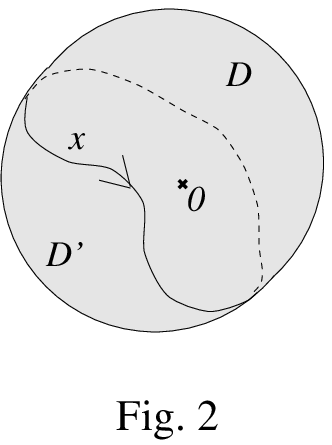}
\vskip 0.1cm
 
The ambiguities (\ref{inF}) are the periods of the closed 
form $F$ over the cycles of the 2$^{\rm nd}$ integer homology
$\,H_2(\NR^3\setminus\{0\})=\NZ$. They take discrete values which 
are multiples of $4\pi\m e\m \mu$. The latter value 
is obtained for the unit sphere in $\NR^3$, 
a generator of $H_2(\NR^3\setminus\{0\})$. 
The discrete ambiguities are acceptable in classical mechanics
where one studies the extrema of the action. In quantum mechanics, 
however, we have to give sense to the Feynman amplitudes 
$\,\ee^{\m i\m S^{Dir}(x)}\m$, hence only the ambiguities in the 
action with values in $2\pi\NZ$ are admissible. Demanding that 
the quantum-mechanical amplitudes be unambiguously defined
reproduces this way the Dirac quantization condition 
$\,e\mu\in\hf\m\NZ$. 
\vskip 0.3cm

For open trajectories $[0,T]\ni t\mapsto x(t)$, 
the amplitudes $\,\ee^{\m i\m e\int x^*A}$ may not, in general,
be unambiguously assigned numerical values. They may be only defined 
as maps between the fibers $\CL_{x(0)}$ and $\CL_{x(T)}$
of a line bundle $\CL$. Geometrically, they give the parallel 
transport in the bundle corresponding to a $U(1)$-connection 
with the curvature form $F$. We shall recover the analogous
situation below when discussing how to give meaning to
the WZ term in the action of the WZW model.
\vskip 0.4cm

\subsection{Wess-Zumino action on surfaces without boundary}

Let us first consider the case of compact Riemann
surfaces without boundary. 

\leavevmode\epsffile[-100 5 320 160]{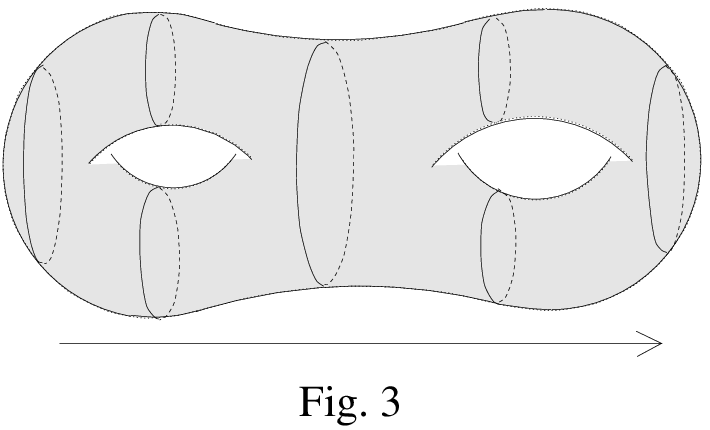}

\noindent Topologically, such surfaces
are characterized by the genus $g_{_\Sigma}$ equal
to the number of handles of the surface. They may be viewed
as world sheets of a closed string created
from the vacuum, undergoing in the evolution $\m g_{_\Sigma}$ 
splittings and recombinations and finally disappearing
into the vacuum, see Fig.\,\,3 where a surface of genus 2
was represented. 
At genus zero, there is only one (up to diffeomorphisms) 
Riemann 
surface, the Riemann sphere $\NC P^1=\NC\cup\{\infty\}$,
see Fig.\,\,4(a).

\leavevmode\epsffile[-60 -20 347 175]{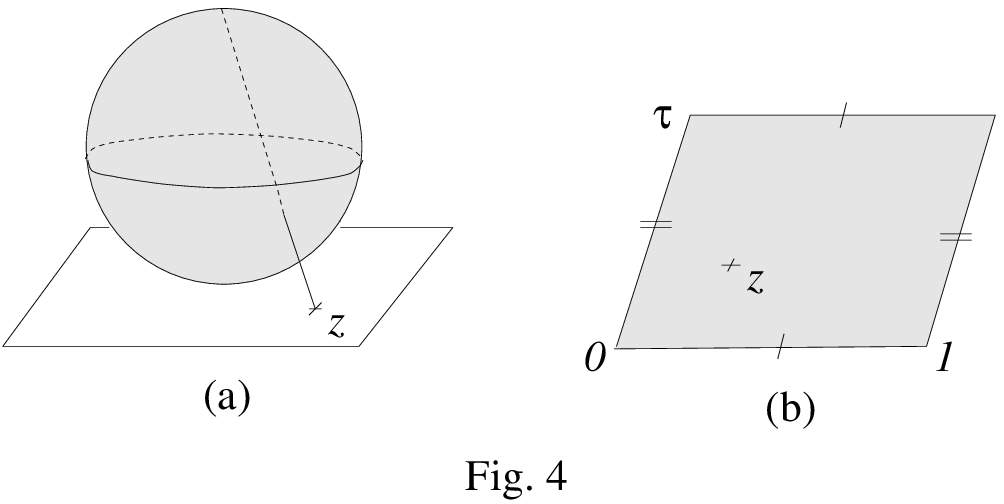}

\noindent At genus one, there is a complex one-parameter family 
of Riemann surfaces: the complex tori
$\,{T}_\tau=\NC/(\NZ+\tau\NZ)\m$ with $\tau$ 
in the upper half plane $H^+\m$ of the complex numbers
such that $\,{\rm Im}\m\tau>0$, see Fig.\,\,4(b).
The tori ${T}_\tau$ and ${T}_{\tau'}$, where $\m\tau'
={a+b\tau\over c+d\tau}\,$ for $\m({_a\atop^c}
\,{_b\atop^d})\m$ in the modular group $SL(2,\NZ)$, 
may be identified by the map $\,z\mapsto z'=(c\tau+d)^{-1}z\m$. 
The space of the diffeomorphism classes (i.e.\,\,the {\bf moduli
space}) of genus one Riemann surfaces is equal to $H^+/SL(2,\NZ)$
and has complex dimension 1. For higher genera,  
the moduli spaces of Riemann surfaces have complex 
dimension $3(g_{_\Sigma}-1)$.
\vskip 0.3cm

Let us return to the discussion of the action of the WZW model.
Assume that $G$ is connected and simply connected
and that $\Sigma$ is a compact Riemann surface without boundary.
Following \cite{4} and mimicking the trick used for a particle 
in a monopole field, one may extend the field $g:\Sigma
\rightarrow G$ to a map $\tilde g:B\rightarrow G$ 
of a 3-manifold $B$ such that $\da B=\Sigma$ and set:
\qq
S^{^{WZ}}(g)\ =\ {_k\over^{4\pi i}}\int_{_B}\hspace{-0.07cm}
\tilde g^*\chi\,.
\label{28}
\qqq
By the Stokes formula, this expression coincides with 
$k\m S^\beta(g)$ whenever the image of $\tilde g$ is contained 
in the domain of definition of a 2-form $\beta$ such that 
$d\beta=\chi$, but it makes sense in the general case. 
The price is that the result depends on the extension 
$\tilde g$ of the field $g$. The ambiguities have the form
of the integrals
\qq
{_k\over^{4\pi i}}\int_{_{\tilde B}}\hspace{-0.07cm}
\tilde g^*\chi
\label{per}
\qqq
over 3-manifolds $\tilde B$ without boundary with
$\tilde g:\tilde B\rightarrow G$, see Fig.\,\,5. 

\leavevmode\epsffile[-35 -20 310 150]{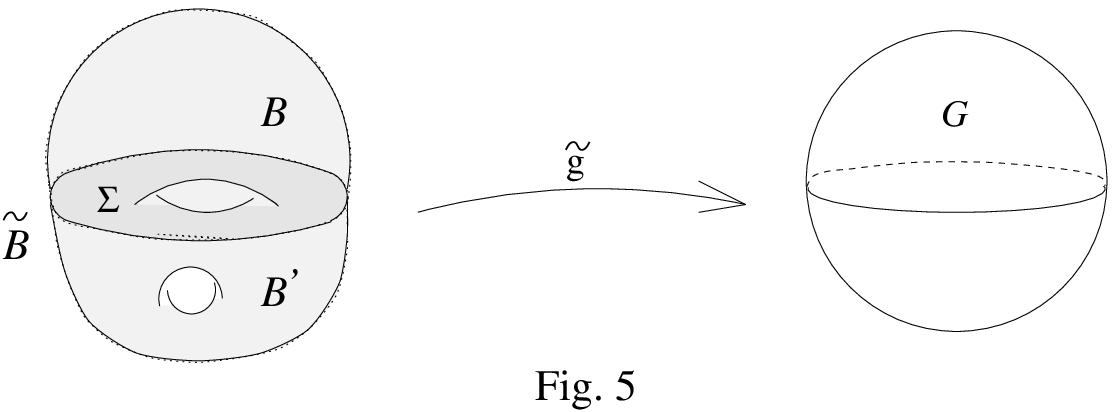}

\noindent They are proportional 
to the periods of the 3-form $\chi$ over the integer 
homology $H_3(G)$. Such discrete contributions do not 
effect the classical equations of motion $\delta S=0$. 
In quantum mechanics, however, where we deal 
with the Feynman amplitudes\footnote{there is no $i$ 
in front of $S$ since we work with the Euclidean 
action} $\,\ee^{-S(g)}\m$, \m only ambiguities in 
$2\pi i\NZ$ are allowed. Hence we have to find conditions
under which the periods (\ref{per}) lie in $2\pi i\NZ$.
\vskip 0.3cm

Recall that we normalized the invariant form $\tr$ on the Lie 
algebra $\Ng$ of $G$ so that the long roots in $\Ntt$ have 
length squared 2. When $\,G=SU(2)\cong\{\m x\in\NR^4\,\vert\,
\vert x\vert^2=1\m\}\m$, \,the 3-form $\chi$ equals 
then to $4$ times the volume 
form of the unit 3-sphere. Since the volume of the latter
is equal to $2\pi^2$, we infer that
\qq
{_1\over^{4\pi i}}
\int_{_{SU(2)}}\hspace{-0.06cm}\chi\ =\ -2\pi i\,.
\qqq
For the other simple, simply connected groups, the roots
$\alpha$ determine the sub-algebras $su(2)_{_\alpha}\subset\Ng$, 
obtained by identifying the corresponding coroot $\alpha^\vee=
{2\,\alpha\over{\hbox{tr}\,\alpha^2}}$ and the step generators
$e_{\pm\alpha}$ with the Pauli matrices $\sigma_3$ and
$\sigma_{\pm}$, respectively. By exponentiation, we obtain
the $SU(2)_{_\alpha}$ subgroups of $G$. Clearly,
\qq
{_1\over^{4\pi i}}
\int_{_{SU(2)_{_\alpha}}}\hspace{-0.06cm}\chi\ 
=\ -{_{4\pi i}\over^{\hbox{tr}\,\alpha^2}}\,.
\qqq
The ratio $2\over\tr\,\alpha^2$ is equal to 1 for long roots 
and is a positive integer for the others. It appears 
that any of the subgroups $SU(2)_{_\alpha}\cong S^3\m$ 
for $\alpha$ a long root generates $H_3(G)=\NZ$. 
Thus the unambiguous definition of the amplitudes 
$\ee^{-S^{WZ}(g)}$ requires that the coupling
constant $k$, called the {\bf level} of the model, be a 
(positive) integer, in the analogy to the Dirac quantization
of the magnetic charge.
\vskip 0.3cm

It is easy to see that, although the action $S^{WZ}(g)$
cannot be expressed, in general, as a local integral 
over $\Sigma$, the variation of $S^{^{WZ}}$ has such a form:
\qq
\delta S^{^{WZ}}(g)\ =\ {_k\over^{4\pi i}}\int_{_\Sigma}
\hspace{-0.09cm}\tr\,(g^{-1}\delta g)(g^{-1}dg)^2\m.
\label{vWZ}
\qqq
The above formula is a special case of the general, very useful, 
geometric identity: $\,\delta\hspace{-0.05cm}\int 
\hspace{-0.05cm}f^*\alpha=\int\hspace{-0.05cm}\CL_{\delta f}
\alpha\,$, \,where $\CL_\CX$ is the Lie derivative.
Applied to $f=g$ and $\alpha=\chi$ it gives, 
in conjunction with the Stokes formula, 
the above relation. \,It is also important to note the behavior 
of $S^{^{WZ}}$ under the point-wise multiplication of fields, 
a basic property of the WZ term:
\qq
S^{^{WZ}}(g_1g_2)\ =\ S^{^{WZ}}(g_1)\,+\, S^{^{WZ}}(g_2)\,+\,
W(g_1,g_2)\,,
\label{pwWZ}
\qqq
where
\qq
W(g_1,g_2)\,=\,
{_k\over^{4\pi i}}\int_{_{\Sigma}}\hspace{-0.1cm}\tr
\,(g_1^{-1}dg_1)(g_2dg_2^{-1})\m.
\label{WWZ}
\qqq
The relation follows easily from the definition (\ref{28}),
again by applying the Stokes formula.
\vskip 0.3cm

The complete action of the WZW model on a closed Riemann surface 
$\Sigma$ is the sum of the $\gamma$- and the WZ-terms with 
the same coupling constant $k$: $\,S(g)=S^\gamma(g)+S^{^{WZ}}(g)$.
Since $S^\gamma$ is unambiguous, it has the same ambiguities 
as $S^{WZ}$, requiring that $k$ be
a (positive) integer. The relations (\ref{vWZ}) and (\ref{pwWZ}) 
get also contributions from $S^\gamma$ and become:
\qq
&&\delta S(g)\ =\ -\m{_k\over^{2\pi i}}\int_{_\Sigma}
\hspace{-0.09cm}\tr\,(g^{-1}\delta g)\m\da(g^{-1}\de g)\m,
\label{vWZW}\\\cr
&&S(g_1g_2)\ =\ S(g_1)\,+\, S(g_2)\,+\,
{_k\over^{2\pi i}}\int_{_{\Sigma}}\hspace{-0.1cm}\tr
\,(g_1^{-1}\de g_1)(g_2\da g_2^{-1})\m.
\label{pwWZW}
\qqq
The last relation is often called the Polyakov-Wiegmann formula.
From Eq.\,\,(\ref{vWZW}), we obtain the classical equations 
of motion
\qq
\da(g^{-1}\de g)\,=\,0\,\qquad{\rm or, \,\,equivalently,}
\qquad\de(g\da g^{-1})\,=\,0\,.
\label{cleq}
\qqq
They have few solutions with values in $G$ (this would not be
the case if we considered $\Sigma$ with a Minkowski metric).
In all the above formulae, however, we could have taken fields 
$g$ with values in the complexified group $G^\NC$. For such
fields, the general local solutions of Eqs.\,\,(\ref{cleq})
have the form
\qq
g(z,\bar z)\ =\ g_\ell(z)\,g_r(\bar z)^{-1}
\label{lcls}
\qqq
where $g_\ell$ ($g_r$) are local holomorphic (anti-holomorphic)
maps with values in $G^\NC$. Thus Eqs.\,\,(\ref{cleq})
constitute a non-linear generalization of the Laplace equation
in two dimensions whose solutions are harmonic functions 
which are, locally, sums of holomorphic and anti-holomorphic 
ones. In particular, multiplying a solution (\ref{lcls}) by 
a holomorphic map into $G^\NC$ on the left and by an anti-holomorphic 
one on the right we obtain another solution. Similarly, composing 
a solution with a local holomorphic map or inverting it in $G^\NC$ 
after composition with a local anti-holomorphic map of $\Sigma$ 
one produces new solutions. Hence a rich symmetry structure 
of the classical theory. This structure will be preserved 
by the quantization leading to the current and Virasoro algebra 
symmetries of the quantum WZW model.
\vskip 0.4cm

\subsection{Wess-Zumino action on surfaces with boundary}

What if the surface $\Sigma$ has a boundary? Of course
only the WZ term in the action causes problems
due to its non-local character. The term $S^\gamma$ 
is defined unambiguously for any compact surface. It will 
be convenient
to represent $\Sigma$ as $\Sigma'\setminus(\mathop{\cup}
\limits_nD_{n}^{^{^{{\hspace{-0.21cm}}o}}})$
where $D_{n}$ are disjoint unit discs $\{\,z\,\vert\,\vert z\vert\leq 1\}$ 
embedded in a closed surface $\Sigma'$ without boundary, 
see Fig.6. 

\leavevmode\epsffile[-80 -20 337 170]{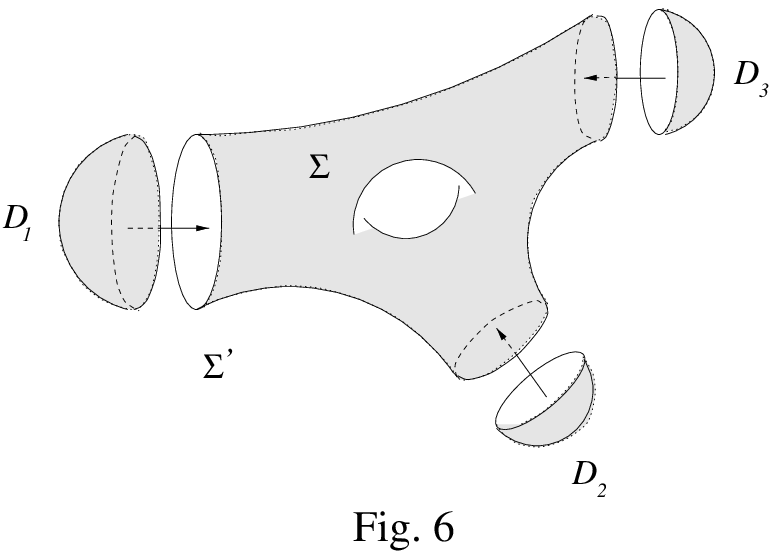}

\noindent Note that the boundaries of $\Sigma$ are then 
naturally parametrized by the unit circles. One way to proceed 
in the presence of boundaries is to extend the field 
$g:\Sigma\rightarrow G$ to a map $g':\Sigma'\rightarrow G$ 
and to consider the action $\,S^{{WZ}}_{_{\Sigma'}}(g')\,$
pertaining to the surface $\Sigma'$, as stressed
by a subscript. We are again confronted with the question
as to how the action on $\Sigma'$ depends on the extension 
of the field. The answer is easy to work out.
If $g''=g'h$ is another extension of $g$
then, by Eq.\,\,(\ref{pwWZ}),
\qq
S^{^{WZ}}_{{\Sigma'}}(g'')\ =\ S^{^{WZ}}_{{\Sigma'}}(g')\,
+\,S^{^{WZ}}_{{\Sigma'}}(h)\,+\,W_{{\Sigma'}}(g',h)\,.
\qqq
It will be convenient to localize 
the changes in the discs $D_{n}$ 
by rewriting the last formula as
\qq
S^{^{WZ}}_{{\Sigma'}}(g'')\ =\ S^{^{WZ}}_{{\Sigma'}}(g')\,+
\sum\limits_n\left(S^{^{WZ}}_{{S^2}}
(h_n)\,+\,W_{_{D_{_n}}}(g',h)\right),
\label{trpro}
\qqq
where $h_n$, mapping spheres (compactified planes) $S^2_n$
to $G$, extend the maps $h\vert_{_{D_{_n}}}$ by unity and 
$W_{_{D_{_n}}}$ are as in Eq.\,\,(\ref{WWZ}) but with 
the integration restricted to $D_{n}$. To account for
the change (\ref{trpro}), we shall define 
the following equivalence relation between the pairs $(g',z)$ 
where $g':D\rightarrow G$ and $z\in\NC$\m :
\qq
(g',\m z)\ \sim\ (g'h,\m z\,\ee^{-S^{{WZ}}_{S^2}(h)-
W_{_{D}}(g',h)})
\non
\qqq
for $h:S^2\rightarrow G$ equal to unity outside the
unit disc $D\subset S^2$. The set of equivalence classes
forms a complex line bundle $\CL$ over the loop group $LG$
of the boundary values of the maps $g'$. Comparing 
to Eq.\,\,(\ref{trpro}), we infer that for 
$g:\Sigma\rightarrow G$ with $\Sigma$ as above, 
the amplitude $\ee^{-S^{{WZ}}_{{\Sigma}}(g)}$ makes sense 
as the element of a tensor product of the line bundles $\CL$,
one for each boundary component of $\Sigma\m$,
\qq
\ee^{-S^{WZ}_{{\Sigma}}(g)}\ \in\ \mathop{\otimes}\limits_n
\CL_{_{g|_{_{\da D_{_n}}}}}\,,
\non
\qqq
where $\CL_{_h}$ denotes the fiber of $\CL$ over the loop
$h\in LG$. Hence the WZ amplitudes 
$\,\ee^{-S^{WZ}_{{\Sigma}}(g)}\,$
take values in line bundles instead of having numerical values,
exactly as for the amplitudes giving the parallel transport 
in a $U(1)$-gauge field mentioned before.
\vskip 0.3cm

The line bundle $\CL$ is an interesting object.
It carries a hermitian structure given by the absolute value 
of $z$. The fibers of $\CL$ over $g$ and $\check g$ 
where $\check g$ ia a reversed loop, $\check g(\ee^{i\varphi})
=g(\ee^{-i\varphi})$, may be naturally 
paired so that, for $g:\tilde\Sigma
\rightarrow G$, where $\tilde\Sigma$ is obtained from two
surfaces $\Sigma$ and $\Sigma'$ by gluing them along some 
boundary components, see Fig.\,\,7,
\qq
\langle\,\ee^{-S^{WZ}_{_\Sigma}(g|_{_\Sigma})},\,
\ee^{-S^{WZ}_{_{\Sigma'}}
(g|_{_{\Sigma'}})}\,\rangle\ =\ \ee^{-S_{_{\tilde\Sigma}}(g)}\,.
\label{glue}
\qqq

\leavevmode\epsffile[-100 -20 317 180]{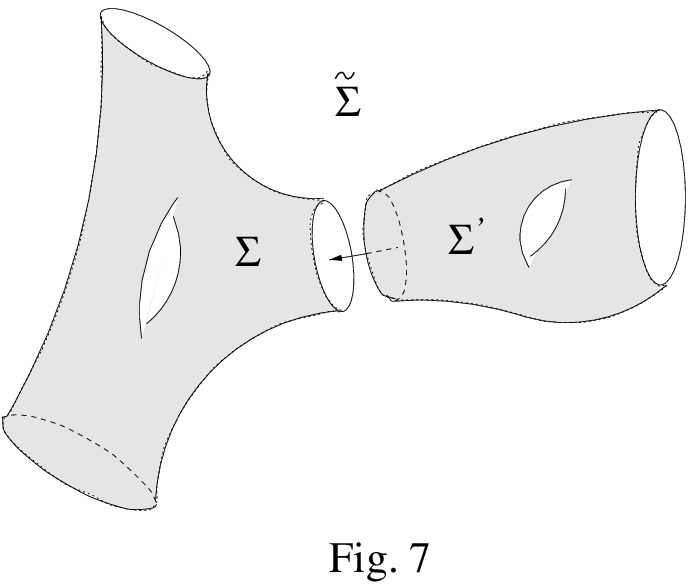}

\noindent$\CL\,$ may be also equipped with a product structure 
such that
\qq
\ee^{-S_{_\Sigma}(g_1)}\cdot\ee^{-S_{_\Sigma}(g_2)}\ =\ 
\ee^{-S_{_\Sigma}(g_1g_2)}\,\ee^{\m W_{_\Sigma}(g_1,g_2)}\,,
\label{prKM}
\qqq
compare to Eq.\,\,(\ref{pwWZ}). Under the product, 
the elements of unit length in $\CL$ 
form a group $\hat{G}$ which is a {\bf central extension} of the 
loop group $LG$ by the circle group $U(1)$:
\qq
1\rightarrow U(1)\rightarrow\hat{G}\rightarrow LG\rightarrow 1\,.
\qqq
The second arrow sends $\ee^{\m i\varphi}$ to the equivalence 
class of $(1,\m\ee^{\m i\varphi})$. The extensions for $k>1$ 
are powers of the universal one corresponding to $k=1$. On the 
infinitesimal level, one obtains the central extensions  
of the loop algebra $L\Ng$ of the maps of the circle
$S^1$ to the Lie algebra $\Ng$ by the real line:
\qq
0\rightarrow\NR\rightarrow\hat{\Ng}\rightarrow L\Ng\rightarrow 0\,.
\label{exs}
\qqq
The Lie algebra $\hat{\Ng}$, called the {\bf current} or 
the affine Kac-Moody {\bf algebra}, may be described explicitly 
in terms of the complexified generators $t^a_n$ corresponding 
to the loops $t^a\ee^{in\varphi}$ in $L\Ng^\NC$ and the central
element $K$ satisfying the commutation relations:
\qq
[t^a_n,t^b_m]\,=\, i\m f^{abc}\m t^c_{n+m}\,+\,{_1\over^2}
\m K\m n\,\delta^{ab}\delta_{n+m,0}\,.
\label{KM}
\qqq
The algebra $\hat\Ng$ is the same 
for all levels $k$ but the central element 
$K\in\hat\Ng$ is the image of $k\in\NR$ under the second arrow 
in the exact sequences (\ref{exs}). Note that the generators $t^a_0$
span a subalgebra $\Ng\subset\hat\Ng$. As we shall see, the group
$\hat G$ and the algebra $\hat{\Ng}$ play in the WZW theory a similar 
role to that of $G$ and $\Ng$ for the particle on the group.
\vskip 0.4cm

\subsection{Coupling to gauge field}

We may couple the WZW model to the gauge field 
$i\m A\equiv i(A^{10}+A^{01})$, a 1-form with values 
in the Lie algebra $\Ng$ (or, more generally, 
$\Ng^\NC$), where $A^{10}=t^a\m A^a_z dz$ and $A^{01}=t^a\m
A^a_{\bar z}d\bar z\m$ 
are, respectively, a 1,0- and a 0,1-form (the chiral components
of the gauge field). In most what follows, we shall treat the gauge 
field as external, i.e. non-dynamical.
Nevertheless, the coupling will allow to test the variation 
of the quantum system under the changes of the gauge field 
background and, finally, will facilitate the exact solution 
of the model. For a surface without boundary, we define
\qq
S(g,\m A)\ =\ S(g)\,+\,
{_{ik}\over^{2\pi}}\int_{_\Sigma}
\hspace{-0.08cm}\tr\,[A^{10}(g^{-1}\de g)\,+\,
(g\da g^{-1})\,A^{01}
\,+\,g\m A^{10} g^{-1}A^{01}]\,.\hspace{0.7cm}
\label{tog}
\qqq 
Under the local gauge transformations $h:\Sigma
\rightarrow G$, the gauge fields transform in the standard way:
\qq
A^{10}\,\mapsto\,{}^h\hspace{-0.09cm}A^{10}\,=\, h\m A^{10}\, h^{-1}\,
+\,h\m\da h^{-1}\,,\qquad A^{01}\,\mapsto\,{}^h\hspace{-0.09cm}
A^{01}\,=\, h\m A^{01} h^{-1}\,+\,h\de h^{-1}\,.
\non
\qqq
The reaction of the action to the chiral changes of the gauge 
is encoded in the identity
\qq
S(g,\m A)\ =\ S(h_1g\m h_2^{-1},\m{}^{h_2}
\hspace{-0.09cm}A^{10}+
{}^{h_1}\hspace{-0.1cm}A^{01})\,+\,S(h_2,\m A^{10})\,
+\,S(h_1^{-1},\m A^{01})
\label{trprop}
\qqq
which follows in a direct manner from the Polyakov-Wiegmann
formula (\ref{pwWZW}). For the later convenience,
we have chosen a modified way of coupling to the gauge field, 
as compared to the more standard way with the addition
of the term $-A^{10}A^{01}$ in the brackets on the right hand
side of Eq.\,\,(\ref{tog}). The latter way would render 
the action invariant with respect to the diagonal 
(i.e.\,\,non-chiral) gauge transformations with $h_1=h_2$.
\vskip 0.3cm

For surfaces with boundary, we define $\ee^{-S(g,\m A)}$ 
by the same prescription but on the level of the amplitudes
with values in the product of line bundles. The definition
(\ref{prKM}) of the product implies then the transformation
rule:
\qq
\ee^{S(g,\m A)}\ =\  
\ee^{-S(h_1^{-1},\m A^{01})}\cdot\m\ee^{-S(h_1g\m h_2^{-1},
\m{}^{^{h_2}}\hspace{-0.11cm}A^{10}\m+\m{}^{^{h_1}}
\hspace{-0.11cm}A^{01})}\cdot\m
\ee^{-S(h_2,\m A^{10})}
\label{trpropb}
\qqq
which extends the property (\ref{trprop}) to the case with 
boundaries.
\vskip 0.4cm

\nsection{Quantization of the WZW model}
\subsection{Quantum amplitudes}

The Feynman quantization prescription instructs us that
in the quantum WZW model we should sum the amplitudes 
of different classical configurations. 
This leads to formal functional integrals such as, for example,
\qq
\CA_{_\Sigma}\ =\ \int\ee^{-S_\Sigma(g)}\,\, Dg\,,
\non
\qqq
where one integrates over the maps $g:\Sigma\rightarrow G$
and $Dg$ stands for the local product $\prod_{_\xi}\hspace{-0.08cm}
dg(\xi)$ of the Haar measures. If $\Sigma$ is closed, then the above 
integral should take a numerical value $\CZ_{_\Sigma}$ called 
the {\bf partition function} 
(because of the statistical physics analogy). For $\Sigma$ with 
boundary, it should define, instead, a Hilbert space state. 
Let $\Gamma(\CL)$ denote the space of sections of the line bundle 
$\CL$ over the loop 
group $LG$. $\,\Gamma(\CL)$ plays the role of the space of states 
of the quantized theory. If $\Sigma$ has a boundary, we should 
fix in the functional integration the boundary
values $\un{g}=(g_n)$ of fields $g:\Sigma\rightarrow G\m$:
\qq
\CA_{_\Sigma}(\un{g})\ =\ \int\limits_{g|_{_{\da D_{_n}}}=\,g_n}
\ee^{-S_\Sigma(g)}\,\, Dg\,.
\non
\qqq
The result, in its dependence on $\un{g}$, 
should give an element of the tensor product $\,\mathop{\otimes}
\limits_n\Gamma(\CL)\m$ of the state spaces: the {\bf quantum amplitude}
corresponding to the surface $\Sigma$. More generally,
we shall consider the quantum amplitudes in the presence
of external gauge field:
\qq
\CA_{_\Sigma}(\un{g};\m A)\,\equiv\,
\ =\ \int\limits_{g|_{_{\da D_{_n}}}=\,g_n}
\ee^{-S_\Sigma(g,\m A)}\,\, Dg
\label{FfiA}
\qqq
again with the values in $\,\mathop{\otimes}
\limits_n\Gamma(\CL)\m$. We would like to give a rigorous
meaning to such objects. In general, the functional
integrals require complicated renormalization procedures
which, besides, work only in some cases (of the so called 
renormalizable theories) and even then, in most instances,
have been implemented only on the level of formal
perturbation series. The WZW models are perturbatively 
renormalizable. In this case, however, one may follow
a shortcut by exploiting formal symmetry properties 
of the functional integrals and showing that they 
fix uniquely the quantum amplitudes. This will be 
the line of thought adopted below, although
we shall only describe the essential points of the
argument and make detours to introduce other important
notions.
\vskip 0.3cm

Let us start by discussing the formal structure 
of the space of states $\Gamma(\CL)$. The scalar product 
and the bilinear form
\qq
{\rm\bf(}\psi,\m\psi'{\rm\bf)}\ =\ \int_{_{LG}}(\psi(g),
\m\psi'(g))\,\, Dg\,,
\qquad{\bf\langle}\m\psi,\m\psi'\m{\bf\rangle}
=\ \int_{_{LG}}\langle\m\psi(g),\m
\psi'(\check{g})\m\rangle\,\,Dg\,,
\label{fbf}
\qqq
which employ the hermitian structure and the duality (\ref{glue})
on the line bundle $\CL$, should turn $\Gamma(\CL)$ 
into a Hilbert space $\CH$
and should allow the identification of $\CH$ with its dual.
The space $\Gamma(\CL)$ carries also two commuting actions
of the group $\,\hat{G}\,$: $\,
\psi\,\mapsto\,{}^{\hat h}\hspace{-0.02cm}\psi\,$ and
$\,\psi\,\mapsto\,\psi^{\hspace{0.01cm}\hat h}\m$.
They are defined by:
\qq
{}^{^{\hat h}}\hspace{-0.06cm}\psi(g)\ 
=\ \hat h\cdot\psi(h^{-1}g)\,
\qquad
\psi^{^{\hat h}}(g)\ =\ \psi(gh)\cdot{\hat h}^{-1}\,,
\qqq
where $g$ and $h$ are elements of the loup group $LG$
and $h$ is the projection of $\,\hat h\in\hat{G}$. 
Formally, these actions preserve the scalar product and 
the bilinear form in $\Gamma(\CL)$. 
On infinitesimal level, they give rise to two commuting
actions of the current algebra $\hat\Ng$ 
in $\Gamma(\CL)$. We shall
denote by $J^a_n$ and $\tilde J^a_n$ the operators 
in $\Gamma(\CL)$ corresponding to the left and right
action of the generators $t^a_n$ of $\hat\Ng$. 
The central generator $K$ acts 
in $\Gamma(\CL)$ as multiplication by $k$. Of course, 
$J^a_n$ and $\tilde J^a_n$ satisfy the commutation 
relation (\ref{KM}).
\vskip 0.3cm

As stressed by Segal in \cite{Segal}, there are two 
important properties of the quantum amplitudes
$\CA_{_\Sigma}$ which are crucial for their rigorous 
construction. The first one, is the gluing property 
\qq
\CA_{_{\tilde \Sigma}}(g_n,g_{n'})\ =\ 
\int \langle \CA_{_\Sigma}(g_n,g_{n_0})\m,
\,\CA_{_{\Sigma'}}(\check g_{n_0},
g_{n'})\rangle\ Dg_{n_0}
\label{glueq}
\qqq
which states that for a surface $\tilde\Sigma$ glued along
boundary components of two pieces $\Sigma$ and $\Sigma'$,
as in Fig.\,\,7, the functional integral may be computed 
iteratively, by first keeping the values of $g$ fixed 
on the gluing circle and integrating over them only
after the integration over the fields on $\Sigma$ 
and $\Sigma'$, see Eq.\,\,(\ref{glue}).
Using the bilinear form on $\Gamma(\CL)$ 
applied to the glued channel, we may write this relation 
as the identity
\qq
\CA_{_{\tilde \Sigma}}\ =\  
{\bf\langle}\,\CA_{_\Sigma}\m,\,\CA_{_{\Sigma'}}\,{\bf\rangle}\,.
\label{glueA}
\qqq
It is often more convenient to view the quantum
amplitudes $\CA_{_\Sigma}$ as operators from the tensor product
of some of the boundary spaces $\CH$ into the others.
This is always possible because of the linear isomorphism between
$\CH$ and its dual. Then Eq.\,\,(\ref{glueA}) may be simply  
rewritten with the use of the product of operators:
\qq
\CA_{_{\tilde \Sigma}}\ =\  
\CA_{_\Sigma}\,\CA_{_{\Sigma'}}\,.
\label{glueB}
\qqq
One may also glue two boundary components in a single 
connected surface $\Sigma$. The amplitude for the resulting
surface $\tilde\Sigma$ is then obtained from that 
of $\Sigma$ by pairing the two corresponding factors
in the product of the Hilbert spaces or, in the operator
interpretation, by the partial trace applied to the
glued channel:
\qq
\CA_{_{\tilde \Sigma}}\ =\ \tr_{_\CH}\,\CA_{_\Sigma}\,.
\label{glueC}
\qqq
In fact, as pointed out in \cite{Segal}, the last relation 
encompasses also the previous 
one if one introduces the amplitudes for 
the disconnected Riemann surfaces defining them 
as the tensor product of the amplitudes of the components.
Clearly, similar gluing relation should hold 
for the amplitudes in external gauge field. 
\vskip 0.3cm

The second important property of the quantum amplitudes 
follows formally from the transformation 
property (\ref{trpropb}) of the classical amplitudes 
under the chiral gauge transformations 
$h_{1,2}:\Sigma\rightarrow G$. It reads:
\qq
{}^{^{\hat h_1}}\hspace{-0.11cm}
\CA_{_\Sigma}(A)^{^{\m\hat h_2}}\ =\ \CA_{_\Sigma}(
{}^{h_2}\hspace{-0.11cm}A^{10}+\m{}^{h_1}
\hspace{-0.11cm}A^{01})
\label{trprq}
\qqq
for $\,\hat h_1^{-1}=\ee^{-S_{_\Sigma}(h_1^{-1},\m A^{01})}\,$ 
and $\,\hat h_2=\ee^{-S_{_\Sigma}(h_2,\m A^{10})}\m$. 
\,The identity (\ref{trprq}) expresses the covariance 
of the quantum amplitudes under the chiral gauge 
transformations. It is at the basis of the rich
symmetry structure of the quantized WZW theory.
\vskip 0.4cm

\subsection{The spectrum}

To give a rigorous construction of the Hilbert space $\CH$
of the WZW model, whose vectors represent quantum states 
of a closed string moving on the group manifold, 
one may resort to the representation theory of the current 
algebras. The algebra $\hat{\bf g}$ possesses a distinguished 
family of irreducible unitary representations labeled by pairs 
$\hat R=(k,R)$, where $k$, a non-negative integer called 
the level, is the value taken in the representation
by the central generator $K$ of $\hat\Ng$ and where $R$ is 
an irreducible representation of $G$ (and of $\Ng$). 
The irreducible unitary representations $\hat R$ 
act in spaces $V_{_{\hat R}}$ possessing a (unique) subspace 
$V_{_R}\subset V_{_{\hat R}}$ annihilated by all 
the generators $t^a_n$ with $n>0$ and carrying the irreducible 
representation $R$ of the subalgebra $\Ng\subset\hat\Ng$ 
generated by $t^a_0$. They are characterized by this property. 
Not all irreducible representations $R$ of $G$ appear 
for the fixed level $k$ but only the ones corresponding to the
the so called integrable HW's which satisfy the condition
\qq
\tr\,(\phi^\vee\lambda_{_R})\,\leq\, k\,,
\label{iHW}
\qqq
where $\phi=\phi^\vee$ is the highest root of $\Ng$, i.e. 
such a root that $\phi+\alpha$ is not a root for any positive 
root $\alpha$. Given $k$, there is only a finite 
number of integrable HW's. 
For $\hat{su(2)}$, the integrable HW's correspond 
to spins $j=0,{1\over 2},1,\dots,{k\over2}$.  
If $\lambda_{_R}$ satisfies the condition (\ref{iHW})
then so does $\lambda_{_{\ov{R}}}$ and the space
$V_{_{\hat{\ov{R}}}}$ is canonically isomorphic to 
$\ov{V_{_{\hat{R}}}}$. The scalar product 
on $\m V_{_{\hat{R}}}$ may then be viewed as a bilinear pairing
between $V_{_{\hat{R}}}$ and $V_{_{\hat{\ov{R}}}}$.
\vskip 0.3cm

The rigorous definition of the Hilbert space of states $\CH$
for the WZW model of level $k$ makes the two notions of the level
coincide:
\qq
\CH\ \ =\ \mathop{\oplus}\limits_{\hat R\ {\rm of\ level}\ k}
\left(V_{_{\hat R}}\otimes V_{_{\hat{\ov{R}}}}\right)^{-}\,,
\label{HSrd}
\qqq
where the symbol $\,(\dots)^-$ stands for the Hilbert space 
completion. This is the loop group analogue of the decomposition 
(\ref{decomp}) of $\m L^2(G,dg)$. 
The operators $J^a_n$ and $\tilde J^a_n$ representing the action 
of the generators $t^a_n$ in, respectively, $V_{_{\hat R}}$ 
and $V_{_{\hat{\ov{R}}}}$ satisfy the unitarity conditions
$\,{J^a_n}^\dagger=J^a_{-n}\m,\,\m\tilde J^a_n{}^\dagger=
\tilde J^a_{-n}$. It is not difficult to motivate
the above choice of the Hilbert space. \,One may, indeed, 
realize the space (\ref{HSrd}) as a space of sections 
of $\CL$. Formally, this may be done by the assignment 
(compare to the relation (\ref{hwq}))
\qq
V_{_{\hat R}}\otimes V_{_{\hat{\ov{R}}}}\supset 
V_{_R}\otimes V_{_{\ov{R}}}\,\ni\,e^i_{_R}\otimes\ov{e^j_{_R}}
\ \ \ \mapsto\ \ \ \zeta
\int\ov{g_{_R}^{^{ij}}(0)}
\,\,\ee^{-S_{_D}(g)}\,\, Dg\,,
\label{HWq}
\qqq
where the normalization constant $\zeta$ will be fixed later. 
The functional integral on the right hand side, 
as a function of $g\vert_{_D}=h$ is the corresponding section 
of $\CL$. One may argue that the above 
integral is given, up to normalization, by its semi-classical 
value, 
\qq
\ov{(g_{cl})_{_R}^{^{ij}}(0)}\,\,\ee^{-S_{_D}(g_{cl})}
\label{HWcl}
\qqq
where $g_{cl}:D\rightarrow G^\NC$ is the solution of the 
classical equations (\ref{cleq}) with the boundary condition
$g_{cl}\vert_{_{\da D}}=h$. As shown in \cite{7}, the expression 
(\ref{HWcl}) defines a non-singular section of $\CL$
only if the HW of $R$ is integrable at level $k$ (recall
that the action $S(g_{cl})$ is proportional to $k$).
One obtains then a rigorous embedding of the space 
$\,\mathop{\oplus}_{\hat R}V_R\otimes V_{\ov{R}}\,$
into $\Gamma(\CL)$, and, identifying the actions of the 
current algebra $\hat{\Ng}$ in $\Gamma(\CL)$ and  
in $\,\mathop{\oplus}_{\hat R}V_{\hat R}
\otimes V_{\hat{\ov{R}}}\m$, \,also of the latter space.
The formal scalar product and the formal bilinear form 
(\ref{fbf}) on $\Gamma(\CL)$ correspond to the scalar product 
and the bilinear form on $\CH$ induced by the scalar
product on the representation spaces $V_{_{\hat{R}}}$
and the bilinear pairing between $\m V_{_{\hat{R}}}$
and $V_{_{\hat{\ov{R}}}}$ (the latter induces the pairing
between the $\hat R$ and $\hat{\ov{R}}$ summands in $\CH$).
\vskip 0.3cm

The action of the pair of the current algebras in $\CH$
leads to the (projective) action in $\CH$ of the algebra 
of conformal symmetries. Let us discuss how this occurs.
The spaces $V_{\hat R}$ of the irreducible unitary representations
$\hat R$ of $\hat\Ng$ appear to carry also the unitary representations
of the {\bf Virasoro algebra} $Vir$, the central extension of the 
algebra of the vector fields $Vect(S^1)$ on the circle,
\qq
0\rightarrow\NR\rightarrow  Vir\rightarrow Vect(S^1)\rightarrow 
0\,.
\label{vir}
\qqq
The complex generators $\ell_n$ of $\m Vir$ corresponding to 
the vector fields $\,i\m\ee^{\m in\varphi}{\da_\varphi}\,$ satisfy
the commutation relations
\qq
[\ell_n,\m\ell_m]\ =\ (n-m)\m\ell_{n+m}\,+\,{_1\over^{12}}\m
C\m(n^3-n)\m\delta_{n+m,0}\,,
\label{Vir}
\qqq
where $C$ is the central generator, the image of $1$ under
the second arrow in the exact sequence (\ref{vir}).
The action of the generators $\ell_n$ in the spaces of the 
representations $\hat R$ of $\m\hat\Ng\m$ gives rise to the 
set of operators $L_n$ and $\tilde L_n$ acting in the space 
of states $\,\oplus_{_{\hat R}}V_{_{\hat R}}\otimes
V_{_{\hat{\ov{R}}}}\m.$ They implement a projective action
of $\,Vect(S^1)\oplus Vect(S^1)\m$, \,the Lie algebra
of Minkowskian conformal transformations. 
$\,Vect(S^1)\oplus Vect(S^1)\,$ is, indeed, the
Lie algebra of the infinitesimal transformations preserving
the conformal class of the Minkowski metric $dx^2-dt^2=dx^+dx^-$,
where $\,x^\pm=x\pm t\,$ are the light-cone coordinates
on the cylinder with periodic space-coordinate $x$.
\vskip 0.3cm

Explicitly, the operators $L_n$'s and $\tilde L_n$'s 
are given in terms of the operators $J^a_n$ 
and $\tilde J^a_n$, generating the actions of $\hat\Ng$, 
by the so called Sugawara construction:
\qq
L_n\ =\ {_1\over^{k+h^\vee}}\sum\limits_{m=-\infty}^\infty
J^a_{n-m}J^a_{m}\quad\ {\rm for}\ \ n\not=0\,,\qquad
L_0\ =\ {_2\over^{k+h^\vee}}\sum\limits_{m=0}^\infty 
J^a_{-n}J^a_n
\label{sug}
\qqq
and similarly for $\tilde L_n$. Above, $h^\vee$ (the {\bf dual Coxeter 
number}) stands for the value of the quadratic Casimir in the adjoint 
representation of $\Ng$ and is equal to $N$ for the $SU(N)$
group. The operators $L_n$ and $\tilde L_n$
satisfy the relations (\ref{Vir}) with $C$ acting as the multiplication
by $c={k\,{\rm dim}(G)\over{k+h^\vee}}$, the value 
of the {\bf Virasoro central charge} of the WZW theory. Besides,
\qq
[L_n,\m J^a_m]\ =\ -m J^a_{n+m}
\non
\qqq
and similarly for $[\tilde L_n,\m\tilde J^a_m]$. The operators
$L_n$ (and $\tilde L_n$) satisfy the unitarity conditions 
$L_n^{\m\dagger}=L_{-n}$. In particular $L_0$ is a self-adjoint 
operator, bounded below on $V_{_{\hat R}}$ by 
the {\bf conformal dimensions} $\Delta_{_R}={c_{_R}\over k+h^\vee}$, 
the eigenvalue of $L_0$ on the subspace $V_{_R}\subset V_{_{\hat R}}$.
The latter subspace is annihilated by all $L_n$ with $n>0$. 
The Hamiltonian of the
WZW theory is $\,H=L_0+\tilde L_0-{c\over 12}\,$ whereas
$L_0-\tilde L_0$ defines the momentum operator $P$. The 
tensor product of the HW vectors in the subspace 
$\,V_{_{\hat 1}}\otimes\ov{V}_{_{\hat 1}}\subset\CH\,$
corresponding to the trivial representation $R=1$
gives the {\bf vacuum state} $\Omega$ of the theory annihilated 
by $L_0$ and $\tilde L_0$.
\vskip 0.3cm

A certain role in what follows will be played by the
characters of the representations $V_{_{\hat R}}$
defined as traces of loop group operators 
acting in $V_{_{\hat R}}$. To avoid domain problems,
one often considers only the endomorphisms 
$g_{_{\hat R}}$ of $V_{_{\hat R}}$ representing the action 
of the elements $g\in G$ (or in $G^\NC$)
and obtained by the integration of the action of the
generators $t^a_0$. One then defines 
\qq
\chi_{_{\hat R}}(\tau,g)\ =\ \tr_{_{V_{_{\hat R}}}}
\ee^{\m2\pi i\tau\m(L_0-{c\over 24})}\, g_{_{\hat R}}\,,
\label{hatch}
\qqq
where $\tau$ is a complex number in the upper half plane:
${\rm Im}\m\tau>0$. The presence of the factor $\,\ee^{\m2\pi 
i\tau\m L_0}\,$ renders the trace finite. The characters
$\chi_{_{\hat R}}(\tau,g)$ are class functions of $g$
and may be explicitly computed. Their decomposition 
into characters of $G$ encodes the multiplicities 
of the eigenvalues of the Virasoro generator
$L_0$ in the subspaces of $V_{_{\hat R}}$ transforming
according to a given representation of $G$. 
\vskip 0.3cm

The central charge $\m c\m$ entering the commutation 
relations of the Virasoro generators is an important 
characteristic of a conformal field theory.
It appears also in the rigorous version of the quantum amplitudes 
$\CA_{_\Sigma}$. It enters into them in a somewhat subtle way, 
measuring their change under the local rescalings $\eta\mapsto 
\ee^{2\sigma}\eta$ of the metric of $\Sigma$ (recall that 
the amplitudes of the classical configurations $\ee^{-S(g)}$ 
were invariant under such rescalings). Under the change 
$\eta\mapsto\ee^{2\sigma}\eta$ with $\sigma$ vanishing 
around the boundary,
\qq
\CA_{_\Sigma}\ \ \mapsto\ \ \ee^{\m{c\over 12\m\pi}\int_{_\Sigma}
\hspace{-0.06cm}[\hf\m\da_\alpha\sigma\m\da_\beta\sigma\m
\eta^{\alpha\beta}\m+\m\sigma\m r\m+\m\mu(\ee^{2\sigma}-1)]
\sqrt{\eta}}\ \CA_{_\Sigma}
\label{scin}
\qqq
due to renormalization effects, with $\mu$ depending on 
the renormalization prescription. We shall use the prescription 
corresponding to $\mu=0$. The same transformation rule is obeyed 
by the amplitudes $\CA_{_\Sigma}(A)$ in external gauge field. 
Hence, the quantum amplitudes are only projectively 
invariant under the conformal
rescalings of the metric $\eta$. This is an example 
of the standard effect leading to projective actions  
of symmetries in quantum theory. Due to this effect, some
care will have to be taken when making sense out of 
formal properties of the quantum amplitudes, like
the gluing property (\ref{glueq}). We shall always assume
that the metric $\eta$ of $\Sigma$, which, together 
with the orientation of $\Sigma$, defines its complex structure,
is of the special form around the boundary.
Namely, that, in terms of the complex coordinate of 
the unit discs $D_{n}$ holomorphically embedded into
the surface $\Sigma'$ without boundary such that 
$\,\Sigma=\Sigma'\setminus(\mathop{\cup}\limits_n 
D_{n}^{^{^{{\hspace{-0.21cm}}o}}})\m$, it is equal to 
the cylindrical metric $\vert z\vert^{-2}\vert dz\vert^2$. 
Upon gluing of surfaces along boundary components, such metrics
will automatically give smooth metrics on the resulting
surfaces. 
\vskip 0.3cm

In particular, the metrics on the unit discs $D$
will have the form $\ee^{2\sigma}\vert dz\vert^2$ 
with $\sigma=-\ln{\vert z\vert}$ around the boundary
of $D$. Unless otherwise stated, we shall also assume 
that $\sigma=0$ around the center of $D$. 
Consider the Riemann sphere $\NC P^1=\NC\cup\{\infty\}$ 
composed from the two copies of the disc $D$ glued along 
the boundary. The choice 
\qq
\zeta\ =\ \CZ_{\NC P^1}^{-\hf}
\label{ncst}
\qqq
for the normalizing constant will make precise 
the assignment (\ref{HWq}). This choice guarantees 
that the change of $\zeta$ under the rescalings of the metric 
on $D$ will cancel the change of the functional integral 
$\,\int\ov{g^{ij}_{_R}}\,\ee^{-S(g)}\,Dg\m$.
\vskip 0.4cm

\subsection{Correlation functions}

The formalism of Green functions encoding the action of field 
operators constitutes a traditional tool in quantum field theory. 
In the Minkowski space, the Green functions allow
to express easily the scattering matrix elements
(at least for the massive theories, via the LSZ formalism) 
whereas in the Euclidean space they
coincide with correlation functions of continuum
statistical models, providing a bridge between quantum
field theory and statistical mechanics. In the context
of CFT, the correlation functions defined on a general 
Riemann surface $\Sigma$ without boundary constitute 
a somewhat easier objects to deal with than 
the quantum amplitudes $\CA_{_\Sigma}$ for surfaces with
boundary. Besides, even considered on the simplest Riemann
surface, the Riemann sphere $\NC P^1$,
they already contain the full information about the model.
Formally, the correlation functions of the WZW model
are given by the functional integrals
\qq
<\prod\limits_n g^{i_nj_n}_{_{R_n}}(\xi_n)>_{_{\hspace{-0.08cm}\Sigma}}
\hspace{-0.1cm}(A)\ =\ \CZ_{_\Sigma}(A)^{-1}
\int\prod\limits_n{g_{_{R_n}}^{i_nj_n}}\ \ee^{-S_{_\Sigma}
(g,\m A)}\,\,Dg\,,
\label{cfu}
\qqq
where $\xi_n$ are disjoint points in a Riemann surface $\Sigma$ 
without boundary. For the Riemann surface without boundary
$\Sigma'$ obtained by gluing unit discs $D_{n}$ to a surface $\Sigma$ 
with boundary\footnote{the metric $\eta$ on $\Sigma'$ 
is assumed to come from metrics on $\Sigma$ and on the discs 
$D_{n}$ of the type described above}, \m see Fig.\,\,6,
and for the points $\xi_n$ placed at the centers of the discs 
$D_{n}$, the correlation functions without the gauge field 
may be expressed, with the use of the assignment (\ref{HWq}) 
and of the gluing property (\ref{glueA}), by the scalar products
of the quantum amplitudes $\CA_{_\Sigma}\m$ with special vectors
in the Hilbert space of states:
\qq
<\prod\limits_n g^{i_nj_n}_{_{R_n}}(\xi_n)>_{_{\hspace{-0.08cm}\Sigma'}}
\ =\ \CZ_{_\Sigma}^{-1}\,\,{\bf(}\mathop{\otimes}
\limits_{n}(e^{i_n}_{_{R_n}}\otimes\ov{e^{i_n}_{_{R_n}}})
\m,\,\m\CA_{_\Sigma}\m{\bf)}\,.
\non
\qqq
The normalization factor is given by the partition
function of the surface with boundary $\Sigma$ defined by
\qq
\CZ_{_{\Sigma}}\ =\ \CZ_{_{\Sigma'}}\prod_n\zeta_n
\qqq
with $\zeta_n$ as in Eq.\,\,(\ref{ncst}).
The combination of the partition functions 
on the right hand side does not change under local rescalings 
of the metric inside the discs $D_{n}$.
\vskip 0.3cm

On the level of correlation functions,
the symmetry properties of the theory are encoded
in the so called Ward identities. For example,
the behavior (\ref{trprop}) of the action 
under the chiral gauge transformations with
for $h_1=h$ and $h_2=1$ implies formally that
\qq
&&\CZ_{_{\Sigma}}(A)\,<\mathop{\otimes}\limits_n g_{_{R_n}}
(\xi_n)>_{_{\hspace{-0.08cm}\Sigma}}\hspace{-0.1cm}(A)\cr\cr
&&\hspace{0.6cm}=\ \ee^{-S(h^{-1},\m A^{01})}\ 
\mathop{\otimes}\limits_n h^{-1}_{_{R_n}}(\xi_n)\ 
\CZ_{_{\Sigma}}(A^{10}+{}^{h}\hspace{-0.09cm}A^{01})
\,<\mathop{\otimes}\limits_n g_{_{R_n}}(\xi_n)
>_{_{\hspace{-0.08cm}\Sigma}}\hspace{-0.1cm}
(A^{10}+{}^{h}\hspace{-0.09cm}A^{01})\,,\hspace{1cm}
\label{WIcfl}
\qqq
where we view $\,\otimes g_{_{R_n}}\,$ 
as taking value in $\,\mathop{\otimes}\limits_nEnd(V_{_{R_n}})$
and collecting all the matrix elements $\,\prod g^{^{i_nj_n}}
_{_{R_n}}\m$. \,Similarly, for $h_1=1$ and $h_2=h$, we obtain
the mirror relation:
\qq
&&\CZ_{_{\Sigma}}(A)\,<\mathop{\otimes}\limits_n g_{_{R_n}}
(\xi_n)>_{_{\hspace{-0.08cm}\Sigma}}\hspace{-0.1cm}(A)\cr\cr
&&\hspace{0.6cm}=\ \ee^{-S(h,\m A^{10})}
\ \CZ_{_{\Sigma}}({}^{h}\hspace{-0.09cm}A^{10}+A^{01})\,
<\mathop{\otimes}\limits_n g_{_{R_n}}(\xi_n)
>_{_{\hspace{-0.08cm}\Sigma}}\hspace{-0.1cm}
({}^{h}\hspace{-0.09cm}A^{10}+A^{01})
\ \mathop{\otimes}\limits_n h_{_{R_n}}(\xi_n)\,.\hspace{0.8cm}
\label{WIcfr}
\qqq
These are the Ward identities expressing the symmetry
of the correlation functions under the chiral gauge 
transformations. 
\vskip 0.3cm

It is useful and customary to introduce more general 
correlation functions with insertions of {\bf currents} testing
the reaction of the functions (\ref{cfu})
to infinitesimal changes of the gauge fields.
On the surface $\Sigma'$ they are defined by
\qq
\CZ_{_{\Sigma'}}<J^a(z_m)\prod\limits_n g^{^{i_nj_n}}_{_{R_n}}
(\xi_n)>_{_{\hspace{-0.08cm}\Sigma'}}
\,=\,-\pi\m{{\delta}\over{\delta A^a_{\bar z}(z_m)}}
{\bigg\vert}_{_{A=0}}\CZ_{_{\Sigma'}}\hspace{-0.1cm}(A)
<\prod\limits_n g^{i_nj_n}_{_{R_n}}
(\xi_n)>_{_{\hspace{-0.08cm}\Sigma'}}\hspace{-0.1cm}(A)\,,
\hspace{0.6cm}
\qqq
where $z_m$ is the complex coordinate of the disc $D_m$, or by 
\qq
\CZ_{_{\Sigma'}}<\tilde J^a(\bar z_m)\prod\limits_n 
g^{^{i_nj_n}}_{_{R_n}}(\xi_n)>_{_{\hspace{-0.08cm}\Sigma'}}
\,=\,-\pi\m{{\delta}\over{\delta A^a_{z}(z_m)}}
{\bigg\vert}_{_{A=0}}\CZ_{_{\Sigma'}}(A)
<\prod\limits_n g^{i_nj_n}_{_{R_n}}(\xi_n)
>_{_{\hspace{-0.08cm}\Sigma'}}(A)\,.\hspace{0.6cm}
\qqq
Its is not very difficult to show, expanding the
Ward identities (\ref{WIcfl}) and (\ref{WIcfr})
to the first order in $h$ around $1$, 
that the insertions of $J^a(z)$ ($\tilde J^a(\bar z)$)
are analytic (anti-analytic) in $z\not=0$ but that
they have simple poles at $z=0$, the location point 
of one of the insertions $g_{_{R}}(\xi)$, 
with the behavior
\qq
J^a(z)\,\m g_{_{R}}(\xi)\ =\ -\m{1\over{z}}\,\m t^a_{_R}\,
g_{_{R}}(\xi)\ +\ \dots\,,\qquad
\tilde J^a(\bar z)\,\m g_{_{R}}(\xi)\ =\ {1\over{\bar z}}\,\m
g_{_{R}}(\xi)\, t^a_{_R}\ +\ \cdots\,,\hspace{0.8cm}
\label{fope}
\qqq
where the dots denote non-singular terms. The latter are 
related to the action of the current algebra generators 
in the space of states by the following relations involving
the contour integrals\footnote{oriented counter-clockwise}:
\qq
{_1\over^{2\pi i}}\hspace{-0.2cm}\int\limits_{\vert z_m\vert=\rho}
\hspace{-0.3cm}
<J^a(z_m)\prod\limits_n g^{i_nj_n}_{_{R_n}}(\xi_n)
>_{_{\hspace{-0.08cm}\Sigma'}}
\, z_m^p\m dz_m&=& 
-\m\CZ_{_\Sigma}^{-1}\,\,{\bf(}\mathop{\otimes}
\limits_{n}({J^a_{p}}^{^{\hspace{-0.05cm}\#}}\m e^{i_n}_{_{R_m}}
\otimes
\ov{e^{i_n}_{_{R_n}}})\m,\,\m\CA_{_\Sigma}\m{\bf)}\,,
\label{cfl}\hspace{1.3cm}\\
-\m{_1\over^{2\pi i}}\hspace{-0.2cm}\int\limits_{\vert z_m\vert=\rho}
\hspace{-0.3cm}
<\tilde J^a(z_m)\prod\limits_n g^{i_nj_n}_{_{R_n}}(\xi_n)
>_{_{\hspace{-0.08cm}\Sigma'}}
\,\bar z_m^p\m d\bar z_m&=& 
-\m\CZ_{_\Sigma}^{-1}\,\,{\bf(}\mathop{\otimes}
\limits_{n}(e^{i_n}_{_{R_n}}\otimes
\tilde J^a_{p}{}^{^{\hspace{-0.05cm}\#}}\m
\ov{e^{i_n}_{_{R_n}}})\m,\,\m
\CA_{_\Sigma}\m{\bf)}\hspace{1.3cm}
\label{cfr}
\qqq
with $\rho<1$ and the superscript $\#$ indicating that the operator
appears only for $n=m$. Eqs.\,\,(\ref{fope}) are examples
of the {\bf operator product expansions}, in this case, the ones
stating that $g_{_{R}}(\xi)$ are {\bf primary fields} of the
current algebra, in the CFT jargon. 
\vskip 0.3cm

Multiple current insertions, see Fig.\,\,8, integrated over contours 
of increasing radii lead to multiple insertions of the current algebra 
generators. For example:
\qq
{_1\over^{2\pi i}}\hspace{-0.2cm}\int\limits_{\vert z_m\vert=\rho_1}
\hspace{-0.3cm}
{_1\over^{2\pi i}}\hspace{-0.2cm}\int\limits_{\vert w_m\vert=\rho_2}
\hspace{-0.3cm}
<J^a(z_m)\,J^b(w_m)
\prod\limits_n g^{i_nj_n}_{_{R_n}}(\xi_n)
>_{_{\hspace{-0.08cm}\Sigma'}}\,\m z_m^{p}\,w_m^q
\,\m dz_m\, dw_m\hspace{3cm}\cr
=\ \CZ_{_\Sigma}^{-1}\,\,{\bf(}\mathop{\otimes}
\limits_{n}(J^a_{p}{}^{^{\hspace{-0.05cm}\#}}
J^b_{q}{}^{^{\hspace{-0.05cm}\#}}\m e^{i_n}_{_{R_n}}
\otimes\ov{e^{i_n}_{_{R_n}}})\m,\,\m\CA_{_\Sigma}\m{\bf)}
\hspace{1cm}
\non
\qqq
for $\rho_1>\rho_2$ whereas for $\rho_1<\rho_2$
the order of $J^a_{p}J^b_{q}$
should be reversed. In particular,
the commutator $\m[J^a_{p},\m J^b_{q}]\m$ corresponds
to the difference of the two double contour integrals.
It follows, that general matrix elements 
of the quantum amplitudes $\CA_\Sigma$ may be read 
of the correlation function in the external gauge. It is then 
enough to find the latter to describe the complete theory.

\leavevmode\epsffile[-80 -20 337 190]{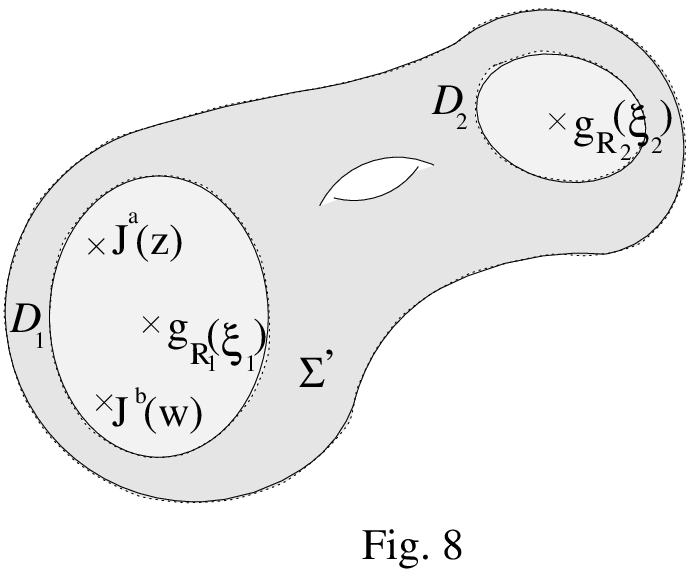}

\vskip 0.1cm

The action of the Virasoro generators may be interpreted 
similarly in terms of the insertions of the {\bf energy-momentum} 
tensor into the correlation functions which test their variation
under the changes of the metric on the surface:
\qq
\CZ_{_{\Sigma'}}\,<T(z_m)\prod\limits_n g^{i_nj_n}_{_{R_n}}(\xi_n)
>_{_{\hspace{-0.08cm}\Sigma'}}
\,&=&\,4\pi\m{{\delta}\over{\delta\eta^{zz}(z_m)}}\,\m
\CZ_{_{\Sigma'}}\,
<\prod\limits_n g^{i_nj_n}_{_{R_n}}(\xi_n)
>_{_{\hspace{-0.08cm}\Sigma'}}\,,
\cr
\CZ_{_{\Sigma'}}\,<\tilde T(\bar z_m)\prod\limits_n 
g^{i_nj_n}_{_{R_n}}(\xi_n)>_{_{\hspace{-0.08cm}\Sigma'}}
\,&=&\,4\pi\m{{\delta}\over{\delta\eta^{\bar z\bar z}(z_m)}}\,\m
\CZ_{_{\Sigma'}}\,<\prod\limits_n g^{i_nj_n}_{_{R_n}}(\xi_n)
>_{_{\hspace{-0.08cm}\Sigma'}}\,.
\qqq
Under the local rescaling of the metric $\eta\mapsto\ee^{2\sigma}
\eta$ with $\sigma$ vanishing around the insertion 
points $\xi_n$, the correlation functions (\ref{cfu}) are invariant. 
This is not any more the case for general $\sigma$ due
to (the ``wave function'') renormalization of the insertions. 
For general $\sigma$, the correlation functions 
pick up the product of local factors equal to 
$\ee^{-2\Delta_{_{R_n}}\hspace{-0.06cm}\sigma(\xi_n)}$, 
where the {\bf conformal
dimension} $\Delta_{_R}$ of the fields $g_{_R}(\xi)$
coincide with the lowest eigenvalues of the Virasoro generator
$L_0$ in the HW representations of the current algebra
discussed before. The partition functions transform under
the metric rescaling according to the rule (\ref{scin}).   
The infinitesimal versions of the above transformation properties
together with the covariance of the whole scheme under
infinitesimal diffeomorphisms of the surface $\Sigma$ may be shown
to imply that the insertions of $T(z_m)$ $(\tilde T(\bar z_m)$) 
are analytic (anti-analytic)\footnote{in the standard metric 
$\vert dz\vert^2$ around the insertion point} in $z_m$ for $z_m\not=0$
with the singular part given by the operator product expansion
\qq
&&T(z)\,\m g_{_R}(\xi)\ =\ {1\over{z^2}}\,{\Delta_{_R}}\,g_{_R}(\xi)
\,+\,{1\over z}\,\da_z\m g_{_R}(\xi)\ +\ \dots\,,\cr\cr
&&\tilde T(\bar z)\,\m g_{_R}(\xi)\ =\ {1\over{{\bar z}^2}}
\,{\Delta_{_R}}\,g_{_R}(\xi)
\,+\,{1\over{\bar z}}\,\da_{\bar z}\m g_{_R}(\xi)\ +\ \dots\,.
\nonumber
\qqq
The latter expansions state that $g_{_R}(\xi)$ are primary fields
of the Virasoro algebra. The insertions of the energy momentum 
tensor encode the action of the Virasoro algebra generators 
in the space of states:
\qq
{_1\over^{2\pi i}}\hspace{-0.2cm}\int\limits_{\vert z_m\vert=\rho}
\hspace{-0.3cm}
<T(z_m)\prod\limits_n g^{i_nj_n}_{_{R_n}}(\xi_n)
>_{_{\hspace{-0.08cm}\Sigma'}}
\, z_m^{p+1}\m dz_m&=& 
\CZ_{_\Sigma}^{-1}\,\,{\bf(}\mathop{\otimes}
\limits_{n}(L_{p}^{^{\hspace{-0.04cm}\#}}\m e^{i_n}_{_{R_n}}\otimes
\ov{e^{i_n}_{_{R_n}}})\m,\,\m\CA_{_\Sigma}\m{\bf)}\,,
\hspace{1.5cm}
\label{cflV}\\
-\m{_1\over^{2\pi i}}\hspace{-0.2cm}\int\limits_{\vert z_m\vert=\rho}
\hspace{-0.3cm}
<\tilde T(\bar z_m)\prod\limits_n g^{i_nj_n}_{_{R_n}}(\xi_n)
>_{_{\hspace{-0.08cm}\Sigma'}}
\,\bar z_m^{p+1}\m d\bar z_m&=& 
\CZ_{_\Sigma}^{-1}\,\,{\bf(}\mathop{\otimes}
\limits_{n}(e^{i_n}_{_{R_n}}\otimes
\tilde L_{p}^{^{\hspace{-0.02cm}\#}}\m\ov{e^{i_n}_{_{R_n}}})\m,\,\m
\CA_{_\Sigma}\m{\bf)}\,.
\hspace{1.5cm}
\label{cfrV}
\qqq
\vskip 0.3cm

The Ward identities of the chiral gauge symmetry together with
the transformation properties under the local rescalings 
of the metric and under diffeomorphisms of the surface,
expanded to the second order in the symmetry generators,
yield the operator product expansions
\qq
J^a(z)\m J^b(w)&=&{{k\m\delta^{ab}}\over{2(z-w)^2}}\,+\,
{{i\m f^{abc}}\over{z-w}}\, J^c(w)\,+\ \dots\,,\cr\cr
T(z)\m T(w)\,\,&=&{{c}\over{2(z-w)^4}}\,+\,
{{2}\over{(z-w)^2}}\, T(w)\,+\,
{1\over{z-w}}\, \da_w T(w)\,+\ \dots\,,\cr\cr
T(z)\m J^a(w)&=&{{1}\over{(z-w)^2}}\, J^a(w)\,+\,
{1\over{z-w}}\, \da_w J^a(w)\,+\ \dots\,,\cr
\label{opem}\\
\tilde J^a(\bar z)\m\tilde J^b(\bar w)&=&{{k\m\delta^{ab}}
\over{2(\bar z-\bar w)^2}}\,+\,
{{i\m f^{abc}}\over{\bar z-\bar w}}\,\tilde J^c(\bar w)\,
+\ \dots\,,\cr\cr
\tilde T(\bar z)\m\tilde T(w)\,\,&=&{{c}
\over{2(\bar z-\bar w)^4}}\,
+\,{{2}\over{(\bar z-\bar w)^2}}\,\tilde T(\bar w)\,+\,
{1\over{\bar z-\bar w}}\, \da_{\bar w}\tilde T(\bar w)\,
+\ \dots\,,\cr\cr
\tilde T(\bar z)\m\tilde J^a(\bar w)
&=&{{1}\over{(\bar z-\bar w)^2}}
\,\tilde J^a(\bar w)\,+\,{1\over{\bar z-\bar w}}\, 
\da_{\bar w}\tilde J^a(w)\,+\ \dots\,,
\nonumber
\qqq
where the dots denote the non-singular terms analytic 
(anti-analytic) in $z$ around $w$. The above expansions 
hold when inserted
into the correlation functions as above with $z$ and
$w$ corresponding to the values of the same local coordinate
for two different points and in the standard metric. 
They encode through the relations (\ref{cfl}), (\ref{cfr}), 
(\ref{cflV}) and (\ref{cfrV}) the commutation relations
of the current and Virasoro generators 
$J_n^a,\m\tilde J^a_n,\m L_n,\m\tilde L_n$
obtained from the above expansions by the deformation of
the integration contours 
\qq
\int\limits_{\vert z\vert=\rho+\epsilon}dz
\int\limits_{\vert w\vert=\rho}dw
-\int\limits_{\vert z\vert=\rho-\epsilon}dz
\int\limits_{\vert w\vert=\rho}dw
\ =\ \int\limits_{\vert w\vert=\rho}dw
\int\limits_{\vert z-w\vert=\epsilon}dz\,,
\qqq
see Fig.\,\,9, and the use of the residue theorem. 

\leavevmode\epsffile[-70 -20 347 155]{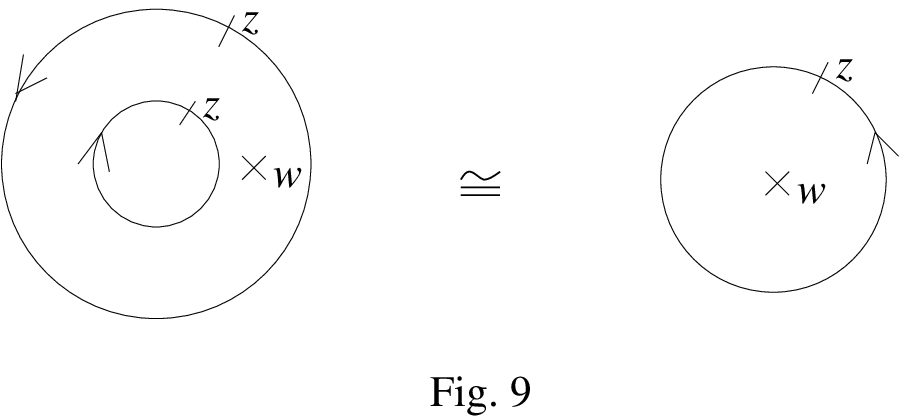}

\noindent The operator expansion
algebra of the insertions into the correlation functions
substitutes then for the operator commutation relations
but allows to encode also more complicated algebraic
relations between the CFT operators, see the last
section. It is the basic technique of two-dimensional 
CFT. 
\vskip 0.3cm

As we have discussed before, the action of the Virasoro generators
in the space of states of the WZW model may be expressed
in terms of the current algebra action, see Eq.\,\,(\ref{sug}). 
This relation may be translated into the language
of the insertions into the correlation functions,
giving rise to the Sugawara construction
of the energy-momentum tensor: 
\qq
T(w)\ =\ \lim\limits_{z\to w}\ {_1\over^{k+h^\vee}}\left(
J^a(z)\m J^a(w)\,-\,{_{k\,{\rm dim}(G)}\over^{2\m(z-w)^2}}\right)
\non
\qqq
and similarly for $\tilde T(\bar w)$.
\vskip 0.5cm

\nsection{Chiral WZW theory and the Chern-Simons states}

As we have seen, the whole information about
the quantum amplitudes of the WZW theory resides 
in the correlation functions (\ref{cfu}) in an
external gauge field. We shall look now more closely
into the gauge-field dependence of these functions.
By (formal) analytic continuation, the chiral
Ward identities (\ref{WIcfl})
and (\ref{WIcfr}) should also hold for the complexified 
gauge fields $A$ with values in $\Ng^\NC$
and for the complexified gauge transformations $h$ with
values in $G^\NC$. As we shall see, they give a powerful 
tool for analysis of the correlation functions.
\vskip 0.3cm

Let us consider first the Ward identity (\ref{WIcfl}).
The holomorphic maps $\Psi$ on the space ${\CA}^{01}$ of 
$\Ng^\NC$-valued 0,1-gauge fields $A^{01}$ with values in 
$\,\mathop{\otimes}\limits_nV_{_{R_n}}\,$
satisfying the equation
\qq
\Psi(A^{01})\ =\ \ee^{-S(h^{-1},\m A^{01})}\ 
\mathop{\otimes}\limits_n h^{-1}_{_{R_n}}(\xi_n)\ 
\Psi({}^{h}\hspace{-0.09cm}A^{01})
\label{CSst}
\qqq
for $h$ in the group $\CG^\NC$ of $G^\NC$-valued gauge 
transformations have an interesting geometric interpretation. 
On one side, they may be viewed as holomorphic sections 
of a vector bundle $W$ with typical fiber 
$\,\mathop{\otimes}\limits_nV_{_{R_n}}\,$
over the orbit space\footnote{this space requires 
a careful definition with a special treatment of
bad orbits} $\CN=\CA^{01}/\CG^\NC$. Mathematically, the orbit space
$\CN$ is the moduli space of the holomorphic $G^\NC$-bundles 
and the mathematicians like to view $\Psi$'s as non-abelian 
generalizations of the classical theta functions.  
Indeed, the latter are holomorphic sections of a line bundle 
over the moduli space (the Jacobian) of the holomorhic 
$\NC^*$-bundles over a Riemann surface. 
\vskip 0.4cm

\subsection{States of the Chern-Simons theory}

Physically, the holomorphic maps $\Psi$ satisfying the Ward
identity (\ref{CSst}) may be identified as the quantum
states of the three-dimensional Chern-Simons (CS) gauge theory
\cite{27}. The classical phase space of the CS
theory on the 3-manifold $\,\Sigma\times\NR\,$ is composed 
of the flat $\Ng$-valued gauge fields $\,i\m A\,$ on $\Sigma$
modulo $G$-valued gauge transformations. The flatness 
condition is
\qq
F(A)\,\equiv\,dA\m+\m A^2\ =\ 0\,.
\label{fla}
\qqq
In the holomorphic quantization {\it \`{a} la} Bargmann, 
the quantum states of the theory are described 
as holomorphic functionals $\Psi$ on the space $\CA^{01}$
with the condition (\ref{fla}) imposed as a quantum constraint:
\qq
F(A)\,\,\Psi(A^{01})\ =\ 0\,,
\label{flaq}
\qqq
with $F(A)$ as in Eq.\,\,(\ref{fla}) but with $A^{01}$ acting 
as the multiplication operator and $A^{10}\m$ 
as the differentiation: 
$\,A^a_{z}=-\m{{2\pi}\over k}\,{{\delta}
\over{\delta A^a_{\bar z}}}$. \m The constraint (\ref{flaq}) 
is closely related to the infinitesimal Ward 
identity:
\qq
(\m F(A)\m+\m{_{4\pi i}\over
^{k}}\sum\limits_n \delta_{\xi_{_n}}\,t^a\m t^a_{_{R_n}})
\,\,\Psi(A)
\ =\ 0
\label{IWI}
\qqq
obtained by expanding the global Ward identity
(\ref{CSst}) to the first order in $h$ around $1$.
The infinitesimal identity (\ref{IWI}) is equivalent 
to its global version (\ref{CSst}). In the absence 
of insertions, it coincides with Eq.\,\,(\ref{flaq}). 
The modifications involving the insertions correspond 
in the CS gauge theory language to the insertions 
of the Wilson lines $\{\xi_n\}\times\NR$ in 
representations $R_n$. 
\vskip 0.3cm

It is a crucial fact that the ($k$-dependent) spaces 
$\m\NW_{_\Sigma}(\un{\xi},\un{R})\m$ of the holomorphic maps 
$\Psi$'s satisfying the Ward identities (\ref{CSst}) 
or (\ref{IWI}) are finite-dimensional, with the dimension 
given by the celebrated Verlinde formula \cite{Verl}.
In particular, only representations $R_n$ with 
HW's $\lambda_{_{R_n}}$ integrable at level $k$ may 
give rise to non-trivial spaces 
$\m\NW_{_\Sigma}(\un{\xi},\un{R})$.
It is instructive to look more carefully into the case of 
the Riemann sphere $\NC P^1$. \m On $\m\NC P^1$ all gauge
fields $A^{01}$ may be written in the form $h^{-1}\de h$ 
or may be approximated by the fields of this form. In other
words, the gauge orbit of $A^{01}=0$ is dense in $\CA^{01}$.
But by Eq.\,\,(\ref{CSst}), 
\qq
\Psi(h^{-1}\de h)\ =\ \ee^{\m S(h)}\ 
\mathop{\otimes}\limits_n h^{-1}_{_{R_n}}(\xi_n)\ \Psi(0)\,,
\label{etse}
\qqq
where $\,\Psi(0)\in\mathop{\otimes}\limits_nV_{_{R_n}}\,$
is an element of a finite-dimensional space. Hence $\Psi(0)$ 
determines $\Psi$ on a dense set of gauge fields $A$, 
so everywhere. In fact, $\Psi(0)$ belongs to the subspace 
$\,(\mathop{\otimes}\limits_nV_{_{R_n}})^{^G}\,$ of
tensors invariant under the diagonal action of $\m G$,
as is easy to see by taking constant $h$ 
in Eq.\,\,(\ref{etse}). We obtain then a natural embedding
\qq
\NW_{_{\NC P^1}}(\un{\xi},\un{R})\ \ \subset\ \ 
(\mathop{\otimes}\limits_nV_{_{R_n}})^{^G}\,.
\qqq
The images of $\m\NW_{_{\NC P^1}}(\un{\xi},\un{R})\m$ 
in the spaces of invariant tensors are, in general, proper 
subspaces of $\,(\mathop{\otimes}\limits_n
V_{_{R_n}})^{^G}$ depending on $k$. The reason 
is that the $\Psi$'s defined
by Eq.\,\,(\ref{etse}) on $A^{01}=h^{-1}\de h$ do not
extend holomorphically to all of $\CA^{01}$
for all invariant tensors $\Psi(0)$.
In particular, the image of $\NW_{_{\NC P^1}}(\un{\xi},\un{R})$, 
which is zero if some of HW's $\lambda_{_{R_n}}$ are not 
integrable at level $k$, becomes the whole 
space of invariant tensors for sufficiently large $k$. 
\vskip 0.3cm

For genus one, i.e. on the complex tori 
$\m{T}_\tau=\NC/(\NZ+\tau\NZ)$, a dense set of  
gauge fields is formed by the gauge orbits 
of the fields
\qq
A^{01}_{u}={_\pi\over^{{\rm Im}\m\tau}}\m{u}\, d\bar z
\non
\qqq
with $\m{u}\m$ in the complexified Cartan algebra $\Nt^\NC
\subset\Ng^\NC$. It is then enough to know the CS
states $\Psi$ only on the gauge fields $A^{01}_u$.
In particular, in the case with no insertions,
the holomorphic functions $\m\psi\m$ defined by
\qq
\psi(u)\ =\ \ee^{-\m{\pi\m k\over2\m{\rm Im}
\m\tau}\m\,\tr\,\m{u}^2}\,\,\Psi(A^{01}_{u})
\non
\qqq
characterize completely the CS states $\m\Psi\m$. 
\vskip 0.3cm

It appears that the functions $\psi(u)$ are 
arbitrary combinations of the characters 
$\chi_{_{\hat R}}(\tau,\ee^{\m2\pi i\m u})$
of the HW representations of the current algebra $\hat\Ng$,
see Eq.\,\,(\ref{hatch}). This fact implies an important property 
of the latter. Recall that the tori ${T}_\tau$
and ${T}_{\tau'}$ for $\tau'=-{1\over\tau}$ may be identified
by the map $\m z\mapsto z'=-z/\tau$. Under this
identification, $A^{01}_{u'}\mapsto A^{01}_u$ if $u'=-u/\tau$.
It follows then that the characters $\chi_{_{\hat{R}'}}(\tau',
\ee^{\m 2\pi i\m u'})$ of the current algebra
are combinations of the characters $\chi_{_{\hat{R}}}(\tau,
\ee^{\m 2\pi i\m u})\m$:
\qq
\chi_{_{\hat{R}'}}(\tau',\ee^{\m 2\pi i\m u'})\ =\ 
\sum\limits_{\hat{R}}\m S_{_{R'}}^{^{R}}\,\,
\chi_{_{\hat{R}}}(\tau,\ee^{\m 2\pi i\m u})\,.
\non
\qqq
Hence the modular transformation $\tau\mapsto-1/\tau$ 
(and more generally, the transformations of $SL(2,\NZ)\m$)
\m may be implemented on the characters of the
current algebra. The symmetric unitary matrices 
$(S_{_{R'}}^{^{R}})$ representing the action 
of the transformation $\tau\mapsto-1/\tau$ may be expressed 
explicitly by the characters $\chi_{_R}$ of the 
group $G$ and by the Weyl denominator of Eq.\,\,(\ref{WD}),
\qq
S^{^{R}}_{_{R'}}\ =\ 1^{_1\over^4}\,\vert T\vert^{-\hf}\,\,
\chi_{_{R'}}(\ee^{\m2\pi i\m\hat\lambda_{_{R}}/\hat k})\,\,
\Pi(\ee^{\m2\pi i\m\hat\lambda_{_{R}}/\hat k})\ =\ 
S^{^{R'}}_{_{R}}\ =\ \ov{S^{^{\ov{R}}}_{_{R'}}}
\label{modm}
\qqq
in the notation: $\hat{\lambda}\equiv\lambda+\rho$ and 
$\hat{k}\equiv k+h^\vee$ with $\rho$ the Weyl vector 
and $h^\vee$ the dual Coxeter number. 
The normalizing factor $\vert\hat T\vert$ is the number 
of the Cartan group elements of the form 
$\ee^{\m 2\pi i\m\hat{\lambda}/\hat{k}}$ with $\lambda$ 
a weight. $1^{_1\over^4}$ is a fourth root of unity.
For the $SU(2)$ group, the above formula reduces to
\qq
S^{^{j}}_{_{j'}}\ =\ \left({_2\over^{k+2}}\right)^\hf\,
\sin{_{\pi\m(2j+1)(2j'+1)}\over^{k+2}}\,.
\qqq
\vskip 0.4cm

\subsection{Verlinde dimensions and the fusion ring}

The dimensions $\m\hat N_{_{\un{R}}}\m$ of the spaces 
$\m\NW_{_\Sigma}(\un{\xi},\un{R})\m$ are independent of the 
complex structure of $\Sigma$ and the locations $\un{\xi}$ 
of the insertion points (but dependent on the level
$k$ of the theory suppressed in the notation). 
They are given by the Verlinde formula which, in the present 
context, is a natural generalization of the classical 
formula for the dimensions $\m N_{_{\un{R}}}\m$ of the spaces 
$\m(\otimes V_{_{R_n}})^{^G}$ of group $G$ invariant tensors. 
The dimensions $\m N_{_R}$ may be computed from the characters
of the representations $\m R_n$:
\qq
N_{_{\un{R}}}\ =\ \int_{_G}\hspace{-0.02cm}\prod\limits_n
\chi_{_{R_n}}(g)\,\, dg
\ =\ \int\prod\limits_n\chi_{_{R_n}}(\ee^{\m 2\pi 
i\m\lambda/k})
\,\m\,\,\vert\Pi(\ee^{\m 2\pi i\m\lambda/k})
\vert^2\,\m d\lambda\,,
\label{wWD}
\qqq
where we have used the relation (\ref{WDcl}). 
For simple, simply connected groups, the last integral 
may be taken over the symplex 
\qq
\Delta_k\ \,=\,\ \{\,\lambda\in\Ntt\,\,\vert\,
\,\tr\,(\alpha^\vee\lambda)\geq 0 
\ {\rm for}\ \alpha>0,\ \tr\,(\phi^\vee\lambda)\leq k\,\}
\label{sympl}
\qqq
whose elements label the conjugacy classes classes $\CC_\lambda$ 
in a one to one way.
Note that the weights in the symplex $\Delta_k$ are exactly 
the HW's integrable at level $k$, see the definition (\ref{iHW}). 
The numbers $N_{_{R\,\m R_1\ov{R}_2}}$ coincide with 
the dimensions $N_{_{R\, R_1}}^{^{R_2}}$ of the multiplicity
spaces in the decomposition (\ref{tpd}) of the tensor 
product of representations, i.e. with the structure
constants of the character ring $\m\CR_{_G}$ of the group $G$, see 
Eq.\,\,(\ref{strf}). 
For example for the $SU(2)$ group, $\,N_{_{j\,\m j_1}}^{^{\m j_2}}
=1\,$ if $\,\vert j-j_1\vert\leq j_2\leq j+j_1\,$ and $\,j+j_1+j_2\,$
is an integer and $\,N_{_{j\,\m j_1}}^{^{\m j_2}}=0\,$ otherwise.
The ring $\m\CR_{_G}$ comes with 
an additive $\NZ$-valued form $\omega$ 
given by the integral over $G$. $\,\omega$ assigns to the combination 
$\sum n_i\chi_{_{R_i}}$ of characters the coefficient of the
character $\chi_{_1}=1$ of the trivial representation $R=1$.
The dimensions $N_{_{\un{R}}}$ are the values of $\omega$ 
on the product of the characters $\chi_{_{R_n}}$ in $\,\CR_{_G}$.
\vskip 0.3cm

The dimensions $\m\hat N_{_{\Sigma,\m\un{R}}}\m$ of the spaces 
$\m\NW_{_\Sigma}(\un{\xi},\un{R})\m$ are given by the formula
\qq 
\hat N_{_{\Sigma,\m\un{R}}}\ =\ {_1\over^{\vert\hat T\vert}}
\sum\limits_{{\rm weights}\atop\lambda\in\Delta_k}
\prod\limits_n\chi_{_{R_n}}(\ee^{\m 2\pi i\m\hat{\lambda}/
\hat{k}})
\,\m\,\,\vert\Pi(\ee^{\m 2\pi i\m\hat{\lambda}/\hat{k}})
\vert^{^{2-g_{_\Sigma}}}\,,
\label{wWDd}
\qqq
in the notations from the end of the last subsection
and with $g_{_\Sigma}$ denoting the genus of the surface 
$\Sigma$. The above equation is 
a rewrite of the original Verlinde formula:
\qq
\hat N_{_{\Sigma,\m\un{\xi}}}\ =\ \sum\limits_{\hat R}
\prod\limits_n (S_{_{R_n}}^{^{R}}/S_{_1}^{^{R}})\ 
(S_{_1}^{^{R}})^{^{2-g_{_\Sigma}}}
\label{orV}
\qqq
which may be easily obtained from Eq.\,\,(\ref{wWDd})
with the use of the explicit expression (\ref{modm})
for the modular matrix $(S_{{R'}}^{R})$.
For the particular case of $\m\Sigma=\NC P^1$, Eq.\,\,(\ref{wWDd})
is clearly a deformation of Eq.\,\,(\ref{wWD}). More exactly,
the sum in Eq.\,\,(\ref{wWDd}) is a Riemann sum approximation 
of the integral in Eq.\,\,(\ref{wWD}).
The genus zero 3-point dimensions $\,\hat N_{_{R\,\m R_1\ov{R}_2}} 
\equiv\hat N_{_{R\, R_1}}^{^{\m R_2}}$ give the structure
constants of a commutative ring $\m{\hat\CR}_{_G}$
which is additively generated by the representations $R$ 
with the HW's integrable at level $k$. 
The ($k$-dependent) ring $\m{\hat\CR}_{_G}$ is called 
the {\bf fusion ring} of the WZW model. 
For the $SU(2)$ group and all spins $\leq{k\over 2}$, 
$\,{\hat N}_{_{j\,\m j_1}}^{^{\m j_2}}
=1\,$ if $\,\vert j-j_1\vert\leq j_2\leq j+j_1\,$ 
and $\,j+j_1+j_2\,$ is an integer $\leq k$ and 
$\,{\hat N}_{_{j,\m j_1}}^{^{\m j_2}}=0\,$ otherwise.
The fusion ring is a deformation 
of the character ring $\m\CR_{_G}$. More exactly,
\qq
{\hat\CR}_{_G}\ \cong\ \CR_{_G}/\hat\CI\,,
\qqq 
where $\hat\CI$ is the ($k$-dependent) ideal in $\CR_{_G}$ 
composed of the functions vanishing on the Cartan group elements 
$\,\ee^{\m2\pi i\m\hat{\lambda}/\hat{k}}\,$ for weights 
$\lambda\in\Delta_k$. The isomorphism identifies
the image of $\chi_{_R}$ in $\CR_{_G}/\hat\CI$ with the generator 
of $\,{\hat\CR}_{_G}$ corresponding to $R$ for representations
$R$ with integrable HW's. The coefficient at the generator corresponding
to the trivial representation defines an additive $\NZ$-valued form
$\hat\omega$ on ${\hat\CR}_{_G}$. For all $R_n$ with integrable HW's, 
the genus zero Verlinde dimensions $\m\hat N_{_{\un{R}}}\m$
are given by the values of $\hat\omega$ 
on the image in the fusion ring of the product
of the characters $\chi_{_{R_n}}$. For fixed 
representations, $\m\hat N_{_{R\, R_1}}^{^{R_2}}
=N_{_{R\, R_1}}^{^{R_2}}\m$ for sufficiently high $k$.
The fusion ring may be also identified as the character
ring of the quantum deformation $\CU_q(\Ng)$ of the
enveloping algebra of $\Ng$ for $q=\ee^{\pi i/(k+h^\vee)}$,
an example of the intricate relations between the
WZW model and the quantum groups.
\vskip 0.4cm

\subsection{Holomorphic factorization}

Consider now the Ward identity (\ref{WIcfr}) for the mirror
chiral gauge transformations.
The anti-holomorphic maps $\Phi$ of the space $\CA^{10}$
of the $\Ng^\NC$-valued 1,0-gauge fields $A^{10}$ with values
in $\mathop{\otimes}\limits_n{V_{_{\ov{R}_n}}}$ such that
\qq
\Phi(A^{10})\ =\ \ee^{-S(h,\m A^{10})}\ 
\mathop{\otimes}\limits_n{h^{\,t}_{_{\ov{R}_n}}(\xi_n)}\ 
\Phi({}^{h}\hspace{-0.09cm}A^{10})
\non
\qqq
are the complex conjugates of the holomorphic maps $\Psi$
satisfying the relation (\ref{CSst}):
\qq
\Phi(A^{10})\ =\ \ov{\Psi(-(A^{10})^*)}\,,
\qqq
where the star denotes the anti-linear involution of the 
complexified Lie algebra $\Ng^\NC$ leaving $\Ng$ invariant
(it coincides with the hermitian conjugation for $\Ng=su(N)$).
It follows that the correlation functions, in their
dependence of the external gauge field, are sesqui-linear
combinations of the elements of the space $\NW_{_\Sigma}
(\un{\xi},\un{R})$ of holomorphic solutions of 
Eq.\,\,(\ref{CSst}):
\qq
\CZ_{_{\Sigma}}(A)\,<\mathop{\otimes}\limits_n g_{_{R_n}}
(\xi_n)>_{_{\hspace{-0.08cm}\Sigma}}\hspace{-0.1cm}(A)\ 
=\ H^{\alpha\beta}\,
\,\m\Psi_\alpha(A^{01})\otimes\ov{\Psi_\beta(-(A^{01})^*)}\,,
\label{sesq}
\qqq
where the states $\Psi_\alpha$ form a basis 
of $\NW_{_\Sigma}(\un{\xi},
\un{R})$ and the right hand side should be summed over
$\alpha$ and $\beta$. The equality involves the natural identification
of the vector spaces $\,(\mathop{\otimes}\limits_nV_{_{{R}_n}})
\otimes(\mathop{\otimes}\limits_nV_{_{{\ov{R}}_n}})
\m\cong\m\mathop{\otimes}\limits_nEnd(V_{_{{R}_n}})\m$.
\,The partition function $\CZ(A)$ has to be given
by similar expressions pertaining to the case without
insertions. In particular, on the complex torus ${T}_\tau$,
and in the vanishing gauge field, 
\qq
\CZ_{_{{T}_{_\tau}}}\ =\ \sum\limits_{\hat R,\,\hat {R}'}
H^{^{\hat R\,\hat{R}'}}\ \ch_{_{\hat R}}(\tau,1)\,\,
\ov{\ch_{_{\hat{R}'}}(\tau,1)}\,.
\label{tpf0}
\qqq
\vskip 0.3cm

The matrices $(H^{\alpha\beta})$ should be specified for
a given choice of the basis $(\Psi_\alpha)$ for each complex 
structure on the surface and for each configuration of the insertions 
points so that, if we did not have means to compute them,
the above formulae would mean little progress
towards the solution of the WZW theory. Fortunately, 
there exist effective ways to determine the coefficients
$H^{\alpha\beta}$.
\vskip 0.5cm

\subsection{Scalar product of the Chern-Simons states}

It was argued in \cite{41}, see also
\cite{Witten} for a formal functional integral argument,
that the matrices $(H^{\alpha\beta})$ appearing
in Eq.\,\,(\ref{sesq}) are inverse to the matrices 
$(H_{\alpha\beta})$ with matrix elements  
\qq
H_{\alpha\beta}\ =\ (\m\Psi_\alpha,\m\Psi_\beta\m)
\non
\qqq
given by the scalar product of the CS states.
According to the rules of holomorphic 
quantization, the latter is given by the functional integral
\qq
(\m\Psi,\m\Psi'\m)\ =\ \int\limits_{\CA^{01}}
(\Psi(A^{01}),\Psi'(A^{01})
)_{_{\otimes V_{_{R_n}}}}\,
\ee^{-\m{k\over 2\pi}\m\Vert A\Vert^2}
\ DA
\label{gfin}
\qqq
over the $\Ng$-valued gauge fields $\,i\m A=i(-(A^{01})^*+A^{01})\,$
with \,$\Vert A\Vert^2\equiv i\int_{_\Sigma}\hspace{-0.05cm}
\tr\,(A^{01})^*\m A^{01}\m.$ \,This is again a formal
expression. The point is, however, that the $DA$-integral 
may be calculated exactly by reducing it to doable
Gaussian (i.e. free field) functional integrals. 
Ones this is done, the exact solution for the correlation
functions follows by Eq.\,\,(\ref{sesq}). Note that
the above solution for coefficients $H^{\alpha\beta}$
guarantees that the right hand side of Eq.\,\,(\ref{sesq})
is independent of the choice of a basis of the CS
states. Let us briefly sketch how one achieves the reduction 
of the integral (\ref{gfin}) to the free field ones.
\vskip 0.3cm

In the first step, the integral (\ref{gfin}) may be rewritten
by a trick resembling the Faddeev-Popov treatment of
gauge theory functional integrals. The reparametrization of 
the gauge fields 
\qq
A^{01}\s=\s{}^{h^{-1}}\hs{-0.11cm}A^{01}(n)
\label{chva}
\qqq
by the chiral $G^\NC$-valued gauge transforms of a (local) slice 
$\,n\mapsto A^{01}(n)\,$ in the space $\,\CA^{01}\,$ cutting each 
gauge orbit in one point, see Fig.\,\,10,

\leavevmode\epsffile[-90 -20 327 160]{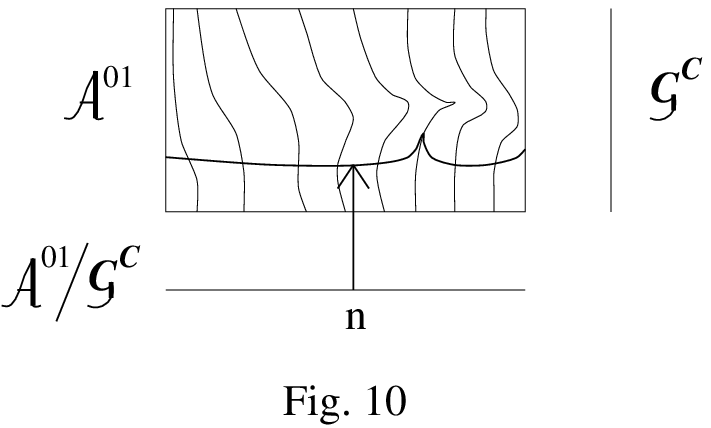}

\noindent permits to rewrite the functional
integral expression for the norm squared of a CS 
state in the new variables as
\qq
\Vert\Psi\Vert^2\m=\m\int(\m\Psi(A^{01}(n),
\otimes(hh^*)_{_{R_n}}^{-1}
\Psi(A^{01}(n)\m)_{_{V_{_{_{R_n}}}}}\hspace{-0.07cm}
\ee^{(k+2h^\vee)\m S(hh^*,\m A^{01}(n))}\ D(hh^*)
\ d\mu(n)\m.\hs{0.7cm}
\label{HWZW}
\qqq
To obtain the expression on the right hand side,
we have used Eq.\,\,(\ref{CSst}). The term $\,2\m h^\vee\m S(hh^*)\,$ 
in the action comes from the Jacobian of the change of variables 
(\ref{chva}) contributing also to the measure $\,d\mu(n)\,$
on the local slice in $\CA^{01}$.
\vskip 0.3cm

Unlike in the standard Faddeev-Popov setup, the integral
over the group of gauge transformations did not
drop out since the integrand in Eq.\,\,(\ref{gfin}) is invariant
only under the $\,G$-valued gauge transformations. Instead,
we are left with a functional integral (\ref{HWZW})
similar to the one for the original
correlation functions, see Eq.\,\,(\ref{cfu}), except 
that it is over fields $\,hh^*\,$. These fields may be considered
as taking values in the contractible hyperbolic space 
$\,G^\NC/G\m$. $\,D(hh^*)\,$ is the formal product 
of the $\,G^\NC$-invariant measures on $\,G^\NC/G\m$. \,The 
gain is that the functional integral (\ref{HWZW}) for the
hyperbolic WZW model correlation functions is doable.
For example for $\m G=SU(2)\m$ and for $\m\Sigma=\NC P^1$, 
where we may set $A^{01}(n)=0$ (in this case the gauge orbit 
of $A^{01}=0$ gives already a dense open subset $\CA^{01}$),
\qq
S(hh^*)\ =\ -\m{_i\over^{2\pi}}\int\da\phi\wedge\de\phi\s
-\s{_i\over^{2\pi}}\int(\da+\da\phi)\bar v\wedge(\de+\de\phi)v
\non
\qqq
in the Iwasawa parametrization
\s\s$
h\m=\m({_{\ee^{\phi/2}}\atop^0}\,{_0\atop^{\ee^{-\phi/2}}})
\,({_1\atop^0}\,{_v\atop^1})\, u\,\,$ of the 3-dimensional 
hyperboloid  $\,SL(2,\NC)/SU(2)\,$ by $\,\phi\in\NR\,$ and 
$\,v\in\NC\m$, \m with $\m u\in SU(2)\m$. \,Although the action
is not quadratic in the fields, it is quadratic 
in the complex field $\m v$ so that the $v$ integral 
can be explicitly computed. Somewhat miraculously,
the resulting integral appears to depend on the remaining
field $\phi$ again in a Gaussian way 
so that the integration may be carried out further. 
The same happens for more complicated groups and on surfaces
with handles, except that the procedure requires more steps. 
At the end, one obtains explicit finite-dimensional
integrals. Hence, the integral (\ref{HWZW}) belongs 
to the class of functional integrals that may be explicitly 
evaluated. The Gaussian functional integrals encountered 
in its computation require mild renormalizations 
(the zeta-function or similar regularization of determinants, 
Wick ordering of insertions) but these are well 
understood. They are responsible for the mild non-invariance 
of the WZW correlation functions under the local rescalings 
of the metric, leading to the values of the Virasoro 
central charge and of the conformal dimensions discussed
above. 
\vskip 0.3cm

On the complex tori ${T}_\tau$ with no insertions and in
the constant metric $|dz|^2$, the scalar product of
the CS states takes a particularly simple form:
the current algebra characters $\chi_{_{\hat R}}(\tau,\m\cdot\m)$
appear to give an orthonormal basis of the space
of states. It follows that the toroidal partition function
in the constant metric is
\qq
\CZ(\tau)\ =\ \sum\limits_{\hat R}|\chi_{_{\hat R}}(\tau,1)|^2\,,
\label{diag}
\qqq
see Eq.\,\,(\ref{tpf0}). The exact normalization
of the constant metric on ${T}_\tau$ is not important
since at genus one constant rescalings of the metric, exceptionally,
do not effect the partition functions. The latter fact has
an important consequence. It implies that the partition function
$\CZ(\tau)$ has to be a modular invariant:
\qq
\CZ(\tau)\ =\ \CZ({_{a\tau+b}\over^{c\tau+d}})
\non
\qqq
for $\m({_a\atop^c}\,{_b\atop^d})\in SL(2,\NZ)$.
This is indeed the property of the right hand side 
of Eq.\,\,(\ref{diag}) since the matrices implementing 
the modular transformations on the characters of the current algebra 
are unitary.
\vskip 0.3cm

Explicit finite-dimensional integral formulae for the scalar 
product (\ref{gfin}) have been obtained for general
groups at genus zero and one and, for \,$G=SU(2)\m$, \m for higher 
genera. It is clear that the case of general group and genus $>$1 
could be treated along the same lines, but the explicit calculation 
has not been done. It should be also said that the general 
proof of the convergence of the resulting finite-dimensional 
integrals is also missing, although several special cases 
have been settled completely.
\vskip 0.4cm

\subsection{Case of $\,G=SU(2)$ at genus zero}

To give a feeling about the form of the explicit
expressions for the scalar product of the CS 
states, let as describe the result for $G=SU(2)$
and $\Sigma=\NC P^1$ with insertions at points 
$z_n$ in the standard complex coordinate $z$. 
In this case, as we have discussed above, 
the CS states correspond to invariant 
tensors $v$
in $\,(\mathop{\otimes}\limits_nV_{_{j_n}})^{^{SU(2)}}\m$,
\,where we label the irreducible representations of $SU(2)$
by spins. The spin $j$ representation acts 
in the space $\,V_{_j}\,$ spanned by the vectors 
$\,(\sigma_-)_{_j}^\ell\m
v_{_{j}}\,$ with $\m\ell=0,1,\dots,2j\m,$ where
$\m v_{_j}\m$ is the normalized HW vector annihilated
by $\,(\sigma_+)_{_j}$, with $\sigma_i$ denoting the Pauli
matrices. For the scalar product of the CS 
states, the procedure described in the previous subsection
gives the following integral expression:
\qq
\Vert v\Vert^2\ =\ f(\sigma,{\un z},{\un j},k)
\int\limits_{\NC^J}\Big\vert\,(\, v\m,\s\omega(\un{z},\un{y})
\,)_{_{\otimes V_{_{R_n}}}}
\,\ee^{-{2\over k+2}\m U(\un{z},\un{y})}\m\Big\vert^2\,\,
\prod\limits_{a=1}^Jd^2y_a\,.
\label{scP0}
\qqq
Let us explain the terms on the right hand side. First,
\qq
f(\sigma,{\un{z}},{\un{j}},k)\ =\ 
\ee^{-{1\over 2\pi i(k+2)}\int\da\sigma\de\sigma}
\,\left({_{{\det}'(-\Delta)}\over^{{area}}}
\right)^{\hspace{-0.06cm}3/2}
\prod\limits_n\ee^{\m2\m{j_n(j_n+1)\over k+2}
\m\sigma(z_n)}
\label{explf}
\qqq
carries the dependence on the metric 
\s$\ee^{2\sigma}\vert dz\vert^2\,$
on $\,\NC P^1\m$, \,with $\,{\rm det}'(-\Delta)\,$ denoting 
the zeta-function regularized determinant of the 
(minus) Laplacian on $\NC P^1$ with omission of
the zero eigenvalue. The $\sigma$-dependence of 
$({\rm det}'(-\Delta)/{area})$ is given by a term
$\,\ee^{\m{1\over4\pi i}\int\da\sigma\de\sigma}\m$
leading altogether to the value ${3\m k\over k+2}$ of
the Virasoro central charge of the WZW theory (recall
that this is the inverse of the scalar product
that enters the WZW correlation functions).
Similarly, the conformal weight $\Delta_j$ of the
fields $g_j(\xi)$ of the $SU(2)$ WZW theory may be
read from Eq.\,\,(\ref{explf}) to be ${j(j+1)\over k+2}$.
\,Next, $\m\omega(\un{z},\un{y})\m$ is a meromorphic 
$\,\mathop{\otimes}\limits_nV_{_{j_n}}$-valued function
of $\un{z}$ and $\m\un{y}=(y_1,\dots,y_{_J})\m$, \m where 
$J=\sum_{_n}\hspace{-0.05cm}j_n\m$:
\qq
\omega(\un{z},\un{y})\ =\ \prod\limits_{r=1}^J\Big(\sum\limits_n
{_1\over^{y_r-z_n}}\s (\sigma_-)_{_{j_n}}
\Big)\,\mathop{\otimes}\limits_n v_{_{j_n}}\,.
\qqq
Finally, $\,U(\un{z},\un{y})\s$ is a multivalued function
\qq
U(\un{z},\un{y})\ =\ \sum\limits_{n<m}
j_n j_{m}\s\m\ln(z_n-z_m)
\s-\s\sum\limits_{n,r}j_n\s\m\ln(z_n-y_r)\s+\s\sum\limits_{r<s}
\m\ln(y_r-y_s)\,.
\qqq
\vskip 0.3cm

The integral (\ref{scP0}) is over a positive density with
singularities at the coinciding $\m y_r\m$'s and the question 
arises as to whether it converges.
A natural conjecture states that the integral
is convergent if and only if the invariant
tensor $\m v\m$ is in the image of the space of 
states $\,\NW(\NC P^1,\un{z},\un{j})$ explicitly
described as the set of $\,v\in(\mathop{\otimes}
\limits_nV_{_{j_n}})^{^{SU_2}}\,$ such that
\qq
(\otimes v_{j_n}\m,\,\prod\limits_{n}
(\sigma_+)_{_{j_n}}^{p_n}
\,\ee^{\m z_n\m(\sigma_+)_{_{j_n}}}\s v\m)_{_{\otimes V_{_{j_n}}}}
\hspace{-0.15cm}=\,0\ \,\ \quad{\rm for}\quad\ \sum p_n
\leq J-k-1\m.
\qqq
In particular, for two or three points, the image does not 
depend on the location of the insertions and gives the whole
space of invariant tensors if $\sum j_n\leq k$ and zero
otherwise. In this case, the integrals in Eq.\,\,(\ref{scP0}) 
may indeed be computed in a closed form confirming
the conjecture. Numerous other special cases have been
checked. In general, the ``only if'' part of the conjecture 
is easy but the ``if'' part remains to be verified. 
\vskip 0.4cm

\subsection{Knizhnik-Zamolodchikov connection}

There is another way to construct the matrices 
$(H^{\alpha\beta})$ entering the formula (\ref{sesq}) for
the correlation functions. Let us describe it briefly.
\vskip 0.3cm

The spaces $\NW_{_\Sigma}(\un{\xi},\un{R})$ of the CS 
states depend on the complex structure of the surface $\Sigma$ 
and on the insertion points. They form, in a natural way,
a holomorphic vector bundle $\CW$ whose base is the space of 
complex structures on a given smooth surface $\Sigma$ and of
non-coinciding insertions $\un{\xi}$ (modulo 
diffeomorphisms). The scalar product of the CS 
states equips this bundle with a hermitian 
structure. In turn, a hermitian structure on a holomorphic 
vector bundle determines a unique connection
such that the covariant derivatives of the structure
and of the holomorphic sections vanish. Although the scalar 
product of the CS states has been rigorously
defined in the general case only modulo the convergence 
of finite-dimensional integrals (see the end of the
last subsection), the connection on the bundle 
of the states may be easily constructed with full
mathematical rigor. It appears to be projectively
flat (i.e. with a curvature that is a scalar 2-form
on the base space), a crucial fact. In other words,
the parallel transport of a CS state around a closed loop 
in the space of complex structures and insertions at most
changes its normalization.
\vskip 0.3 cm

For the genus zero case, there is only one complex 
structure modulo diffeomorphisms.
If we fix it as the standard complex structure 
on $\NC P^1$, then we are only left with the freedom to move 
the insertion points $\un{z}$. The bundle $\CW$ is in this  
case a subbundle of the trivial bundle with the fiber
formed by the invariant group tensors
$(\mathop{\otimes}\limits_nV_{_{R_n}})^{^G}$
and the connection may be extended to this trivial bundle.
The covariant derivatives of the sections of the latter
are given explicitly by the equations:
\qq
\nabla_{_{\bar z_n}}v\ =\ \da_{_{\bar z_n}}v\s,\quad\quad
\nabla_{_{z_n}}v\ =\ 
\da_{_{z_n}}v\s+\s{_2\over^{k+h^\vee}}\sum\limits_{m\not= n}
{_{t^a_{_{R_m}}\hspace{-0.09cm}t^a_{_{R_n}}}\over^{z_m-z_n}}\,v
\label{KZ}
\qqq
which go back to the work \cite{5} of Knizhnik-Zamolodchikov 
on the WZW theory. In fact the above connection is
flat as long as the insertion points stay away from
infinity and the article \cite{5} studied their
horizontal sections such that $\nabla v=0$.
The higher genus generalizations of these equations
were first considered by Bernard \cite{10}\cite{11} 
and we shall call the connection on the bundle $\CW$
the Knizhnik-Zamolodchikov (KZ) connection
for genus zero or the Knizhnik-Zamolodchikov-Bernard (KZB)
one for higher genera.
\vskip 0.3cm

In general, the KZB connection can be made flat by some
choices (as in the case of genus zero, where the curvature 
has been concentrated at infinity). For a flat connection,
one may choose locally a basis $(\Psi_\alpha)$ of 
horizontal sections. The gain from such 
a choice of the basis of the CS 
states is that the coefficients $H^{\alpha\beta}$ 
in Eq.\,\,(\ref{sesq}) become then independent 
of the complex structure or the positions of the insertions. 
Indeed, since $H_{\alpha\beta}
=(\Psi_\alpha,\m\Psi_\beta)$ and the KZB connection
preserves the scalar product of the states, the 
above scalar products are constant for horizontal
$\Psi_\alpha$. Since the horizontal sections are, in particular, 
holomorphic, Eq.\,\,(\ref{sesq}) gives then
a holomorphic factorization of the correlation functions
into sesqui-linear combinations of the {\bf conformal blocks}
holomorphic in their dependence on the complex structure 
and positions of insertions. Such a finite factorization
is the characteristic feature of {\bf rational} CFT's.
As we see, the conformal blocks of the WZW theory
are given by the horizontal sections $\Psi_\alpha$ 
of the bundle $\CW$ of the CS states. 
For example, at genus zero with no insertions,
the conformal blocks are formed by the characters 
of the current algebra and Eq.\,\,(\ref{diag})
provides a particular realization of the holomorphic 
factorization. 
\vskip 0.3cm

Since the KZB connection, although flat, has nevertheless
a non-trivial holonomy, the conformal blocks are, in general, 
multivalued. The coefficients $H^{\alpha\beta}$ 
in Eq.\,\,(\ref{sesq}) may then be fixed, up to normalization, 
by demanding that the correlation
functions  be uni-valued. The overall 
normalization may be fixed, in turn, by considering
the limits when the insertion points coincide. 
This was the strategy used in the original work \cite{5}
to compute the 4-point correlation function of the spin
$\hf$ field $g_{\hf}(\xi)$ of the $SU(2)$ WZW model 
on the Riemann sphere. The horizontality relations 
for the conformal blocks reduce in this case to
the hypergeometric equation and the calculations 
of the conformal blocks and of their monodromy are easy
to perform. For general genus-zero conformal blocks, one 
obtains generalizations of the hypergeometric equation
whose solutions may be expressed by contour integrals
\cite{14}\cite{40}. The latter are, essentially, 
the holomorphic versions of the integrals (\ref{scP0}) so 
that the two strategies to obtain explicit solutions for
the correlation functions, one based on the study
of the monodromy of the conformal blocks and the other
one involving a calculation of the scalar product of the
CS states, are closely related.
\vskip 0.3cm

There appears to be a very rich structure behind
the connection (\ref{KZ}) and its generalizations.
It is closely related to the integrable systems of
mechanics and statistical mechanics, see 
e.g.\,\,\cite{GaTr}\cite{FrResh}.
The holonomy of the connection gives representations
of the braid groups which played an important role in
the construction of the Jones-Witten invariants of knots
and 3-manifold invariants \cite{27}. The perturbative 
solutions of the horizontality equations enter the Vasiliev
invariants of knots \cite{Konts}. The KZ connection is also 
closely connected to quantum groups \cite{Feldqg}
and to Drinfel'd's quasi-Hopf algebras \cite{Drinf}, 
to mention only some interrelated topics.
\vskip 0.5cm

\nsection{Coset theories}

There is a rich family of CFT's
which may be obtained from the WZW models by a simple
procedure known under the name of a {\bf coset construction}
\cite{42}\cite{43}. On the functional integral level,
the procedure consists of coupling the \s$G$-group WZW
theory to a gauge field $\m i\m B=i(B^{10}+\m B^{01})\m$ 
with values in a subalgebra
$\,\Nh\subset\Ng\,$. The field $B$ is then integrated over with
gauge-invariant insertions \cite{45}\cite{46}\cite{47}\cite{49}.
Let $H\subset G$ be the subgroup of $G$ corresponding to $\Nh$. 
Choose elements $\m t_n\m$ in the space $\,(End(V_{_{R_n}},
V_{_{r_n}}))^H\,$ of the intertwiners of the action of $\m H\m$ 
in the irreducible $\,G$- and $\,H$-representation spaces.
The simplest correlation functions of the $\,G/H\,$ coset 
theory take the form
\qq
&&<\prod\limits_{i=1}^n \tr_{_{V_{_{r_n}}}}\hs{-0.1cm}(t_n\m 
g_{_{R_n}}\hspace{-0.06cm}(x_n)\m t_n^\dagger)
>_{_{\hspace{-0.08cm}\Sigma}}\cr
&&\hspace{2cm}=\ {_1\over^{{\CZ^{^{G/H}}_{_\Sigma}}}}
\int\,\prod\limits_{n} 
\tr_{_{V_{_{r_n}}}}\hs{-0.1cm}(t_n\m g_{_{R_n}}\hspace{-0.06cm}
(x_n)\, t_n^\dagger)\ \m\ee^{-k\m S(g,\m B)-
{k\over 2\pi}\m\Vert B\Vert^2}\,\, Dg\,\m DB
\,,\hspace{0.8cm}
\label{cfGH}
\qqq
where $\,\CZ^{^{G/H}}_{_\Sigma}\m
=\m\int\ee^{-k\m S(g,\m B)-{k\over 2\pi}\m\Vert B\Vert^2}
\,Dg\, DB\,$ is the partition function of the $G/H$ theory.
Note that the $\,g$-field integrals are the ones of the
WZW theory and are given by Eq.\s\s(\ref{sesq}).
Consequently,
\qq
&&\CZ^{^{G/H}}_{_\Sigma}\,<\prod\limits_{n} 
\tr_{_{V_{_{r_n}}}}\hs{-0.1cm}(t_n\m g_{_{R_n}}\hspace{-0.06cm}
(x_n)\m t_n^\dagger)
>_{_{\hspace{-0.08cm}\Sigma}}\cr
&&\hspace{2cm}=\ H^{\alpha\beta}\int(\m\otimes t_n
\Psi_\beta(B^{01})\m,\,\otimes t_n\Psi_\alpha
(B^{01})\,)_{_{\otimes V_{_{r_n}}}}\, 
\ee^{-{k\over 2\pi}\m\Vert B\Vert^2}
\,\,DB\,.\hs{0.4cm}
\label{aneq}
\qqq
The composition with $\,\otimes t_n\,$ defines a linear map $T$
between the spaces of the group $G$ and the group $H$  
CS states, i.e. $T:\NW_{_\Sigma}(\un{\xi},\un{R})\rightarrow
\NW_{_\Sigma}(\un{\xi},\un{r})\,$ with
\qq
(T\m \Psi)(B^{01})\ =\ \mathop{\otimes}\limits_n t_n\,\Psi(B^{01})\,.
\qqq
Indeed, it is straightforward to check that the right hand
side satisfies the the group $H$ version of the Ward identity
(\ref{CSst}). Eq.\s\s(\ref{aneq}) may be rewritten with the 
use of the map $T$ as
\qq
\CZ^{^{G/H}}_{_\Sigma}<\prod\limits_{n} 
\tr_{_{V_{_{r_n}}}}\hspace{-0.1cm}(t_n\m g_{_{R_n}}
\hspace{-0.06cm}(x_n)\m t_n^\dagger)
>_{_{\hspace{-0.08cm}\Sigma}}\ 
=\ H^{\alpha\beta}\ (\m T\Psi_\beta\m,\, T\Psi_\alpha\m)
\ =\ \tr_{_{\NW_{_\Sigma}(\un{\xi},\un{R})}}\hspace{-0.05cm} 
T^\dagger T\,.
\label{GHcf}
\qqq
Let $\,(T^\mu_\alpha)\,$ denote the (``branching'') matrix
of the linear map $\m T\m$ in the bases 
$\,(\Psi_\alpha)\m$ and $\,(\psi_\mu)\,$ of, respectively,
$\,\NW_{_\Sigma}(\un{\xi},\un{R})\,$ and
$\,\NW_{_\Sigma}(\un{\xi},\un{r})\m$, i.e.
$T\m\Psi_\alpha=T^\mu_\alpha\m\psi_\mu\m$. \,Then
\qq
\CZ^{^{G/H}}_{_\Sigma}<\prod\limits_{n} 
\tr_{_{V_{_{r_n}}}}\hspace{-0.1cm}(t_n\m g_{_{R_n}}
\hspace{-0.06cm}(x_n)\m t_n^\dagger)
>_{_{\hspace{-0.08cm}\Sigma}}\ 
=\ H^{\alpha\beta}\,\,\m\overline{T^\mu_\beta}\,\,\m
h_{\mu\nu}\,\, T^\nu_\alpha\,,
\non
\qqq
where $\,h_{\mu\nu}=(\psi_\mu,\m\psi_\nu)\m$.
\,Since the above relations hold also for the partition function,
it follows that the calculation of the coset theory
correlation functions (\ref{cfGH}) reduces to that of the scalar 
products of group \s$G\s$ and group \s$H\s$ CS
states, both given by explicit, finite-dimensional integrals.
\vskip 0.3cm

Among the simplest examples of the coset theories
is the case with the group $\,G=SU(2)\times SU(2)\,$ 
with level $\m(k,1)\m$ (for product groups, the levels may
be taken independently for each group) and with $\m H\m$
being the diagonal $\m SU(2)\m$ subgroup. The resulting
theories coincide with the unitary {\bf minimal series} of 
CFT's with the Virasoro central charges 
$\,c=1-{6\over{(k+2)(k+3)}}\m$, \m first considered by
Belavin-Polyakov-Zamolodchikov \cite{3}. 
The Hilbert spaces of these theories are built
from irreducible unitary representations
of the Virasoro algebras with $\,0<c<1\m$. 
The simpliest one of them with $\,k=1\,$ and $\,c={1\over2}\,$ 
describes the continuum limit of the Ising model at critical 
temperature or the scaling limit of the massless $\,\phi^4_2\,$ 
theory. In particular, in the continuum limit the spins in 
the critical Ising model are represented 
by fields \s$\tr\,g_{\hf}(\xi)\s$ where \s$g\s$ takes 
values in the first \s$SU(2)\m$. \s The corresponding correlation 
functions may be computed as above. One obtains this way
for the 4-point function on the complex plane (or the Riemann  
sphere) an explicit expression in terms of the hypergeometric 
function.
\vskip 0.3cm

The $\,G/H\,$ coset theory with $\m H=G\m$ is a prototype
of a two-dimensional topological field theory. As follows from
Eq.\,\,(\ref{GHcf}), the correlation functions of the fields 
$\,\tr\, g_{_R}(\xi)\,$ are equal to the dimension of the
spaces $\,\NW_{_\Sigma}(\un{\xi},\un{R})\,$, normalized 
by the dimension of \s$\NW_{_\Sigma}(\emptyset,\emptyset)\,$ 
and are given by the Verlinde formula (\ref{wWDd}). 
In particular, they do not depend on the position 
of the insertion points, a characteristic feature 
of the correlation functions in topological field theories.
\vskip 0.6cm

\nsection{Boundary conditions in the WZW theory}

Discussing in Sects.\,\,3.4 and 4.1 above the classical 
and the quantum amplitudes for the WZW model on surfaces 
with boundary, we have admitted an arbitrary behavior 
of the fields on the boundary. In physical situations, 
one often has to constrain this behavior by imposing 
the boundary conditions (BC) on the fields. 
The simplest example is provided 
by the Dirichlet or Neumann BC's for the free 
fields which fix to zero, respectively, the tangent or 
the normal derivative of the field (the absorbing
versus the reflecting condition). Such conditions leave
unbroken an infinite-dimensional set of symmetries of
the free field theory. We shall be interested in BC's 
in the WZW model with a similar property. 
\vskip 0.3cm

The theory of boundary CFT's was pioneered
by Cardy \cite{Cardy} and Cardy-Lewellen \cite{CardLew}.
It found its applications e.g.\,\,in the theory of
isolated impurities in condensed matter physics 
(in the so called Kondo problem, a traditional 
playground for theoretical ideas) \cite{Affleck}.
In string theory, the use of the Neumann BC for free
open strings has a long tradition \cite{GWSch}.
The realization that one should also consider 
free open strings with the Dirichlet BC came much later 
and gave rise to a theory of Dirichlet- 
or D-{\bf branes} \cite{Polch}: the end of an open string, 
some of whose coordinates are restricted by the Dirichlet BC, 
moves on a surface (brane) of a lower dimension.
The D-branes provide the basic tool in the analysis 
of the non-perturbative effects in string theory:
of the stringy solitons and of the strong-weak coupling 
dualities \cite{Polch}. The general theory 
of boundary CFT's is slowly becoming an important 
technique of string theory (see e.g.\,\,\cite{ReckSch}). 
\,Here, for the sake of illustration, 
we shall discuss a particular class of BC's for the WZW 
theory. These conditions constrain the group $G$ valued 
field $g$ to stay over the boundary components in fixed 
conjugacy classes of $G$. Such BC's were discussed in 
\cite{AlSch}, see also \cite{KlimSev}. They clearly generalize
the Dirichlet BC of the free fields, contrary to the claim 
in \cite{AlSch} (based on the conventions of reference 
\cite{Okada}) associating them to the Neumann BC. Our 
presentation, along similar lines, clarifies, hopefully, 
some of the discussions of the above papers. 
\vskip 0.5cm

\subsection{The action}

As before, we shall represent a Riemann surface 
$\Sigma$ with boundary as $\Sigma'\setminus(\mathop{\cup}
\limits_mD_{m}^{^{^{{\hspace{-0.21cm}}o}}})$,
where $D_{m}$ are disjoint unit discs embedded in a closed 
surface $\Sigma'$ without boundary,
see Fig.\,\,6. We have seen in Sect.\,\,3.4 that the classical 
amplitudes $\,\ee^{-S(g)}\,$ of the fields 
$\,g:\Sigma\rightarrow G\m$ of the WZW model take 
values in a line bundle $\m\CL\m$ rather than being numbers. 
The line bundle $\CL$ over the loop group $LG$ is not trivial but
it may be trivialized over certain subsets of $LG$, for 
example the ones formed by the loops taking values 
in special conjugacy classes $\CC_{\lambda}$. 
It will then be possible to give numerical values to 
the amplitudes $\m\ee^{-S_{_{\Sigma}}(g)}\m$ for 
$\m g:\Sigma\rightarrow G\m$ satisfying the BC's
\qq
g(\da D_{m})\ \subset\ \CC_{\lambda_m}\,.
\label{bc}
\qqq
In order to achieve this goal, we shall fix the 2-forms
\qq
\omega_{_\lambda}\ =\ 
\tr\,(g^{-1}dg)\m(1-Ad_g)^{-1}(g^{-1}dg)\ =\ 
\tr\,\,(g_0^{-1}dg_0)\m
\ee^{\m2\pi i\lambda/k}\m(g_0^{-1}dg_0)\m
\ee^{-2\pi i\lambda/k}
\label{om}
\qqq
on the conjugacy classes $\,\CC_{\lambda}$ composed of the
elements $\,g\m=\m g_0\,\ee^{\m 2\pi i\lambda/k}g_0^{-1}\m$
(the operator $(1-Ad_g)$ is invertible on the vectors tangent 
to $\CC_{\lambda}$). It is easy to check by a direct
calculation that $\,d\m\omega_{_\lambda}\,$ coincides 
with the restriction of the 3-form $\m\chi={1\over 3}\m\tr
\,(g^{-1} dg)^3\m$ to $\CC_\lambda$. 
\vskip 0.3cm

Since the conjugacy classes in a simply connected group $G$
are simply connected, any field $g:\Sigma\rightarrow G$
satisfying the BC's (\ref{bc})
may be extended to a field $g':\Sigma'\rightarrow G$
in such a way that $g'(D_{m})\subset\CC_{\lambda_m}$
and then to a field $\tilde g$ on a 3-manifold $B$ 
such that $\da B=\Sigma'$, see Fig.\,\,11. 

\leavevmode\epsffile[-20 -20 367 157]{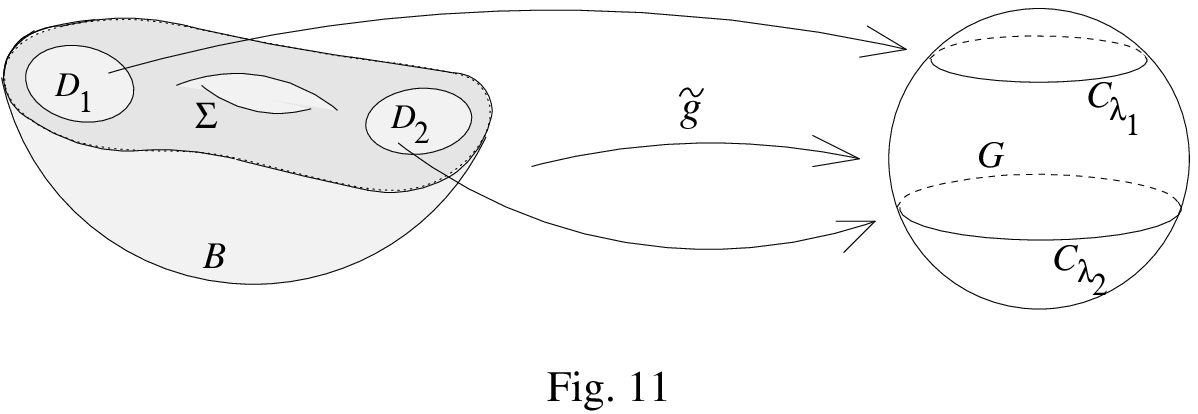}

\noindent Having done this, 
we define the Wess-Zumino part of the action as
\qq
S^{^{WZ}}_{\Sigma}(g)\ =\ {_k\over^{4\pi i}}
\int_{_{B}}\hspace{-0.08cm}\tilde g{}^*\chi\ 
-\ {_k\over^{4\pi i}}\sum\limits_m
\int_{_{D_{_m}}}\hspace{-0.1cm}\tilde g|_{_{D_{_m}}}^{\, *}
\omega_{_{\lambda_m}}\m.
\non
\qqq
The ambiguities in this definition are the values
of the integrals 
\qq
{_k\over^{4\pi i}}
\int_{_{\tilde B}}\hspace{-0.08cm}\tilde g^*\chi\ -
\ {_k\over^{4\pi i}}\sum\limits_m
\int_{_{S^2_m}}\hspace{-0.1cm}\tilde g|_{_{S^2_m}}^{\, *}
\omega_{_{\lambda_m}}
\label{WZbca}
\qqq
for 3-manifolds $\tilde B$ with $\da\tilde B=
\mathop{\cup}\limits_m  S^2_m$ and for maps 
$\tilde g:\tilde B\rightarrow 
G$ such that $\tilde g(S^2_m)\subset\CC_{\lambda_m}$,
see Fig.\,\,12. 

\leavevmode\epsffile[-7 -20 377 185]{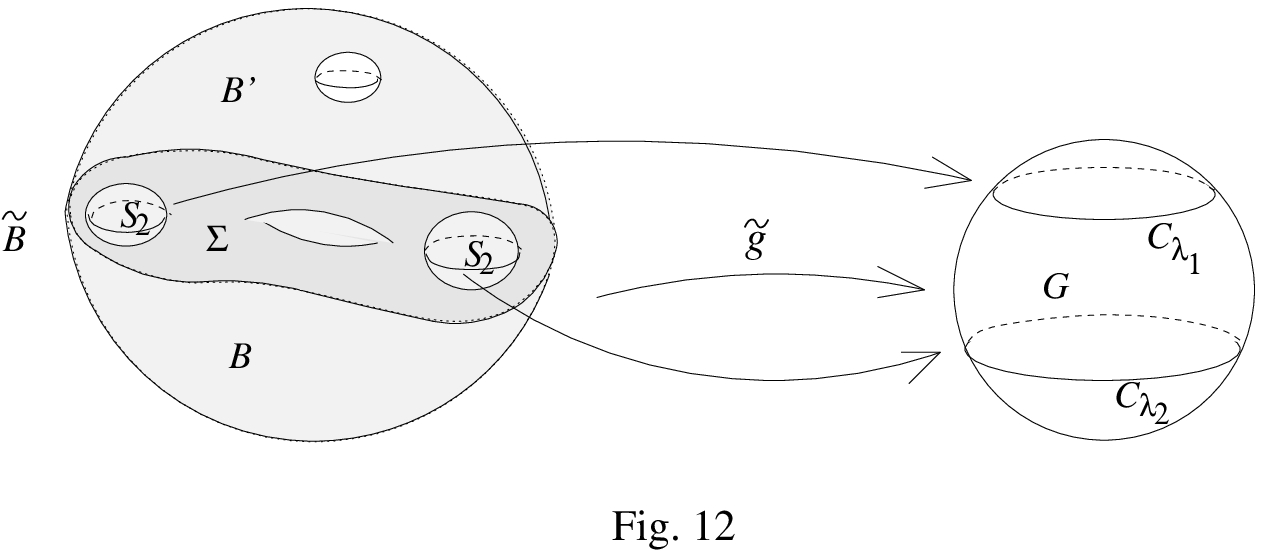}

\noindent In other words, they are proportional to the periods
of $(\chi,\omega)$ over the cycles of the relative
integer homology $H_3(G,\mathop{\cup}\limits_m
\CC_{\lambda_m})$, as noticed in \cite{KlimSev}.
\vskip 0.3cm

It is not difficult to get a hold on these
ambiguities. Let us glue the unit 3-balls $B_m$ 
to $\tilde B$ along the boundary spheres $S^2_m$ to obtain
a 3-manifold $\tilde B'$ without boundary and let us extend 
$\tilde g$ to a map $\tilde g':\tilde B'\rightarrow G$.
The expression (\ref{WZbca}) may be now rewritten as 
\qq
{_k\over^{4\pi i}}
\int_{_{\tilde B'}}\hspace{-0.08cm}\tilde g'{}^*\chi\ -
\ {_k\over^{4\pi i}}\sum\limits_m\Big(\int_{_{B_m}}
\hspace{-0.09cm}{\tilde g'_m}{}^*\chi\ -\,
\int_{_{\da B_m}}\hspace{-0.09cm}{\tilde g'_m}{}^*
\omega_{\lambda_m}\Big),
\non
\qqq
where $\,\tilde g_m'\,$ are the restrictions of $\tilde g'$
to $B_m$ and they satisfy $\,\tilde g'_m(\da B_m)\m\subset\m
\CC_{\lambda_m}\m$. \,As we have discussed in Sect.\,\,3.3,
the first term, involving the integral over the 3-manifold 
without boundary $\tilde B'$, takes values in $2\pi i\NZ$ 
as long as $k$ is an integer. 
\vskip 0.3cm

Let us consider the other terms. For $G=SU(2)\m=\m
\{\m x_0+i x_i\sigma_i\,\vert\, x_0^2+x_i^2=1\}\cong S^3$, 
the conjugacy classes corresponding to $\,\lambda=j\m
\sigma_3\m$, with $\,0\leq{2j}\leq k\m$, are the 2-spheres 
with $\,x_0=\cos{2\pi j\over k}\,$ fixed (except for $j=0$ or 
${k\over 2}$ corresponding to the center elements). They 
are boundaries of two 3-balls $B_j$ 
and $B'_j$ with $x_0\geq\cos{2\pi j\over k}$ 
and $x_0\leq\cos{2\pi j\over k}$, see Fig.\,\,13.

\leavevmode\epsffile[-132 -20 260 215]{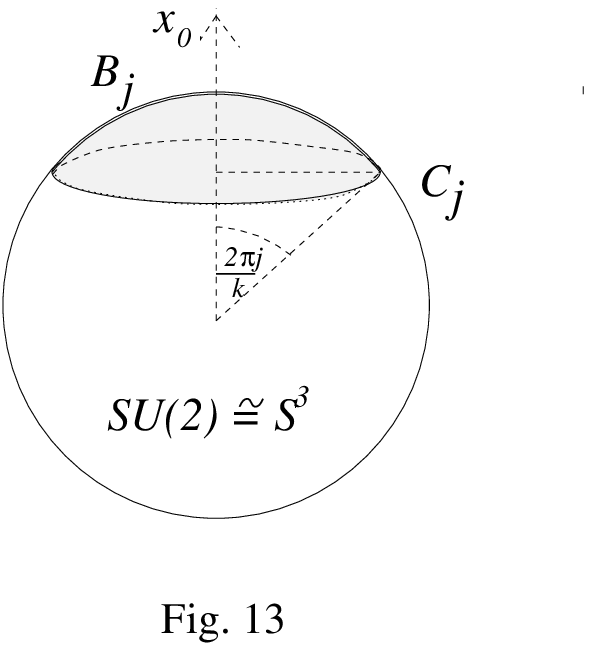}

\noindent A direct calculation shows that 
\qq
{_k\over^{4\pi i}}\Big(\int_{_{B_j}}\hspace{-0.1cm}
\chi\,-\,\int_{_{\da B_j}}\hspace{-0.1cm}\omega\,
\Big)\ =\ -\m 4\pi i\m j\,.
\qqq
If we used $B'_j$ instead of $B_j$, the result would be
$\m 4\pi i\m({k\over2}-j)$. We infer that $j$, between $0$
and ${k\over 2}$, must be an integer or a half-integer for 
the ambiguity to belong to $2\pi i\NZ$.
This result has already been stated in \cite{AlSch}. 
\vskip 0.3cm
 
For the other groups, the restrictions come from the
the 2-spheres in $\CC_\lambda$ of the form 
\qq
\{\, g_0\,\ee^{\m2\pi i\lambda/k}\m g_0^{-1}\,\,
\vert\,\,g_0\in SU(2)_\alpha\,\}\,,
\qqq
where $\,SU(2)_\alpha$ is the $SU(2)$ subgroup 
of $G$ corresponding to a root $\alpha$, see Sect.\,\,3.3.
Decomposing $\,\lambda=(\lambda-\hf\m\alpha^\vee\,\tr\,(\alpha
\m\lambda))+\hf\m\alpha^\vee\,\tr\,(\alpha\m\lambda)\m$,
\,we observe that the first term commutes
with the generators $\alpha^\vee,\,e_{\pm\alpha}$ of the Lie 
algebra of $\,SU(2)_\alpha$ and plays a spectator role. 
The calculation of the ambiguity is now essentially 
the same as for $G=SU(2)$ with $j$ replaced by 
$\hf\,\tr\,(\alpha\m\lambda)$ and an overall factor 
$2\over\hbox{tr}\,\alpha^2$ due to the different
normalization of $\tr\,\alpha^2$. We infer the condition
\qq
{_2\over^{\hbox{tr}\,\alpha^2}}\,\tr\,(\alpha\m\lambda)\ =\ 
\,\tr\,(\alpha^\vee \lambda)\ \in\ \NZ\,.
\non
\qqq
Since the conjugacy classes $\CC_\lambda$ are in one to one
correspondence with $\lambda$ in the symplex (\ref{sympl}), 
the admissible conjugacy classes are in one to one 
correspondence with the HW's $\lambda$ 
integrable at level $k$, see the definition (\ref{iHW}).
\vskip 0.3cm

The full action $S_{_\Sigma}(g)$ of the boundary WZW model 
is still obtained by adding to the WZ
action $S^{{WZ}}$ the $S^\gamma$ term of Eq.\,\,(\ref{sgam}).
The coupling to the gauge field is given again by
Eq.\,\,(\ref{tog}). The behavior of the complete
action under the chiral gauge transformation may be shown
to obey the following BC version of the Eq.\,\,(\ref{trprop}):
\qq
S(g,\m A)\ =\ S(h_1g\m h_2^{-1},\m{}^{h_2}
\hspace{-0.1cm}A^{01}+
{}^{h_1}\hspace{-0.09cm}A^{01})\,+\,S(h_1^{-1}h_2,\m A)\,
-\,{_{i\m k}\over^{2\pi}}\int_{_\Sigma}\hspace{-0.07cm}
\tr\,({}^{h_2}\hspace{-0.09cm}A^{10}\,\m{}^{h_1}
\hspace{-0.1cm}A^{01})\,,
\label{trprobc}
\qqq
provided that $\m g\m$ satisfies the BC's (\ref{bc}) and that
$\,h_1|_{_{\da D_{_m}}}=h_2|_{_{\da D_{_m}}}\m$. Note that
under this conditions, the field $h_1g\m h_2^{-1}$
is constrained on the boundary to the same conjugacy classes
as $g$ and $h_1 h_2^{-1}$ to the trivial one so that
the actions on the right hand side make sense. The above
relation will be employed below to infer the chiral gauge 
symmetry Ward identities for the boundary WZW theory. It may 
be used as the principle that selects the BC's (\ref{bc}).
\vskip 0.3cm

Summarizing: if the field $g:\Sigma\rightarrow G$ satisfies
the BC's (\ref{bc}) with integrable weights $\lambda_m$, 
then the amplitude $\m\ee^{-\m S^{^{WZ}}_{_\Sigma}(g)}\m$,
\m and consequently also $\m\ee^{-\m S_{_\Sigma}(g)}\m$ and 
$\m\ee^{-\m S_{_\Sigma}(g,\m A)}\m$, 
\m may be well defined as complex numbers. Of course, 
the mixed case, where the BC's (\ref{bc}) with integrable
$\lambda_m$ are satisfied only on some boundary components
of $\Sigma$ and no conditions are imposed on the other
(``free'') boundary components can be treated in the same 
way. It results in the amplitudes 
\qq
\ee^{-\m S_{_\Sigma}(g,\m A)}\ \in\ 
\mathop{\otimes}
\limits_{n\ {\rm free}}
\CL_{_{g|_{_{\da D_{_n}}}}}\,.
\non
\qqq
\vskip 0.4cm

\subsection{Quantum amplitudes and correlation functions}

The functional integral definition (\ref{FfiA}) of the quantum 
amplitudes of the WZW model may be naturally 
generalized to the case where on some boundary components 
of $\Sigma$ we impose the BC's (\ref{bc}) with integrable 
weights $\lambda_m$. The resulting amplitudes 
$\,\CA_{_{\Sigma,\un{\lambda}}}\hspace{-0.09cm}
(A)\,$ will be now elements 
of $\mathop{\otimes}\limits_{n\ {\rm free}}\Gamma(\CL)\m$.
\,They may be represented as (partial) contractions of the
amplitudes $\,\CA_{_\Sigma}(A)\,$
with all boundaries free with appropriate states:
\qq
\CA_{_{\Sigma,\un{\lambda}}}\hspace{-0.09cm}(A)\ =\ 
(\,\mathop{\otimes}\limits_m \hat\delta_{_{\lambda_m}}\m,\,
\m\CA_{_\Sigma}(A)\,)\,.
\non
\qqq
The non-normalizable states\footnote{technically, antilinear 
forms on a dense subspace of the Hilbert space $\CH$}
$\hat\delta_{_{\lambda}}$ are given 
by Cardy's formula \cite{Cardy}\m:
\qq
\hat\delta_{_\lambda}\ \ =\ \ 
\sum\limits_{\hat R}\,(S^{^{1}}_{_{R}})^{-\hf}\,
S^{^{R_\lambda}}_{_{R}}
\,\,e^{\hat{i}}_{_{\hat R}} 
\otimes\overline{e^{\hat{i}}_{_{\hat R}}}\,,
\label{Card}
\qqq
where $R_\lambda$ denotes the representation 
of $G$ with the HW $\lambda$, the vectors 
$\m e^{\hat{i}}_{_{\hat R}}\m$ form an orthonormal 
basis of the space $V_{_{\hat{R}}}$ 
and the sum over $\hat{i}$ is understood.
Note the analogy with Eq.\,\,(\ref{Ishcl})
for the delta-function supported by a conjugacy class 
$\CC_\lambda$. The matrix $(S^{^{R_\lambda}}_{_{R}})$
has replaced the one with the elements
$\m(\chi_{_R}(\ee^{\m2\pi i\m\lambda/k})\m)\m$ and 
the representation spaces $V_{_{\hat{R}}}$ of the current 
algebras those of the finite-dimensional group.
The state ${\hat\delta}_{_\lambda}$ should be interpreted 
as a delta-function concentrated on the loops 
in $LG$ contained in the conjugacy 
class $\CC_\lambda$. The non-normalizable states
$\,e^{\hat{i}}_{_{\hat R}}\otimes\overline
{e^{\hat{i}}_{_{\hat R}}}\,$ are called the Ishibashi states
\cite{Ishib} and generalize the (properly normalized)
characters of the group,
see Eq.\,\,(\ref{chdec}).
\vskip 0.3cm

The correlation functions $\,<\otimes g_{_{R_n}}
\hspace{-0.05cm}(\xi_n)>_{_{\hspace{-0.08cm}\Sigma,
\un{\lambda}}}\hspace{-0.19cm}(A)\,$
in the presence of the boundaries with fields constrained 
by the the BC's (\ref{bc}) may be again defined 
by the functional integral (\ref{cfu}) taking numerical 
values. The transformation property (\ref{trprobc}) 
of the action implies now that
\qq
&&\CZ_{_{\Sigma,\un{\lambda}}}
\hspace{-0.09cm}(A)\,
<\mathop{\otimes}\limits_n 
g_{_{R_n}}\hspace{-0.05cm}(\xi_n)>_{_{\hspace{-0.08cm}\Sigma,
\un{\lambda}}}\hspace{-0.09cm}(A)\ \,=\ \, 
\ee^{-S(h_1^{-1}h_2,\m A)
\,+\,{_{i\m k}\over^{2\pi}}\int_{_\Sigma}\hspace{-0.07cm}
\tr\,({}^{^{h_2}}\hspace{-0.1cm}A^{10}\,{}^{^{h_1}}
\hspace{-0.11cm}A^{01})}
\ \CZ_{_{\Sigma,\un{\lambda}}}({}^{h_2}\hspace{-0.09cm}A^{10}+
{}^{h_1}\hspace{-0.1cm}A^{01})\hspace{0.3cm}
\cr\cr
&&\hspace{2.4cm}\cdot\ \mathop{\otimes}\limits_n 
(h_1)_{_{R_n}}^{-1}
\hspace{-0.03cm}(\xi_n)\ \ 
<\mathop{\otimes}\limits_n 
g_{_{R_n}}\hspace{-0.05cm}(\xi_n)
>_{_{\hspace{-0.08cm}\Sigma,\un{\lambda}}}
\hspace{-0.1cm}({}^{h_2}\hspace{-0.09cm}A^{01}+
{}^{h_1}\hspace{-0.1cm}A^{01})\ \ 
\mathop{\otimes}\limits_n (h_2)_{_{R_n}}
\hspace{-0.05cm}(\xi_n)\,.
\label{WIbc}
\qqq
This is the chiral gauge symmetry Ward identity 
for the boundary WZW correlation functions, 
a variant of the identities (\ref{WIcfl}) and (\ref{WIcfr})
in presence of the BC's. Note, however, that 
the identity (\ref{WIbc}) may be factorized as
the latter ones only if we assume that $h_1$ 
and $h_2$ are equal to $1$ on the boundary. The general
case where on the boundary $h_1=h_2$ requires the presence
of both gauge transformations $h_1$ and $h_2$. 
\vskip 0.3cm

It is illuminating to rewrite the Ward identity
(\ref{WIbc}) in a different way. To this end, 
let us define a ``doubled'' Riemann surface 
without boundary $\tilde\Sigma$ by gluing $\Sigma$ 
to its complex conjugate $\ov{\Sigma}$ along 
the boundary components, see Fig.\,\,14. 

\leavevmode\epsffile[-102 -20 320 210]{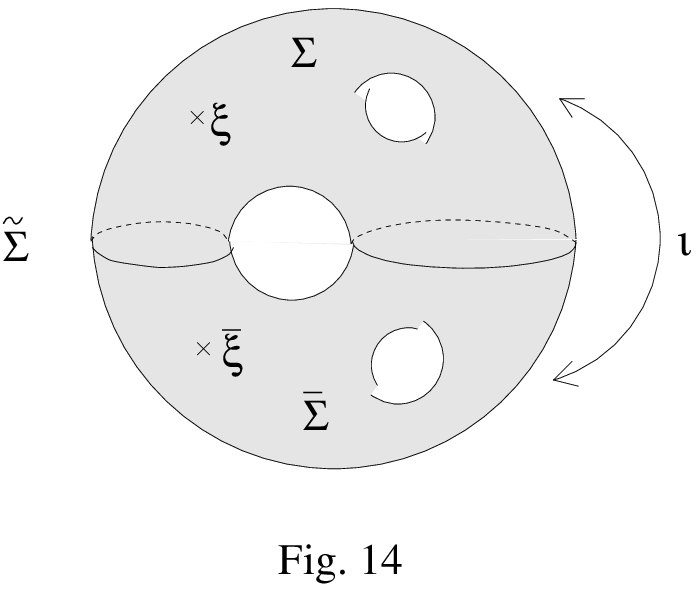}

We shall denote by $\,\iota\,$
the anti-holomorphic involution of $\m\tilde\Sigma$ 
exchanging $\Sigma$ with its complex conjugate: 
$\,\iota(\xi)=\bar\xi\m$. \,Each chiral gauge 
field $\m\tilde A^{01}\m$ on the Riemann surface
$\m\tilde\Sigma\m$ defines a complexified gauge field 
$\,A=(\iota^*A^{01}+A^{01})\vert_{_\Sigma}\,$
on the surface $\Sigma$. Let us define
a holomorphic functional 
$$\Psi_{_{\un{\lambda}}}:\,\tilde\CA^{01}\ 
\longrightarrow\ (\mathop{\otimes}\limits_n
V_{_{R_n}})\otimes(\mathop{\otimes}\limits_n
V_{_{\ov{R}_n}})$$
of the chiral gauge fields on the doubled surface by
\qq
\Psi_{_{\un{\lambda}}}(\tilde A^{01})\ =\ 
\CZ_{_{\Sigma,\un{\lambda}}}
\hspace{-0.09cm}(A)\,<\mathop{\otimes}\limits_n 
g_{_{R_n}}\hspace{-0.05cm}(\xi_n)>_{_{\hspace{-0.08cm}\Sigma,
\un{\lambda}}}\hspace{-0.09cm}(A)\,,
\non
\qqq
where we identify the space $\,(\otimes
V_{_{R_n}})\otimes(\otimes V_{_{\ov{R}_n}})\,$
with $\,\otimes End(V_{_{R_n}})\m$. \,Let $\tilde h$ 
be a $\m G^\NC$-valued gauge transformation on $\tilde\Sigma$.
We shall pose $h_1=\tilde h\vert_{_\Sigma}$ 
and $h_2=\iota^*\tilde h\vert_{_\Sigma}$. Note that $h_1=h_2$
on the boundary of $\Sigma$. It is not difficult
to prove that
\qq
\ee^{-S_{_\Sigma}(h_1^{-1}h_2,\m A)
\,+\,{_{i\m k}\over^{2\pi}}\int_{_\Sigma}\hspace{-0.07cm}
\tr\,({}^{^{h_2}}\hspace{-0.1cm}A^{10}\,{}^{^{h_1}}
\hspace{-0.11cm}A^{01})}\ =\ \ee^{-S_{_{\tilde\Sigma}}
(\tilde h^{-1},\,\tilde A^{01})}\,.
\qqq
The Ward identity (\ref{WIbc}) implies then that
\qq
\Psi_{_{\un{\lambda}}}(\tilde A^{01})\ =\ 
\ =\ \ee^{-S_{_{\tilde\Sigma}}
(\tilde h^{-1},\,{\tilde A}^{01})}\ 
(\mathop{\otimes}\limits_n \tilde h^{-1}_{_{R_n}}
(\xi_n)\m)
\otimes(\mathop{\otimes}\limits_n 
\tilde h^{-1}_{_{\ov{R}_n}}(\bar\xi_n)\m)\  
\Psi_{_{\un{\lambda}}}({}^{\tilde h}\hspace{-0.09cm}
{\tilde A}^{01})\,,
\non
\qqq
i.e.\,\, that $\,\Psi_{_{\un{\lambda}}}\,$ is a CS state
on the doubled surface $\tilde\Sigma$ with the doubled
insertions at points $\xi_n$ and at their complex conjugates 
$\bar\xi_n$, associated, respectively, 
to the representations $R_n$ and to the complex conjugate 
representations $\ov{R}_n$.       
\vskip 0.3cm

We infer that the correlations functions of the boundary
WZW theory on a surface $\Sigma$ may be viewed, 
in their gauge field dependence, as the special states 
$\,\Psi_{_{\un{\lambda}}}\,$ belonging to the space 
$\,\NW_{_{\tilde\Sigma}}(\un{\xi},\un{\bar\xi},\un{R},\un{\ov{R}})\,$ 
of the CS states on the doubled surface $\tilde\Sigma$. The states 
$\,\Psi_{_{\un{\lambda}}}\,$ may be shown, moreover, to be preserved 
by the parallel transport with respect to the KZB connection on the 
restriction of the bundle $\m\tilde\CW\m$
of the CS states on the doubled surface to the subspace 
of the ``doubled'' complex structures and insertions. 
These properties are often summarized by saying 
that the boundary CFT is chiral since its
correlation functions are given by special conformal blocks 
of the WZW theory on $\m\tilde\Sigma$. It would 
be desirable to characterize geometrically the CS states 
$\,\Psi_{_{\un{\lambda}}}\m$. \,Some of their 
special properties are easy to find. For example, 
they are preserved by the antilinear involution 
$\,\Psi\mapsto{}^\iota\Psi\,$ of $\,\NW_{_{\tilde\Sigma}}(\un{\xi}, 
\un{\bar\xi},\un{R},\un{\ov{R}})\,$ induced by the involution 
$\,\iota\,$ of the doubled surface $\tilde\Sigma$ and defined by
\qq
{}^\iota\Psi(\tilde A^{01})\ =\ \ov{\Psi(-(\iota^*\tilde 
A^{01})^*)}\,.
\qqq
A complete geometric characterization of the states
$\,\Psi_{_{\un{\lambda}}}\,$ for different choices $\un{\lambda}$ 
of the BC's seems, however, still missing.

\subsection{Piece-wise boundary conditions}

Up to now, we have imposed the boundary conditions 
forcing the fields to take values in the special conjugacy
classes uniformly on the component circles of $\da\Sigma$.
Since the  conditions are local, it should be also possible
to do this locally on the pieces of the boundary. Suppose 
that the boundary $\da\Sigma$ is divided into 
intervals $I_r$ (the entire boundary circles are 
also admitted). We shall associate integrable weight 
labels $\lambda_r$ to some of these intervals in such a way 
that two labeled intervals in the same boundary component 
are separated by an unlabeled (``free'') one, 
see Fig.\,\,15. 

\leavevmode\epsffile[-80 -20 337 210]{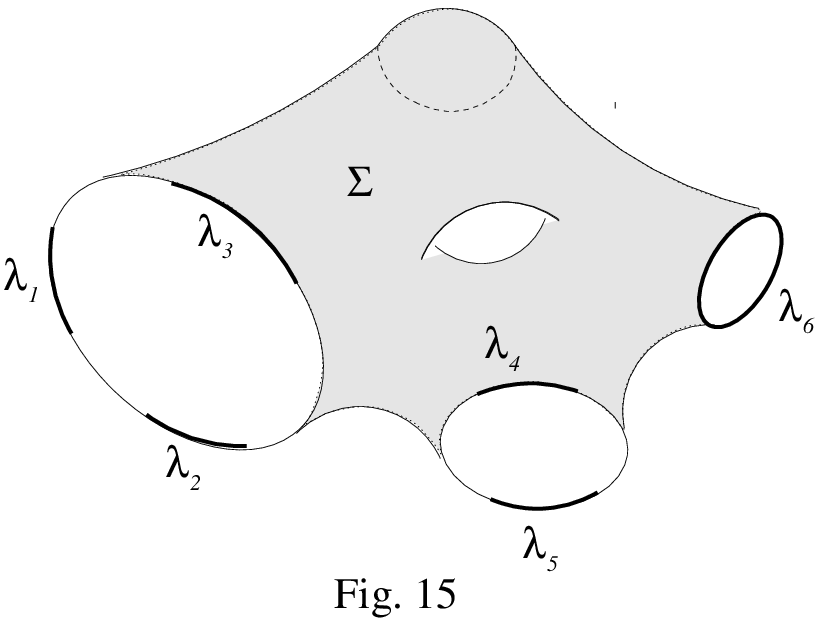}

\noindent We shall now consider the fields 
$g$ on $\Sigma$ which on the labeled intervals
take values in the corresponding conjugacy classes
$\CC_{\lambda_r}$ and are not restricted on the free 
intervals. One may still define the classical amplitudes 
$\,\ee^{-S^{{WZ}}(g)}\,$ for such fields although this 
requires a more local ($\check{{\rm C}}$ech cohomology type) 
technique than the one developed above 
\cite{Carg}. Let us sketch how this is done. 
\vskip 0.1cm

\leavevmode\epsffile[-37 -20 357 220]{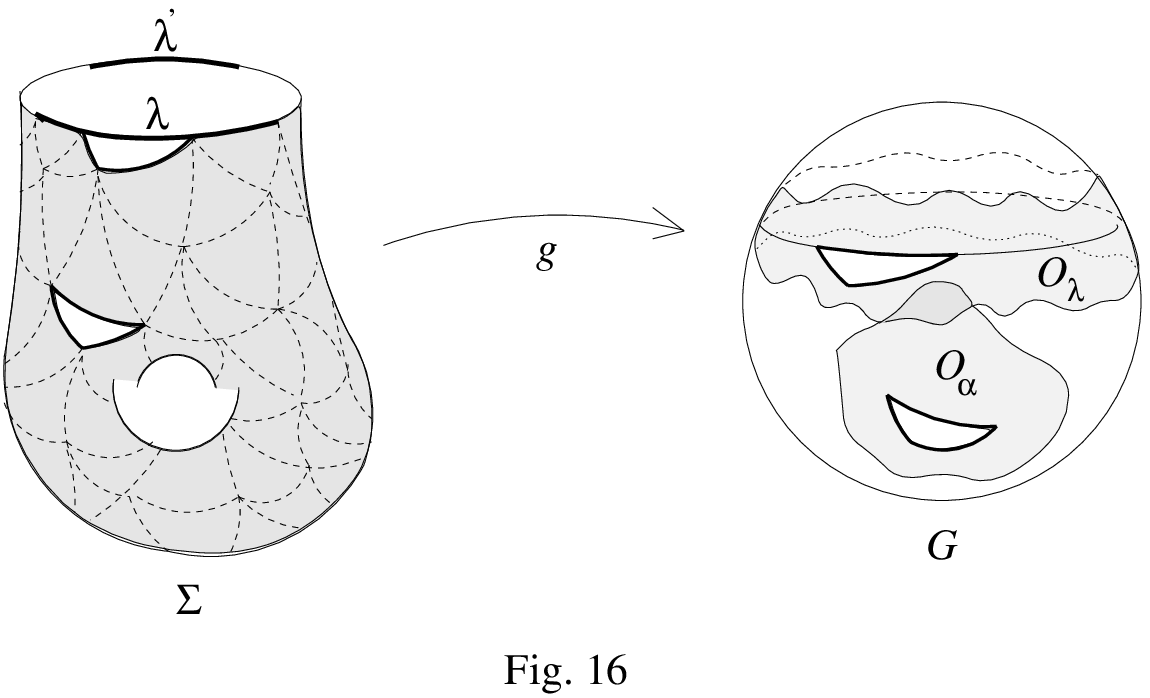}

Recall from the end of Sect.\,\,3.1 that $S^{^{WZ}}(g)$ in the 
first approximation is equal to $\,{k\over 4\pi i}\int g^*\beta$
where $\,d\beta=\chi\,$ is the canonical closed 3-form on $\m G$.
The problem stemmed from the fact that such 2-forms $\beta$ 
exist only locally. However, on the sets of a sufficiently 
fine open covering $(\CO_\alpha)$ of $G$,  we
may choose 2-forms \s$\beta_\alpha\s$ such that 
$d\beta_\alpha=\chi$. Choose a triangulation $\CT$ of $\Sigma$ 
with the triangles $t$, edges $e$ and vertices $v$. If 
$\CT$ is fine enough then each of the simplices $s$ of $\CT$
is mapped by $g$ into an open set, say $\CO_{\alpha_s}$, 
see Fig.\,\,16. The main contribution to the amplitude 
$\,\ee^{-S^{^{WZ}}_{_\Sigma}(g)}\,$
will come from $\,\exp[-{k\over4\pi i}\sum_{_t}
\hspace{-0.06cm}\int_{_t}g^*\beta_{\alpha_t}]\m.$
\,The above expression depends, however, on the choice of 
the forms $\,\beta_\alpha\m$ and of the triangulation. The 
idea is to compensate this dependence by contributions from
simplices of lower dimension. Let 
\s$\eta_{\alpha_{_0}\alpha_{_1}}=
-\eta_{\alpha_{_1}\alpha_{_0}}\s$ be 1-forms defined on 
the non-empty
intersections \s$\CO_{\alpha_{_0}\alpha_{_1}}
\equiv\CO_{\alpha_{_0}}\cap\CO_{\alpha_{_1}}\s$ such that
$$d\m\eta_{\alpha_{_0}\alpha_{_1}}\ =\ \beta_{\alpha_{_1}}-
\beta_{\alpha_{_0}}$$
and let $\,f_{\alpha_{_0}\alpha_{_1}\alpha_{_2}}\m$ 
be functions on the triple 
intersections $\,\CO_{\alpha_{_0}\alpha_{_1}\alpha_{_2}}\m$, 
\m antisymmetric in the indices, satisfying
$$d\m f_{\alpha_{_0}\alpha_{_1}\alpha_{_2}}\ =\ 
\eta_{\alpha_{_1}\alpha_{_2}}-\eta_{\alpha_{_0}
\alpha_{_2}}+\eta_{\alpha_{_0}\alpha_{_1}}$$
and such that on the four-fold intersections 
$$f_{\alpha_{_1}\alpha_{_2}\alpha_{_3}}-
f_{\alpha_{_0}\alpha_{_2}\alpha_{_3}}+f_{\alpha_{_0}
\alpha_{_1}\alpha_{_3}}-
f_{\alpha_{_0}\alpha_{_1}\alpha_{_2}}\in 8\pi^2\m\NZ\,.$$
Such data may, indeed, be chosen. 
We define then
\qq
\ee^{-S_{_\Sigma}^{^{WZ}}(g)}\,=\, 
\exp\Big[-{_k\over^{4\pi i}}\Big(\sum\limits_t\hspace{-0.06cm}
\int\limits_{t}\hspace{-0.06cm}g^*\beta_{\alpha_t}\m
-\m\sum\limits_{e\subset t}
\int\limits_e\hspace{-0.06cm}g^*\eta_{\alpha_e\alpha_t}\m+\m
\sum\limits_{v\in e\subset t}\hspace{-0.06cm}(\pm) 
f_{\alpha_v\alpha_e\alpha_t}(g(v))\Big)\Big]\m,\hspace{0.5cm}
\label{WZ2}
\qqq
where in the last sum the sign is taken according to the 
orientation of the vertices $v$ inherited from the triangles $t$ 
via the edges $e$. One may show that for the surface without 
boundary, the above expression does not depend on the choices 
involved and coincides with the definition given in Sect.\,\,3.3. 
\vskip 0.3cm

In the presence
of boundary circles with unconstrained fields, the above
expression may be used to define the amplitudes with 
values in a line bundle and it provides an alternative
construction of the bundle $\CL$ over the loop group \cite{Carg}.
In the presence of the boundary conditions on the intervals  
$I_r$ we shall still employ the same definition, but with 
some care. Namely, we include neighborhoods $\CO_\lambda$ 
of the conjugacy classes $\CC_\lambda$ into the open covering 
$(\CO_\alpha)$ of $G$. We choose 2-forms $\beta_\lambda$
on $\CO_\lambda$ coinciding with $\omega_\lambda$ 
of Eq.\,\,(\ref{om}) when restricted to $\m\CC_\lambda$. 
The triangulations used in Eq.\,\,(\ref{WZ2}) are 
required to be compatible with the splitting of the boundary. 
To the simplices in the labeled 
intervals $I_r$ we asign the open sets $\CO_{\lambda_r}$,
see Fig.\,\,16. 
The amplitudes resulting from Eq.\,\,(\ref{WZ2}) coincide 
then with those defined in the previous 
section for the special case when the labeled intervals
fill entire circles. In the general case, 
\qq
\ee^{-S^{^{WZ}}_{_\Sigma}(g)}\ \ \in\ \ 
\prod\limits_{{\rm free}\ I_r}
(\CL_{_{I_r}}
)_{_{g\vert_{_{I_r}}}}
\qqq
where $\CL_{_{I_r}}$ is a line bundle
over the space of maps from an interval $I_r$ to $G$
taking on the boundary of $I_r$ the values 
in the conjugacy classes $\CC_{\lambda_r}$ and
$\CC_{\lambda'_r}$
specified by the labels of the neighboring 
intervals\footnote{if $I_r$ is a full circle,
$\CL_{_{I_r}}=\CL$}.
\vskip 0.3cm

The space of sections $\Gamma(\CL_{_{I_r}})$ plays,
as before, the role of the space of states of the WZW theory
but, this time, on the interval and with boundary conditions
specified by the conjugacy classes $\CC_{\lambda_r}$
and $\CC_{\lambda'_r}$.
In the string language, these are states of the open string 
moving on the group with the ends on the {\bf branes} 
$\CC_{\lambda_r}$ and $\CC_{\lambda'_r}$, see Fig.\,\,17. 

\leavevmode\epsffile[-70 -20 310 156]{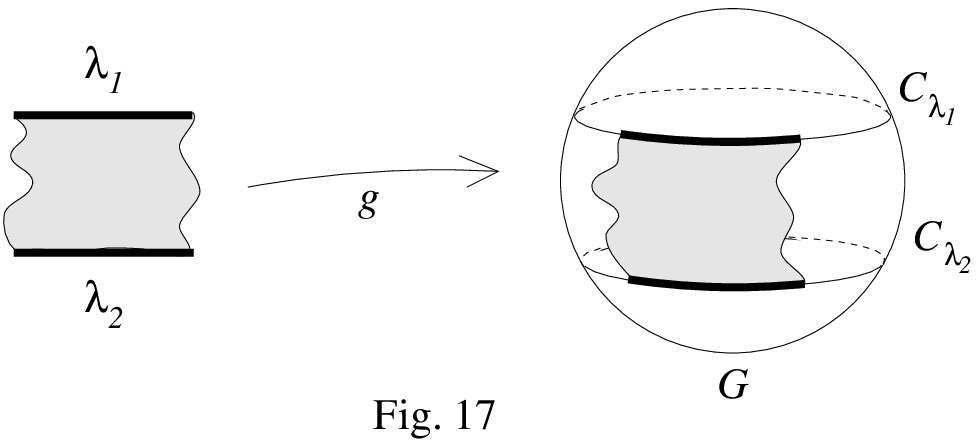}

\noindent One may still define an action of the central extension 
of the loop group in the spaces $\m\Gamma(\CL_{_{I_r}})\m$ 
(a single one) and base on its analysis
a rigorous construction of the open-string Hilbert 
spaces of states $\CH_{_{\lambda\m\lambda'}}\m$, 
\m as we did in Sect.\,\,4.1 for the closed-string
states, see Eq.\,\,(\ref{HSrd}). One obtains
\qq
\CH_{_{\lambda\m\lambda'}}\ =\ 
\mathop{\oplus}\limits_{_{\hat R}}\,
M_{\lambda\m\lambda'}^{^{\,R}}\otimes V_{_{\hat R}}\,.
\label{opHs}
\qqq
The multiplicity spaces may be naturally identified
with the spaces of the genus zero CS states 
$\CW(\NC P^1,\un{\xi},\un{R})$ with three insertion
points in representations $\ov{R}_{\lambda}$, $R_{\lambda'}$
and $R$. In particular, the dimension of the multiplicity
spaces is given by the fusion ring structure constants 
$\m{\hat N}^{\,\ov{R}}_{\ov{R}_{\lambda}R_{\lambda'}}$. 
The spaces $\CH_{_{\lambda\m\lambda'}}$
carry the obvious action of the current algebra $\hat\Ng$
and of the Virasoro algebra, the latter obtained 
by the Sugawara construction (\ref{sug}). The generator 
$L_0-{c\over 24}$ gives the Hamiltonian of the open string 
sectors. The spaces $\CH_{\lambda\lambda}$ with the same 
BC on both sides contain the vacua $\Omega_{_\lambda}$, 
i.e.\,\,the states annihilated by $L_0$ (unique up 
to normalization).
\vskip 0.4cm

\subsection{Elementary quantum amplitudes}

The quantum amplitudes with the general boundary conditions
are given now by the formal functional integrals.
\qq
\CA_{_{\Sigma,\un{I},\un{\lambda}}}\hspace{-0.09cm}(A)\ \ =\ 
\int\limits_{g(I_r)\m\subset\m\CC_{\lambda_r}}
\ee^{-S_{_\Sigma}(g,\m A)}\,\, Dg
\qqq
and, should take values in the space
$\,\mathop{\otimes}
\limits_{{\rm free}\ I_r}\CH_{_{\lambda_r\lambda'_r}}\m$.
\,They should possess a gluing property 
along free boundary intervals with
opposite boundary weight assignment, generalizing
the gluing properties (\ref{glueB}) or (\ref{glueC}). 
As discussed in detail by Segal in \cite{Segal} 
for the closed string sector,
the general amplitudes may be constructed by gluing 
from the elementary ones for the geometries listed 
on Fig.\,\,18. 

\leavevmode\epsffile[-40 -10 345 290]{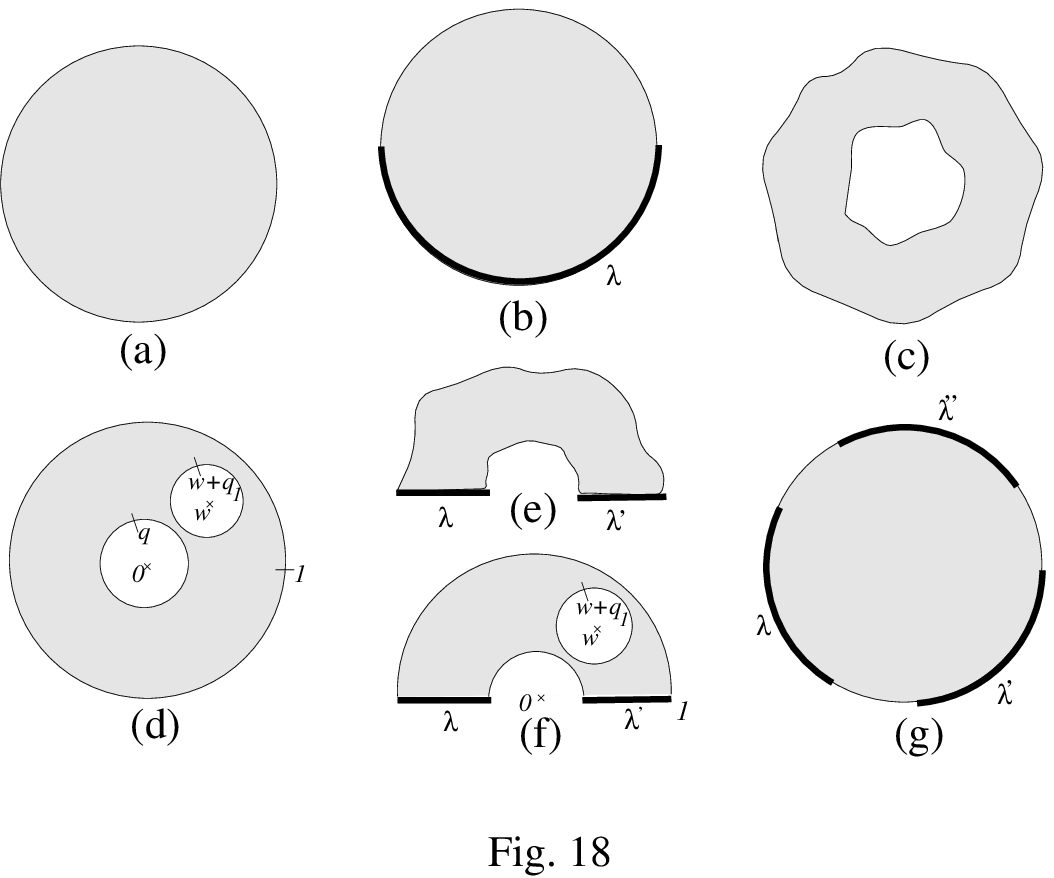}
\vskip 0.1cm

The elementary amplitudes (a) and (b) represent, respectively, 
the vacuum state $\m\Omega\m$ in the closed string space $\CH$, 
and the vacua $\m\Omega_{_\lambda}$ in the open string 
spaces $\CH_{_{\lambda\lambda}}$.
The amplitudes (c) for arbitrary annuli encode the 
action of the pair of Virasoro algebras in $\CH$.
In particular, for a complex number $q\not=0$ inside
the unit disc one may consider the annular regions
$\,A_q=\{\,z\,\m\vert\m\,
\vert q\vert\leq\vert z\vert\leq1\,\}\m$, \m see Fig.\,\,19,  
obtained from $\m\NC P^1$ by taking out the unit discs 
embedded by the maps $z\mapsto qz$ and $z\mapsto z^{-1}$.

\leavevmode\epsffile[-53 -20 310 165]{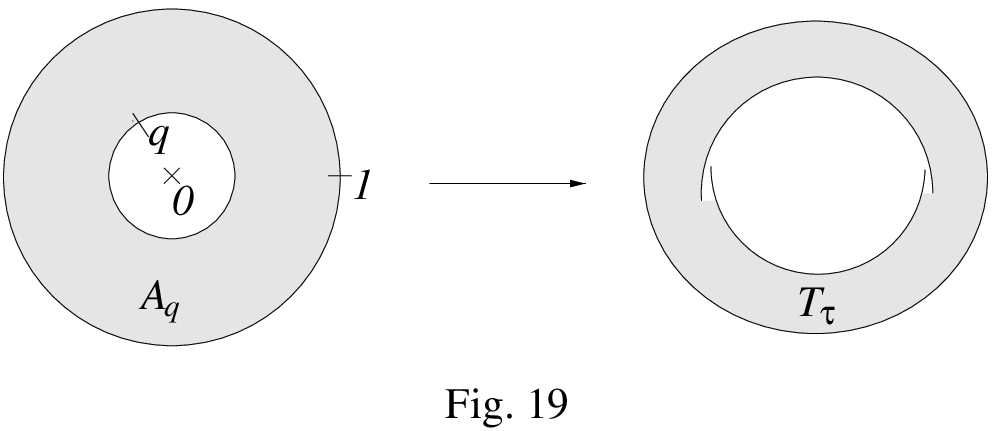}

\noindent Viewing the amplitude of $\m A_q$ as 
an operator from the space $\CH$ associated to the first 
boundary component to $\CH$ associated to the second 
boundary, one has: 
\qq
\CZ_{_{A_q}}^{-1}\,\,\CA_{_{A_q}}\ =
\ q^{\m L_0}\m{\bar q}^{\m\tilde L_0}\,.
\non
\qqq
The gluing of the two boundary circles of $\m A_q$ leads  
to the complex torus $\m{T}_\tau$ where 
$q=\ee^{\m 2\pi i\m\tau}$.
According to the gluing relation (\ref{glueC}), this 
produces the toroidal partition function  
\qq
\CZ({T}_\tau)\ =\ \CZ_{_{A_q}}\ \tr_{_\CH}\,(q^{\m L_0}\m
{\bar q}^{\m\tilde L_0})\,.
\qqq
Upon choosing a flat metric on ${T}_\tau$ and  
working out the partition function\footnote{$\CZ_{_{A_q}}$ 
is a ratio of two partition function on the Riemann sphere 
and may be easily found from the relation (\ref{scin})
to be equal to $\m\vert q\vert^{-{c\over12}}$}
$\CZ_{_{A_q}}$, one obtains finally
\qq
\CZ(\tau)\ =\ \tr_{_{\CH}}\,q^{\m L_0-{c\over 24}}\,
{\bar q}^{\m\tilde L_0-{c\over 24}}
\qqq
which is nothing else but Eq.\,\,(\ref{diag}).
\vskip 0.3cm

The amplitude for a disc $\m P_{w,q,q_1}$ with two round 
holes, as in Fig.\,\,18(d), gives rise to a 3-linear form 
on $\CH$ which may be also viewed as an operator from 
the space $\CH\otimes\CH$ associated to the inner discs 
to $\CH$ corresponding to the outer one. It is customary 
in CFT to rewrite this amplitude as an operator $\Phi(e;\m w)$ 
in $\CH$ labeled by the vectors $e$ in (a dense subspace of) 
$\CH$ and the point $w$ inside the unit disc:
\qq
\Phi\m(e;\m w)\, e'\ 
=\ \CZ_{_{P_{w,q,q_1}}}\,\,\CA_{_{P_{w,q,q_1}}}
\,(q^{-L_0}\m{\bar q}^{\m-\tilde L_0}\m e')\otimes 
(q_1^{-L_0}\m{\bar q_1}^{\m-\tilde L_0}\m e)\,.
\qqq
The combination with the powers of $L_0$ and $\tilde L_0$ 
assures the independence of the expression of $q$ and $q_1$. 
The vectors $e$ can be recovered from the operators
$\m\Phi(e;\m w)\m$ by acting with them on the vacuum
vector
\qq
\lim\limits_{w\to0}\,\Phi(e;\m w)\,\Omega\ =\ e\,.
\qqq
Pictorially, this corresponds to filling up the
central whole of $P_{w,q,q_1}$ by gluing a disc to its
boundary. The operators $\m\Phi(e;\m w)\m$ satisfy an important
relation:
\qq
\Phi\m(e;\m z)\,\,\Phi(e';\m w)
\ =\ \Phi(\m\Phi(e;\m z-w)\,e';\,w)\,.
\label{OPEgl}
\qqq
The above identity holds for $\m0<|w|<|z|\m$ and $\m0<|z-w|<1$.
It results from the two ways that one may obtain the disc with 
three holes by gluing two discs with two holes, see Fig.\,\,20.

\leavevmode\epsffile[-65 -20 320 160]{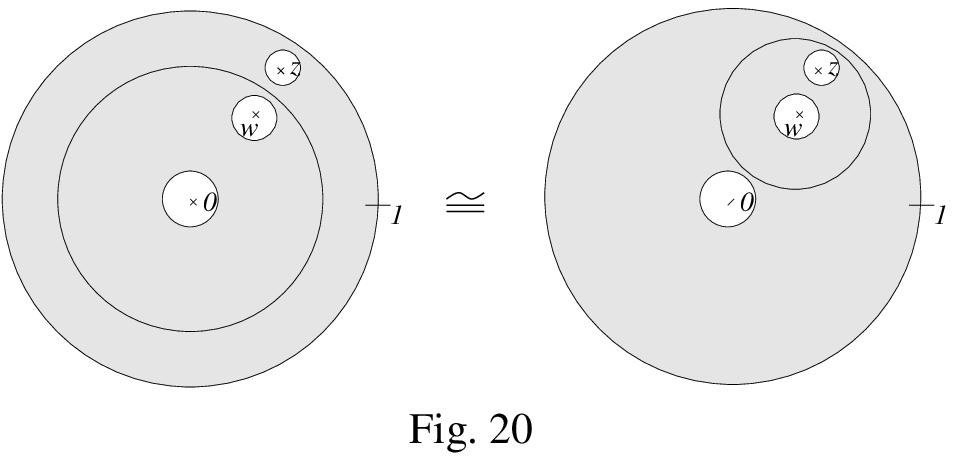}

\noindent The relation (\ref{OPEgl}) may be viewed as a
global form of the operator product expansion.
The local forms may be extracted from it by expanding 
the vector $\,\Phi(e;\m z-w)\,$ into terms homogeneous 
in $\m(z-w)$. \m In particular,
for specially chosen vectors $e$ and $e'$ one obtains
the operator versions of the relations
(\ref{fope}) and (\ref{opem}), hence the name of the latter.
The vector-operator correspondence together with the 
operator product expansion (\ref{OPEgl}) 
are the cornerstones of the non-perturbative 
approach to CFT. 
\vskip 0.3cm

The amplitudes corresponding to the surfaces with
boundary of Fig.\,\,18(e) represent the action 
of the Virasoro algebra in the open string spaces 
$\m\CH_{_{\lambda\lambda'}}$. 
The surfaces (f) give rise, in turn, to the amplitudes 
which, applied to vectors $\m q_1^{-L_0}
{\bar q_1}^{\m-\tilde L_0}\m e\in\CH\m$, \,define 
the action of the closed string sector fields 
$\,\Phi(e,\m w)\,$ in the open string
spaces $\m\CH_{_{\lambda\lambda'}}$. Finally,
the amplitudes of the disc (g) with three
labeled and three free boundary intervals
define 3-linear forms on the corresponding
open string spaces. As before in the closed
string sector, one may interpret them in terms
of boundary operators labeled by vectors
in, say, $\m\CH_{_{\lambda''\lambda}}$
and mapping from $\m\CH_{_{\lambda\lambda'}}$
to $\m\CH_{_{\lambda'\lambda''}}$. 
\vskip 0.3cm

The gluing properties give rise to non-trivial 
relations between various amplitudes. For example,
gluing along the free sides a rectangle
with a local BC imposed on the two other sides, 
see Fig.\,\,21, one obtain a finite cylinder $Z_{_L}$ 
with the BC's imposed on the boundary components.

\leavevmode\epsffile[-44 -20 330 140]{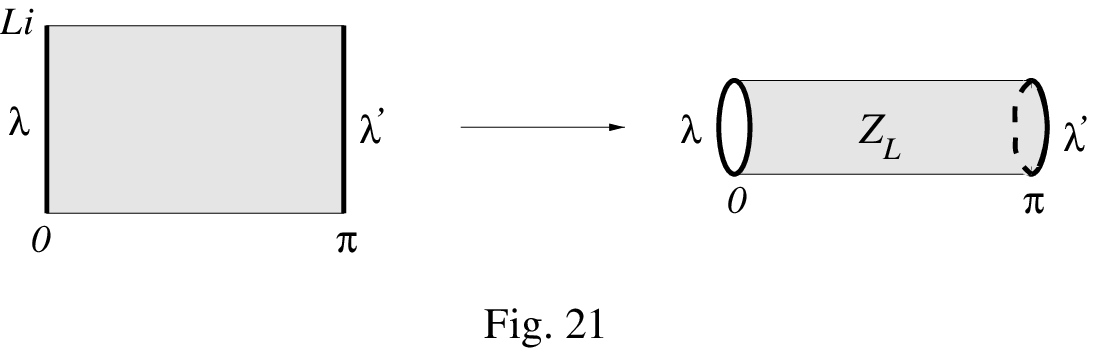}

\noindent Its amplitude $\m\CA_{_{Z_{_{L}}}}$ (in 
the flat metric) may be computed in two ways. On one 
hand side, using the decomposition (\ref{opHs}), 
we infer that
\qq
\CA_{_{Z_{_{L}}}}\ =\ \tr_{_{H_{_{\lambda\lambda'}}}}\,  
q^{L_0-{c\over 24}}\ =\ \sum\limits_{\hat{R}}\,
{\hat N}^{\,\ov{R}}_{\ov{R}_{\lambda}R_{\lambda'}}\,\,
\tr_{_{V_{_{\hat{R}}}}}\,q^{L_0-{c\over 24}}
\ =\ \sum\limits_{\hat{R}}\,
{\hat N}^{\,\ov{R}}_{\ov{R}_{\lambda}R_{\lambda'}}\,\,
\chi_{_{\hat{R}}}(\tau,\m1)\hspace{0.4cm}
\label{Car1}
\qqq
with $\tau={L\m i\over2\pi}$ and $q=\ee^{\m 2\pi i\m\tau}$.
On the other hand, we may express this amplitude
as a matrix element of the close string amplitude 
between the boundary states $\hat{\delta}_{_\lambda}$ and
$\hat{\delta}_{_{\lambda'}}$. With $\,q'=\ee^{\m2\pi i\m\tau'}$
and $\,\tau'=-{1\over\tau}={2\pi i\over L}$, \,we obtain
\qq
\CA_{_{Z_{_{L}}}}\ =\ \Big(\m\hat{\delta}_{_{\lambda}}\m,
\,\m(q')^{\hf(L_0-{c\over24})}\m
(\ov{q'})^{\hf(\tilde L_0-{c\over24})}\,
\m\hat{\delta}_{_{\lambda'}}\m\Big)\,.
\qqq
Upon the substitution of Cardy's expression (\ref{Card}) 
for the boundary states $\hat{\delta}_{_\lambda}$, this 
becomes
\qq
\CA_{_{Z_{_{L}}}}
&=&
\sum\limits_{\hat{R},\m{\hat{R}}'}
\m(S^{^{1}}_{_{R}})^{-\hf}\,
S^{^{\ov{R}_\lambda}}_{_{R}}
\,(S^{^{1}}_{_{R'}})^{-\hf}\,
S^{^{R_{\lambda'}}}_{_{R'}}
\m\,\Big(\m e^{\hat{i}}_{_{\hat R}} 
\otimes\overline{e^{\hat{i}}_{_{\hat R}}}\,,
\,\,(q')^{\hf(L_0-{c\over24})}\m
(\ov{q'})^{\hf(\tilde L_0-{c\over24})}\,
e^{{\hat{i}}'}_{_{{\hat{R}}'}} 
\otimes\overline{e^{{\hat{i}}'}_{_{{\hat{R}}'}}}\m\Big)
\hspace{0.4cm}\cr
&=&
\sum\limits_{{\hat{R}}'}
\,(S^{^{1}}_{_{R'}})^{-1}\,
S^{^{\ov{R}_\lambda}}_{_{R'}}
\,S^{^{R_{\lambda'}}}_{_{R'}}
\,\m\Big(e^{\hat{i}}_{_{\hat R'}}\m, 
\,(q')^{\hf(L_0-{c\over24})}
\,e^{{\hat{i}}'}_{_{{\hat{R}}'}}\Big)\,
\Big(e^{{\hat{i}}'}_{_{{\hat{R}}'}}\m,
\m\,(\ov{{q'}})^{\hf(L_0-{c\over24})}
\,e^{\hat{i}}_{_{{\hat R}'}}\Big)\cr
&=&
\sum\limits_{{\hat{R}}'}
\,(S^{^{1}}_{_{R'}})^{-1}\,
S^{^{\ov{R}_\lambda}}_{_{R'}}
\,S^{^{R_{\lambda'}}}_{_{R'}}
\,\,\tr_{_{V_{_{{\hat R}'}}}}\,(q')^{L_0-{c\over24}}
\ =\ 
\sum\limits_{{\hat{R}}'}
\,(S^{^{1}}_{_{R'}})^{-1}\,
S^{^{\ov{R}_\lambda}}_{_{R'}}
\,S^{^{R_{\lambda'}}}_{_{R'}}
\,\,\chi_{_{{\hat R}'}}(\tau',\m 1)\,.
\qqq
With the use of the modular transformation property
(\ref{modm}), we finally obtain:
\qq
\CA_{_{Z_{_{L}}}}\ =\ 
\sum\limits_{\hat{R},\m{\hat R}'}
\,(S^{^{1}}_{_{R'}})^{-1}\,
S^{^{\ov{R}_\lambda}}_{_{R'}}
\,S^{^{R_{\lambda'}}}_{_{R'}}
\,S^{^{R}}_{_{R'}}\,\,\chi_{_{\hat R}}(\tau,\m 1)\,.
\non
\qqq
By virtue of the Verlinde formula (\ref{orV}),
the last identity coincides with Eq.\,\,(\ref{Car1}).
We have, in fact, inverted here the logic of reference 
\cite{Cardy}, where the consistency of the two ways 
of computing the amplitude $\,\CA_{_{Z_{_{L}}}}$ was used
to obtain the expression (\ref{Card}) for the boundary 
states $\,\hat{\delta}_{_\lambda}$.
\vskip 0.3cm

The whole system of elementary amplitudes represents 
an intriguing algebraic structure which is
common to all (rational) boundary CFT's. Already the case 
of boundary topological field theories, where the amplitudes 
depend only on the surface topology, leads to an interesting
construction that remains to be fully understood. It 
entangles a commutative algebra structure on the closed 
string space of states and a non-commutative algebroid 
in the open string sector. An example of such a structure
was inherent in the work of Kontsevich \cite{Konts2} 
on the deformation quantization of general 
Poisson manifolds, see \cite{Felder}. Certainly,
the two-dimensional CFT did not unveal yet all of its
secrets.

\end{document}